\documentclass[
reprint,
aps,
prb,
twocolumn,
showpacs,
amsmath,
amssymb,
floatfix, superscriptaddress 
]{revtex4-2}
\usepackage{graphicx}
\usepackage{dcolumn}
\usepackage{mathrsfs}
\usepackage{bm}
\usepackage[normalem]{ulem}
\usepackage{overpic}
\usepackage{amsfonts}
\usepackage{amssymb}
\usepackage{bbm}
\usepackage{xcolor}
\usepackage[colorlinks=true,linkcolor=blue, citecolor=red, urlcolor=magenta]{hyperref}
\usepackage{url}
\DeclareFontFamily{U}{stix2bb}{}
\DeclareFontShape{U}{stix2bb}{m}{n} {<-> stix2-mathbb}{}

\NewDocumentCommand{\indicator}{}{\text{\usefont{U}{stix2bb}{m}{n}1}}

\begin{document}

\title{\large
Quantum transport and unified scaling law in graphene with\\[2pt]
polyadic Cantor electrostatic barriers
}

\author{Sudhanshu Shekhar}
\email{shekhar1997sudhanshu@gmail.com, \\sudhanshushekhar@bhu.ac.in}
\affiliation{Department of Physics, Banaras Hindu University, Varanasi 221005, INDIA}
\author{Mohammad Umar}
\email{aliphysics110@gmail.com, mohd.umar@opc.iitd.ac.in
}
\affiliation{Optics and Photonics Centre, Indian Institute of Technology Delhi, New Delhi 110016, INDIA
}
\author{Bhabani Prasad Mandal} 
\email{bhabani.mandal@gmail.com, bhabani@bhu.ac.in}
\affiliation{Department of Physics, Banaras Hindu University, Varanasi 221005, INDIA}
\date{\today}
\begin{abstract}
We study the quantum transport of Dirac electrons in graphene subjected to a polyadic Cantor-structured electrostatic potential. Using the superperiodic potential formalism, we obtain a closed-form expression for the transmission probability. As the Cantor stage increases, the transmission spectrum evolves from a sparse set of superlattice resonances to a near-transparent regime, with the polyadic order setting the rate of this evolution. The angular response depends on the doping configuration, showing distinct behavior in the $n$--$n$--$n$, $n$--$p$--$n$, and Dirac-point cases. In the near-transparent regime, the transmission follows double-logarithmic scaling laws with respect to four independent control parameters: the Cantor stage, the potential height, the initiator length, and the angle of incidence. By combining these individual scaling relations, we establish a unified scaling law governing quantum transport. These results show that the hierarchical self-similarity of the potential governs the transport properties of such systems.
\end{abstract}
\maketitle
%
\section{Introduction}
Graphene has emerged as one of the most prominent platforms for studying quantum transport phenomena owing to its exceptional electronic, mechanical, and optical properties~\cite{novoselov2012roadmap, mbayachi2021graphene, urade2023graphene, tiwari2020graphene}. As a single atomic layer of carbon atoms arranged in a two-dimensional honeycomb lattice with $sp^2$ hybridization~\cite{choi2010synthesis}, graphene hosts low-energy charge carriers that behave as massless Dirac electrons~\cite{novoselov2004electric}. This relativistic-like dispersion leads to unconventional transport properties, most notably Klein tunneling, in which carriers transmit through electrostatic barriers without backscattering at normal incidence~\cite{katsnelson2006chiral, allain2011klein}. Graphene thus provides an ideal setting for exploring the interplay between relativistic quantum dynamics and externally imposed potential landscapes. Motivated by these properties, charge transport in graphene has been studied extensively under a wide range of electrostatic environments. Periodic potentials and graphene superlattices have received particular attention~\cite{park2008anisotropic, gumbs2014revealing, brey2009emerging, ohta2012evidence, li2024transport, ahmad2024transverse}. Transport across $p$--$n$ and $p$--$n$--$p$ junctions formed by electrostatic gating has also been widely investigated~\cite{huard2007transport, williams2007quantum, oostinga2008gate, gorbachev2008conductance, liu2008fabrication, cheianov2006selective, abanin2007quantized, xu2010electron}. Related studies have addressed transport under magnetic modulation, strain, and other modulated potentials~\cite{novoselov2005two, zhang2005experimental, zhang2005electric, berger2004ultrathin, wang2021global, azarova2014transport, pakdel2021transport, zhan2013transfer}. These studies have revealed that quantum interference, pseudospin conservation, and phase coherence play decisive roles in shaping transport characteristics in graphene-based systems.\\
\indent
In parallel, hierarchical and fractal potentials have long been known to exhibit rich spectral and transport properties in wave systems. Among these, the Cantor set constitutes a paradigmatic example of a self-similar structure generated through an iterative removal process. Cantor-inspired potentials have been widely employed to study wave propagation, localization, and transmission scaling in both quantum and classical contexts~\cite{konotop1990wave, sun1991wave, hasan2020tunneling, bertolotti1996transmission, chiadini2003self, hatano2005strong, hasan2018tunneling, monsoriu2005transfer, honda2006rigorous, chuprikov2008transfer, esaki2009wave, sakaguchi2017scaling, ogawana2018transmission, hasan2018new}. A characteristic feature of such systems is the emergence of nontrivial scaling behavior in the transmission/reflection spectra, arising from the hierarchical structure of the Cantor potential~\cite{sakaguchi2017scaling, ogawana2018transmission, chiadini2003self}.\\
\indent
The \textit{standard} Cantor construction is obtained by dividing the initial region into three equal parts and removing the middle third, with the same procedure recursively repeated for all the remaining segments at each iteration. This construction can be generalized to form a broader class of hierarchical architectures. In the general(ized) Cantor (GC) scheme, a fraction $1/\rho$ ($\rho>1$) is removed from the center of each segment at every iteration, while in the generalized Smith--Volterra--Cantor (GSVC or SVC-$\rho$) construction~\cite{narayan2023tunneling, singh2025quantum}, the removed fraction depends on the iteration stage. It is important to note that, for the Cantor system, the number of surviving segments at the first stage is $N=2$, where a single removal is performed. This construction can be generalized to a polyadic Cantor system, where $N$ surviving segments at the first stage are obtained through $N-1$ systematic removals. The same procedure is then successively repeated at each stage for all the remaining segments. This yields dyadic ($N=2$), triadic ($N=3$), and higher-order polyadic Cantor potentials. Thus, a characteristic feature for identifying a Cantor system is the number of surviving segments, $N$, at the first stage~\cite{umar2025polyadic}. Moreover, leveraging the geometry of the Cantor system, a Cantor-structured Dirac comb potential is constructed, and quantum tunneling through it is investigated~\cite{umar2025transmission}.\\
\indent
The application of Cantor and Cantor-like geometry to graphene has opened new avenues for exploring fractal-induced spectral and transport phenomena. Self-similarity has been demonstrated in the transmission spectra of graphene with fractal quasiperiodic barriers~\cite{rodriguez2016self} and in the conductance of Cantor-like graphene structures~\cite{garcia2017self}. Studies have also examined electron transmission in graphene systems modulated by Cantor pre-fractal magnetic barriers and wells~\cite{sun2010transport}, tunable multi-Dirac-point transport and robust zero-$\bar{k}$ gaps in Cantor graphene superlattices~\cite{xu2015electronic}, and transport through Cantor-type graphene structures under magnetoelectric modulation~\cite{rodriguez2017transmission}. Beyond transport through Cantor-based graphene systems, quantum electronic transport in graphene Sierpinski carpets (GSCs) has also been investigated~\cite{bouzerar2020quantum}. Recently, fractal electronic transport in graphene Cantor systems has also been investigated through power-spectrum analysis, possibly for the first time~\cite{rodriguez2025fractal}. The concept of superperiodic potential (SPP), introduced as a generalization of locally periodic potential~\cite{hasan2018super}, has recently been employed to investigate quantum transport phenomena in graphene-based systems~\cite{shekhar2024relativistic,shekhar2025klein}. Fractal geometry also manifests in graphene-based systems beyond scattering phenomena, such as in Sierpinski graphene structures~\cite{pedersen2020graphene} and moiré fractals in twisted bilayer graphene~\cite{aggarwal2024moire}.\\
\indent
Despite these advances, a systematic understanding of the scaling behavior of the transmission profile through graphene modulated by Cantor geometric electrostatic potentials remains incomplete. While self-similar features have been reported~\cite{rodriguez2016self, garcia2017self}, explicit scaling laws governing the transmission probability across different Cantor stages and other control parameters of Cantor geometry have not been fully explored. Polyadic Cantor potentials (PCPs) are especially attractive in this context because they constitute a natural generalization of the Cantor potential, with the polyadic order $N$ introducing an additional structural parameter. Moreover, low-lacunarity PCPs constitute a special class of SPPs~\cite{umar2025polyadic}, allowing the transmission probability to be obtained in closed form within the SPP framework. These features make PCPs a suitable fractal construct for investigating scaling behavior and hierarchical transport phenomena in graphene systems.\\
\indent
In this work, we investigate quantum transport and its scaling laws in graphene subjected to low-lacunarity polyadic Cantor electrostatic barriers, for which the transmission probability is obtained in closed form. We analyze the transmission characteristics as functions of the Fermi energy, the scaling parameter, and the angle of incidence. As the Cantor stage increases, the transmission evolves from a sparse set of superlattice resonances, separated by a suppressed window near the Dirac point, to a near-transparent regime in which the spectrum is close to unity and structured only by narrow interference dips. The polyadic order sets the rate of this evolution. The angular response differs between the $n$--$n$--$n$ ($E>V_0$), $n$--$p$--$n$ ($E<V_0$), and Dirac-point ($E=V_0$) configurations.\\
\indent
The near-transparent regime, reached at higher Cantor stages, is the natural setting in which scaling behavior emerges, and it is the central focus of the present work. We show that the quantity $\ln(-\ln T)$ follows double-logarithmic scaling laws with respect to four independent control parameters: the Cantor stage $S$, the potential height $V_0$, the initiator length $L$, and the angle of incidence $\phi$. In each case, the scaling arises from the dependence of the transmission probability on the corresponding parameter; varying the parameter shifts the double-logarithmic transmission curve while the resonance structure remains unchanged. The four single-parameter scaling laws are then combined into a unified multi-parameter scaling law, which demonstrates that the geometry and the external control parameters together govern the transmission through a single self-similar mechanism. These findings provide new insight into the role of hierarchical geometry in relativistic quantum transport and suggest routes for designing graphene-based devices with tunable transmission properties.\\
\indent
The paper is organized as follows. In Section~\ref{section_02}, we introduce the PCP and discuss its superperiodic structure within the SPP framework. Section~\ref{section_03} analyzes the transmission probability of Dirac electrons in graphene subjected to polyadic Cantor-structured electrostatic barriers, together with the associated scaling laws. We examine the transmission characteristics in different regimes, including the $n$--$n$--$n$, $n$--$p$--$n$, and the Dirac point configurations. We then derive and analyze scaling laws for the transmission probability and formulate a unified multi-parameter scaling law. Finally, Section~\ref{section_04} summarizes the main findings and presents the conclusions of this work.
\section{Polyadic Cantor potential}
\label{section_02}
The PCP system, a member of the fractal potential family, builds on the foundational Cantor potential framework. Fractal systems are typically described by their fractal dimension~\cite{mandelbrot1983fractal, theiler1990estimating}, but since multiple fractals can share the same fractal dimension, it alone is insufficient for fully characterizing fractals. Thus, exploring alternative descriptors, such as lacunarity~\cite{jaggard1997polyadic, jaggard1998scattering, villatoro2008tunneling, umar2025polyadic}, becomes essential. Lacunarity provides insight into the spatial distribution of features within fractals, classifying them as low, medium, or high lacunarity. In this work, we focus on the low-lacunarity PCP system, as it constitutes a special case of the SPP framework.
\subsection{Polyadic Cantor potential system}
The \textit{standard} Cantor potential is a fundamental example within the fractal family, originating from a closed interval $[0,1]$ at stage $S=0$. In quantum potential formalism, this spatial interval is occupied by a potential of height $V_0$, where at each successive \emph{Cantor stage} $S$, a central fraction of $\frac{1}{3}$ of the potential segment is removed. Further, this concept can be generalized to any interval $[0, L]$, where $L$ is referred to as the \emph{parent length} (the width of the potential at stage $S=0$). Extending this idea to the GC potential system, a fraction $\frac{1}{\rho}$, rather than $\frac{1}{3}$, is removed from the middle of the potential segments at each stage. Consequently, at stage $S=1$, the potential is divided into two segments, at stage $S=2$ into four segments, and at stage $S=3$ into eight segments. More generally, each stage $S$ contains $2^S$ potential segments. In the GC system, the \emph{polyadic order} $N=2$ specifies the presence of two potential segments at stage $S=1$, a configuration known as the \emph{dyadic} Cantor potential.\\
\begin{figure}[t]
\includegraphics[width=\columnwidth]{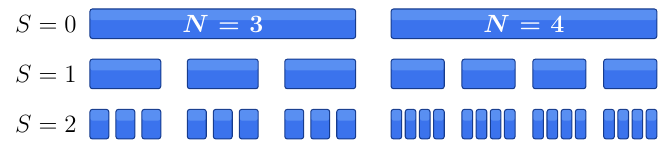}
\caption{Schematic illustration of the PCP geometry for polyadic order $N=3$ (left) and $N=4$ (right).}
\label{pcp_system}
\end{figure}
\indent
The PCP system is the generalization of the GC potential~\cite{umar2025polyadic}, where the polyadic order is $N \geq 2$. Consequently, for $N = 3$, the system consists of three potential segments of equal width at stage $S=1$, and for $N = 4$, it contains four segments at the same stage. This pattern continues for different values of $N$. In this construction, it is crucial to avoid system transparency. Hence, for a given polyadic order $N$, the \emph{scaling parameter} $\rho$ must satisfy $\rho > N-1$. Each stage $S$ consists of $N^{S}$ potential segments of equal length. Figure~\ref{pcp_system} illustrates the PCP system for $N = 3$, and $N = 4$. The system exhibits periodicity at stage $S = 1$, which is lost for stages $S > 1$, where a superperiodic system can be seen~\cite{hasan2018super}. This indicates a transition from a periodic to a superperiodic system as the stage transitions from $S = 1$ to $S > 1$. It is easy to show that the length of the potential segment at each stage is expressed through~\cite{umar2025polyadic}
\begin{equation}
b_{S}=\frac{L}{N^{S}}\left(1-\frac{N-1}{\rho}\right)^{S}.
\label{bs}
\end{equation}
An important feature of this construction is that, at stage $S$, a fraction $\frac{1}{\rho}$ of the length of each potential segment at stage $S-1$ is removed. As a result, this method differs from the conventional PCP construction, where a fraction $\frac{1}{\rho}$ of the initiator length $L$ is removed, rather than a fraction $\frac{1}{\rho}$ of the segment length $b_{S-1}$. For any given polyadic order $N$, a total of $N-1$ portions are removed from each potential segment. These removals are performed so that the remaining segments have equal lengths and are symmetrically distributed about the center.
\subsection{Superperiodic construction of the polyadic Cantor potential}
The SPP~\cite{hasan2018super} is a generalized form of a locally periodic potential. The construction of an SPP begins with a \textit{unit cell} potential, denoted as $V_0$, with width $d_0$. A periodic structure is generated by repeating this unit cell $N_1$ times, with \emph{superperiodic distance} $r_1$ such that $r_1 \geq d_0$, yielding the periodic system $V_1 \equiv (V_0, N_1, r_1)$. The total span of this periodic system is given by $d_1 = d_0 + (N_1 - 1)r_1$. Next, the potential $V_1$ is treated as a new unit cell and is repeated $N_2$ times, with spacing $r_2 \geq d_1$, forming the potential $V_2 \equiv (V_1, N_2, r_2)$. The total span of this system $V_2$ is then $d_2 = d_1 + (N_2 - 1)r_2$. Similarly, $V_2$ is used as a new unit cell to construct $V_3$, which is repeated $N_3$ times with spacing $r_3 \geq d_2$. This gives the potential $V_3 \equiv (V_2, N_3, r_3)$, and the total span of $V_3$ is $d_3 = d_2 + (N_3 - 1)r_3$. 
This process continues iteratively: for each step, the potential $V_q$ is constructed by repeating the previous potential $V_{q-1}$ $N_{q}$ times, with spacing $r_q \geq d_{q-1}$, and the total span at each stage is calculated as $d_q = d_{q-1} + (N_{q} - 1)r_{q}$. This describes the iterative construction of the SPP of \emph{order} $q$, in which each successive potential is obtained by repeating the previous potential structure at progressively larger scales, thereby generating increasingly complex superperiodic structures. The superperiodic formalism of Cantor potentials is explored in detail in~\cite{hasan2018super, singh2023quantum, umar2023quantum, narayan2023tunneling, singh2025quantum, umar2025polyadic, umar2025transmission}. This construction process is summarized in Table~\ref{table01}.\\
\begin{figure}[b]
\includegraphics[width=0.484\textwidth]{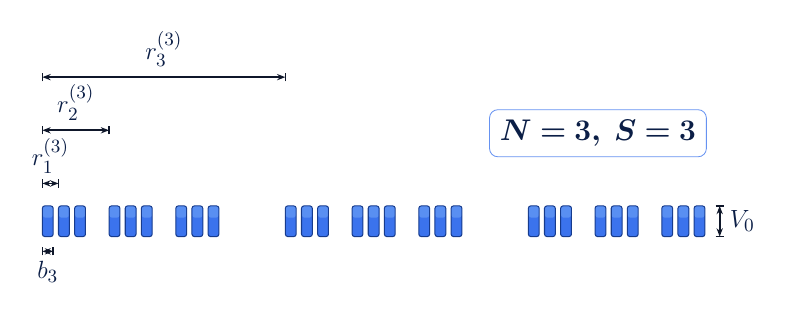}
\caption{Construction of the PCP system of polyadic order $N=3$ at stage $S=3$ through the superperiodic formalism. It starts with a unit cell of width $b_{3}$, and $r_{1}^{(3)}$, $r_{2}^{(3)}$, and $r_{3}^{(3)}$ are the superperiodic distances of the first, second, and third orders, respectively. }
\label{figure_02}
\end{figure}
\begin{table*}[htb]
\caption{\label{table01} Structural parameters of the SPP at successive orders.}
\renewcommand{\arraystretch}{1.25}
\begin{ruledtabular}
\begin{tabular}{ccccc}
SPP order & Unit cell & Superperiodic repetition & Unit cell width & Superperiodic distance \\
\colrule
1 & $V_0$ & $N_1$ & $d_0$ & $r_1$ \\
2 & $V_1\equiv(V_0,N_1,r_1)$ & $N_2$ & $d_1=d_0+(N_1-1)r_1$ & $r_2$ \\
3 & $V_2\equiv(V_1,N_2,r_2)$ & $N_3$ & $d_2=d_1+(N_2-1)r_2$ & $r_3$ \\
4 & $V_3\equiv(V_2,N_3,r_3)$ & $N_4$ & $d_3=d_2+(N_3-1)r_3$ & $r_4$ \\
$\vdots$ & $\vdots$ & $\vdots$ & $\vdots$ & $\vdots$ \\
$q$ & $V_{q-1}\equiv(V_{q-2},N_{q-1},r_{q-1})$ & $N_q$ & $d_{q-1}=d_{q-2}+(N_{q-1}-1)r_{q-1}$ & $r_q$\\
\end{tabular}
\end{ruledtabular}
\end{table*}
\indent
The low-lacunarity PCP system is a special case of the SPP: the PCP at Cantor stage $S$ is an SPP of order $S$, in which the unit cell is repeated $N$ times at each of the $S$ successive levels. The intermediate structures generated along the way are themselves SPPs of order $q = 1, 2, \ldots, S$. We accordingly refer to $q$ as the \emph{superperiodic order} of the SPP or, equivalently, the \emph{hierarchical level} of the PCP, and use the two terms interchangeably throughout this work. Figure~\ref{figure_02} shows the construction of the PCP system of stage $S=3$ for polyadic order $N=3$ through the superperiodic formalism. We start with a potential of height $V_{0}$ and width $b_{3}$. This unit cell potential is further periodically repeated $N_{1}=3$ times at the superperiodic distance $r_{1}^{(S=3)}(L, \rho)\ge b_{S=3}$. Subsequently, this system as a whole is periodically repeated $N_{2}=3$ times at a superperiodic distance $r_{2}^{(3)}(L, \rho)\ge d_{1}=b_{3}+(N_{1}-1)r_{1}^{(3)}$, creating a system of nine potential segments. Now, consider this whole system as a unit cell and periodically repeat this $N_{3}=3$ times at a superperiodic distance $r_{3}^{(3)}(L, \rho)\ge d_{2}=d_{1}+(N_{2}-1)r_{2}^{(3)}$. This system represents the PCP system of stage $S=3$ for $N=3$. This implies that any Cantor stage of the PCP system can be created by the superperiodic formalism. For the PCP construction, the polyadic order fixes $N_q = N$ at every hierarchical level. Mathematically, it is easy to show that the superperiodic distance $r_{q}^{(S)}(L, \rho)$ is expressed through~\cite{umar2025polyadic}
\begin{equation}
r_{q}^{(S)}(L, \rho)=\frac{L}{N^{S+1-q}}\left(1+\frac{1}{\rho}\right)\left(1-\frac{N-1}{\rho}\right)^{S-q}.
\label{rq}
\end{equation}
The superperiodic distance depends on the initiator length $L$, which is the potential width at stage $S = 0$, and the scaling parameter $\rho$. This relationship defines $r_q^{(S)}(L, \rho)$ in such a way that, at every Cantor stage $S$, the potential system remains confined within the interval $[0, L]$. 

It should be noted that the two constructions scale in contrasting manners. In the general SPP, the unit cell width $d_0$ is fixed and the total span $d_q$ increases with each successive order, so that an SPP of order $q$ is contained within one of order $q+1$. In the PCP, the total span is fixed at the initiator length $L$, and increasing the Cantor stage instead refines the structure inward: both the unit cell width $b_S$ and the superperiodic distances $r_q^{(S)}$ depend on $S$ through Eqs.~\eqref{bs} and~\eqref{rq}. Consequently, the order-$q$ SPP substructure of a stage-$S$ PCP is not a stage-$q$ PCP. Successive Cantor stages are therefore not nested in the way successive SPP orders are. This concludes the discussion on the PCP system and its construction using an alternative method, namely the superperiodic method, rather than the traditional segment removal method.
%
\section{Transmission and scaling law}
\label{section_03}
\begin{figure}[b]
\begin{center}
\includegraphics[width=0.485\textwidth]{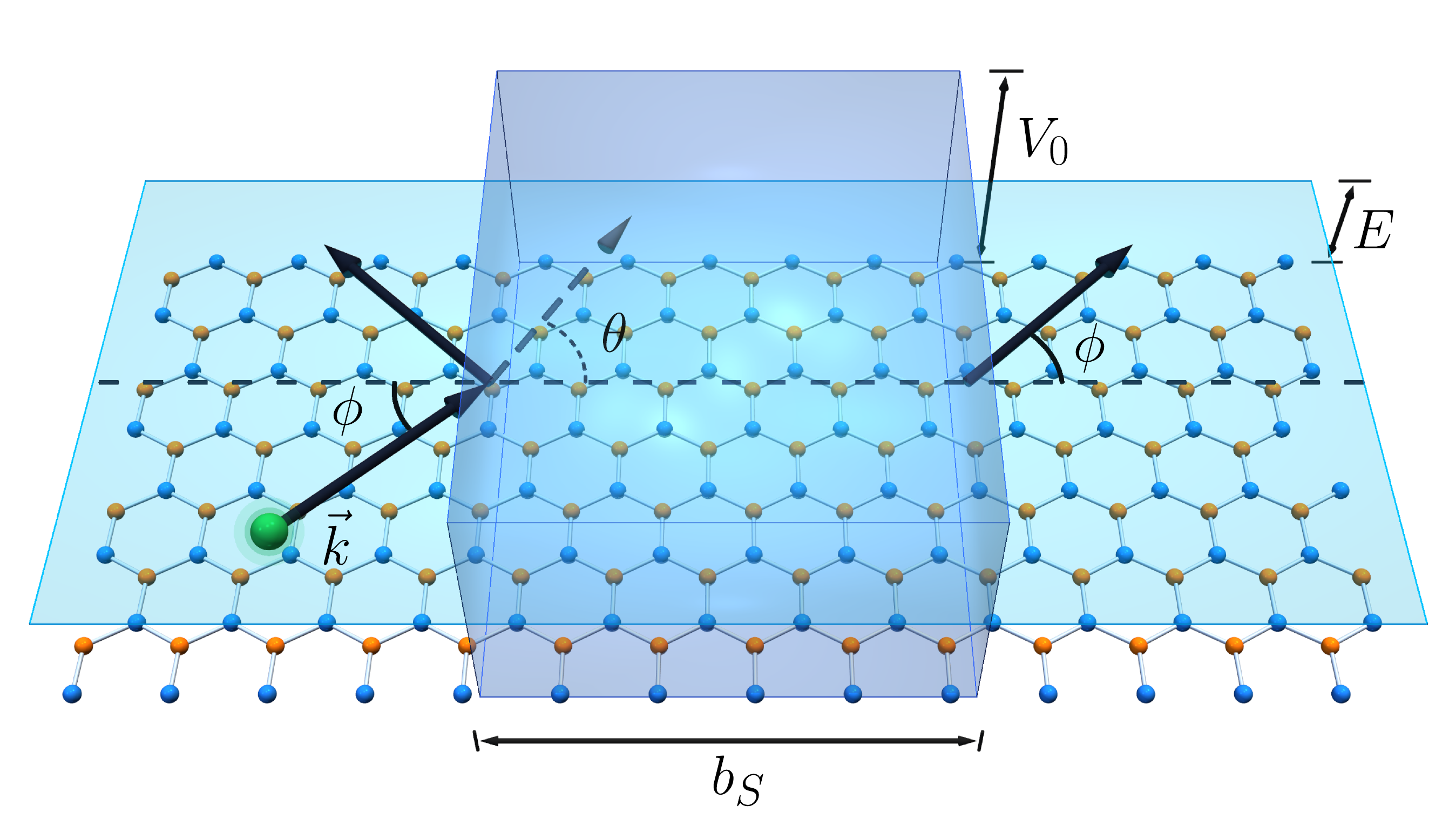} 
\caption{Schematic illustration of a graphene monolayer subjected to an electrostatic barrier of height $V_0$. The vertical height denotes the energy scale, the Fermi level is shown as a horizontal plane, and the arrows show the scattering direction of charge carrier through the barrier.}
\label{fig:graphene_scat}
\end{center}
\end{figure}
In this section, we investigate the transmission behavior of Dirac electrons in graphene through a polyadic Cantor-structured electrostatic potential, and then analyze its scaling behavior. Within the SPP formalism, if the transfer matrix $M$ of the unit cell potential is known,
\begin{equation}
M = \begin{pmatrix}
M_{11} & M_{12} \\
M_{21} & M_{22}
\end{pmatrix},
\end{equation}
the transmission probability through an SPP structure of order $q$ is given by~\cite{hasan2018super}
\begin{equation}
T(N_{1}, N_{2}, \ldots, N_{q}) = \frac{1}{1 + |M_{12}|^{2}\, \displaystyle\prod_{i=1}^{q} U_{N_{i}-1}^{2}(\Gamma_{i})},
\label{t_spp}
\end{equation}
where $U_{n}(\Gamma)$ denotes the Chebyshev polynomial (CP) of the second kind, and the \emph{Chebyshev argument} $\Gamma_{q}$ is defined as one-half of the trace of the transfer matrix at the superperiodic order $q$. This polynomial can also be expressed in terms of a sinusoidal function as
\begin{equation}
U_{n}(\Gamma) = \frac{\sin[(n+1)\gamma]}{\sin\gamma},
\qquad \gamma = \cos^{-1}\Gamma,
\label{cheb_trig}
\end{equation}
which holds for any non-negative integer $n$. Substituting this trigonometric form into Eq.~\eqref{t_spp} and specializing to the polyadic Cantor case, the transmission probability becomes
\begin{equation}
T_{S}(N) = \frac{1}{1 + |M_{12}^{(S)}|^{2}\,
\displaystyle\prod_{q=1}^{S} \frac{\sin^{2}(N\gamma_{q}^{(S)})}
{\sin^{2}\gamma_{q}^{(S)}}},
\label{tn1n2ns}
\end{equation}
where the single-barrier transfer matrix and the Bloch phases depend on the Cantor stage through the geometric quantities $b_S$ and $r_q^{(S)}$. Each factor in the product of Eq.~\eqref{tn1n2ns} has the form $\sin^{2}(Ny)/\sin^{2}y$ with $y = \gamma_{q}^{(S)}$, which is the well-known Laue function from X-ray diffraction theory~\cite{kittel2018introduction, cullity2001elements, warren1990x}. The Chebyshev arguments $\Gamma_{q}^{(S)}$ in Eq.~\eqref{tn1n2ns} represent the Bloch phases of the corresponding fully developed periodic system at hierarchical level $q$, and are given by~\cite{hasan2018super, umar2023quantum, umar2025polyadic}
\begin{equation}
\begin{split}
\Gamma_{q}^{(S)}(k) &= |M_{22}^{(S)}|\,
\cos\!\left[\tau_S - k\!\left\{(N-1)\sum_{p=1}^{q-1}
r_{p}^{(S)} - r_{q}^{(S)}\right\}\right] \\
&\quad \times \prod_{p=1}^{q-1} U_{N-1}(\Gamma_{p}^{(S)}) \\
&\quad + \sum_{h=1}^{q-1} 
\cos\!\left\{k\!\left(N\sum_{p=h}^{q-1} r_{p}^{(S)} 
- \sum_{p=h+1}^{q} r_{p}^{(S)}\right)\right\} \\
&\quad \times U_{N-2}(\Gamma_{h}^{(S)})
\prod_{p=h+1}^{q-1} U_{N-1}(\Gamma_{p}^{(S)}).
\end{split}
\label{eq17}
\end{equation}
where $\tau_S$ represents the argument of the matrix element $M_{22}^{(S)}$. Equation~\eqref{eq17} holds for all $q \ge 2$ directly. To extend its validity to the boundary case $q = 1$: any sum whose lower limit exceeds its upper limit is taken to be zero, and any product whose lower limit exceeds its upper limit is taken to be unity. This concludes the formulation of the transmission coefficient through the PCP system under the SPP formalism. To study the transmission features for a polyadic Cantor-structured electrostatic barrier applied to the monolayer graphene, we next need the transfer matrix for a single electrostatic barrier in graphene.
\indent
Consider a monolayer graphene sheet in the $xy$ plane, subjected to a rectangular electrostatic barrier of height $V_0$ that is infinite along the $y$-direction. The barrier spans a width $b_{S}$ along the $x$-direction, which will serve as the unit cell width for the PCP of stage $S$ in monolayer graphene:
\begin{equation}
V(x) =
\begin{cases} 
V_{0}, &  0 < x < b_S \\
0, & \text{otherwise}.
\end{cases}
\end{equation}
A schematic illustration of electron scattering through the graphene-based potential is shown in Fig.~\ref{fig:graphene_scat}. For an electron incident at an angle $\phi$, the corresponding scattering dynamics are governed by the massless Dirac equation~\cite{katsnelson2006chiral, shekhar2024relativistic}
\begin{equation}\label{eq:Direc_eq}
\left[-i\hbar v_f\,\boldsymbol{\sigma}\cdot\nabla + V(x)\right]\psi(x,y)
= E\,\psi(x,y),
\end{equation}
where $v_f \approx 10^6$ m/s is the Fermi velocity, $\boldsymbol{\sigma} = (\sigma_x, \sigma_y)$ are the Pauli matrices, and $E$ is the Fermi energy of the incident Dirac electron. Solving Eq.~\eqref{eq:Direc_eq} in the three regions gives the two spinor components
\begin{align}
\psi_1(x, y) &=
\begin{cases}
(\mathscr{A} e^{iq_x x} + \mathscr{B} e^{-iq_x x}) e^{iq_y y}, & x < 0 \\
(\mathscr{E} e^{i\widetilde{q}_x x} + \mathscr{F} e^{-i\widetilde{q}_x x}) e^{iq_y y}, & 0 < x < b_S \\
(\mathscr{C} e^{iq_x x} + \mathscr{D} e^{-iq_x x}) e^{iq_y y}, & x > b_S
\end{cases} \\
\psi_2(x, y) &=
\begin{cases}
s (\mathscr{A} e^{iq_x x + i \phi} - \mathscr{B} e^{-iq_x x - i \phi}) e^{iq_y y}, & x < 0 \\
\widetilde{s} (\mathscr{E} e^{i\widetilde{q}_x x + i \theta} - \mathscr{F} e^{-i\widetilde{q}_x x - i \theta}) e^{iq_y y}, & 0 < x < b_S \\
s (\mathscr{C} e^{iq_x x + i \phi} - \mathscr{D} e^{-iq_x x - i \phi}) e^{iq_y y}, & x > b_S.
\end{cases}
\end{align}
Here $(q_x, q_y) = (q_f \cos\phi, q_f \sin\phi)$ are the components of the Fermi wavevector $q_f =  E / \hbar v_f$ outside the barrier, while inside the barrier the longitudinal wavevector is
\begin{equation}
\widetilde{q}_x = \sqrt{\frac{(E - V_0)^2}{\hbar^2 v_f^2} - q_y^2},
\label{vec_inside}
\end{equation}
$\theta = \tan^{-1}(q_y / \widetilde{q}_x)$ is the refraction angle, and $s = \operatorname{sgn}(E)$ and $\widetilde{s} = \operatorname{sgn}(E - V_0)$ denote the band index (electron-like for $+1$, hole-like for $-1$) outside and inside the barrier, respectively. The transfer matrix for the given system is obtained by applying the continuity conditions of the wave function at the boundaries $x = 0$ and $x = b_S$. The corresponding transfer matrix elements are as follows:  
\begin{eqnarray}
M_{11}^{(S)} &=& \frac{e^{iq_x b_S}}{s\widetilde{s}} \left[ s \widetilde{s} \cos \widetilde{q}_x b_S - i \mathcal{A} \sin \widetilde{q}_x b_S \right], \label{m11}\\
M_{12}^{(S)} &=& \frac{e^{-i(2 q_x b_S + \phi)}}{2 s \widetilde{s}} \mathcal{B} \sec\theta \sin \widetilde{q}_x b_S, \label{m12} \\
M_{21}^{(S)} &=& \frac{e^{i(2 q_x b_S + \phi)}}{2 s \widetilde{s}} \mathcal{B} \sec\theta \sin \widetilde{q}_x b_S, \label{m21}\\
M_{22}^{(S)} &=& \frac{e^{-iq_x b_S}}{s\widetilde{s}} \left[ s \widetilde{s} \cos \widetilde{q}_x b_S + i \mathcal{A} \sin \widetilde{q}_x b_S \right], \label{m22}
\end{eqnarray}
where $\mathcal{A} = \sec\theta \sec\phi - s \widetilde{s} \tan\theta \tan\phi$ and $\mathcal{B} = 2 \tan\phi - 2 s \widetilde{s} \sin\theta \sec\phi$. From the transfer matrix elements presented above, it is evident that $M_{11}^{(S)}={\left(M_{22}^{(S)}\right)}^{*}$ and $M_{12}^{(S)}={\left(M_{21}^{(S)}\right)}^{*}$. With this unit-cell transfer matrix established, we will next investigate the transmission behavior of electrons in a graphene monolayer subjected to PCP.
\subsection{Transmission features}
\begin{figure*}[htbp]
\centering
\begin{overpic}[width=0.245\linewidth]{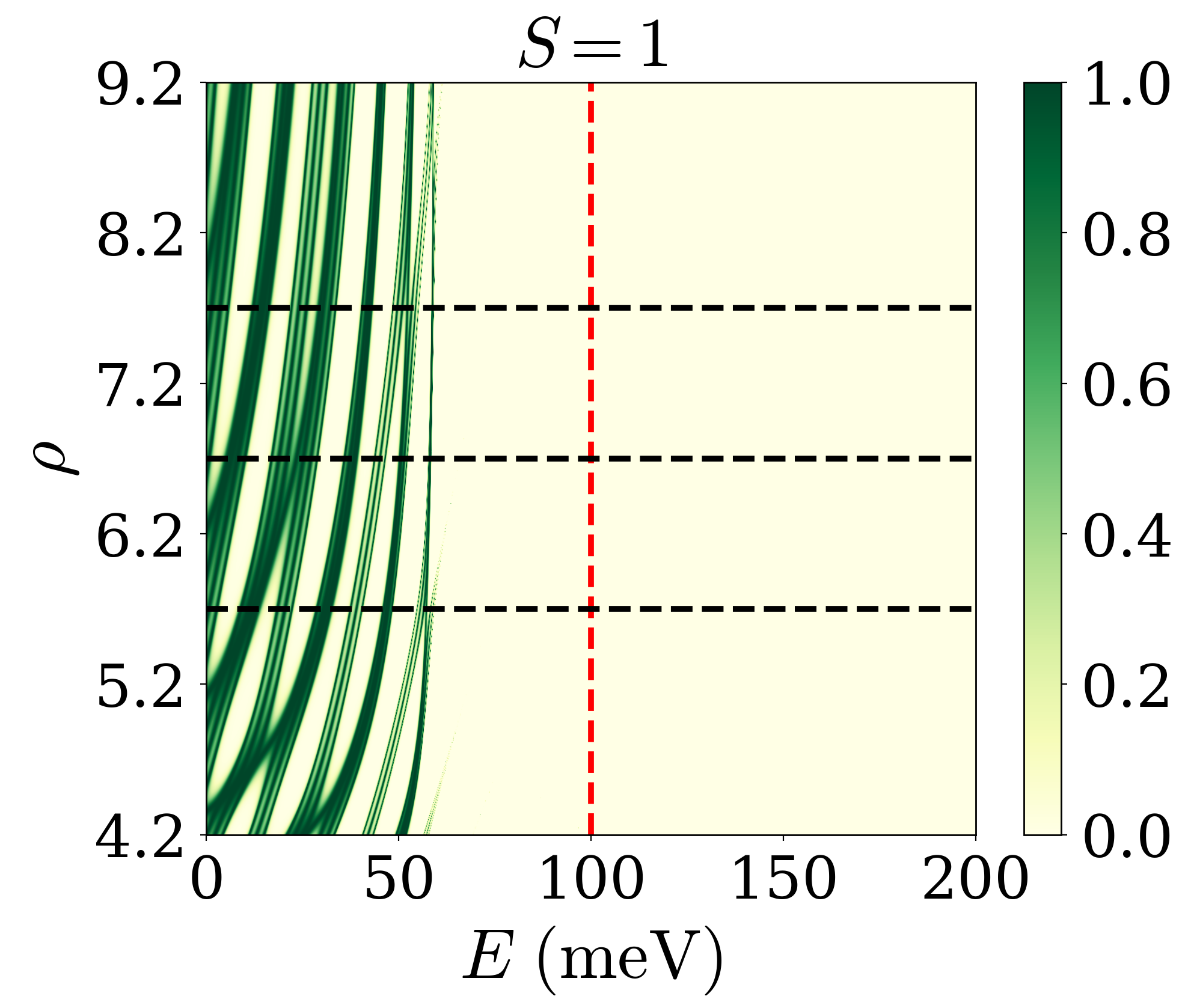}
\put(89,1.5){\bfseries (a)}
\end{overpic}%
\begin{overpic}[width=0.245\linewidth]{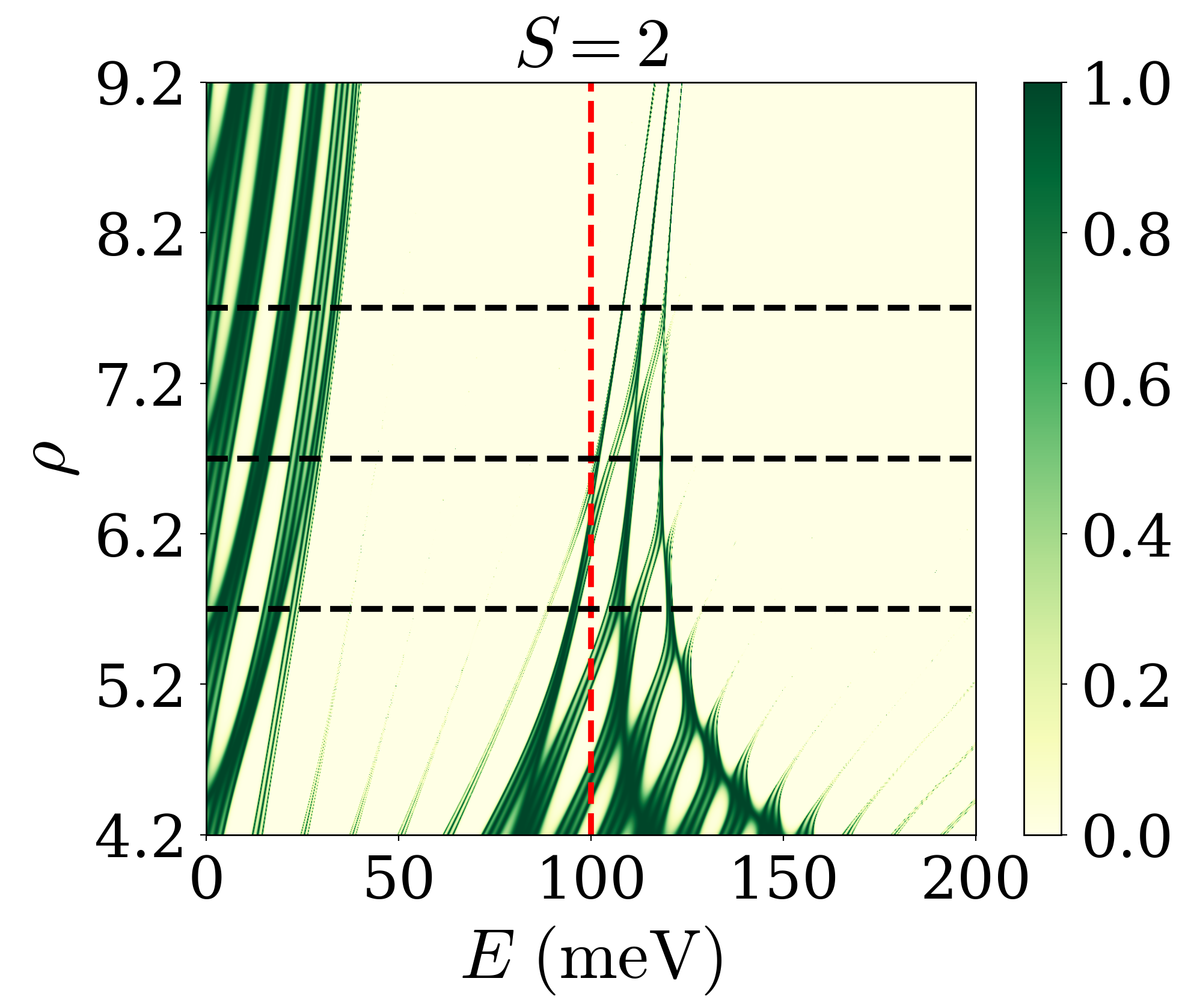}
\put(89,1.5){\bfseries (b)}
\end{overpic}%
\begin{overpic}[width=0.245\linewidth]{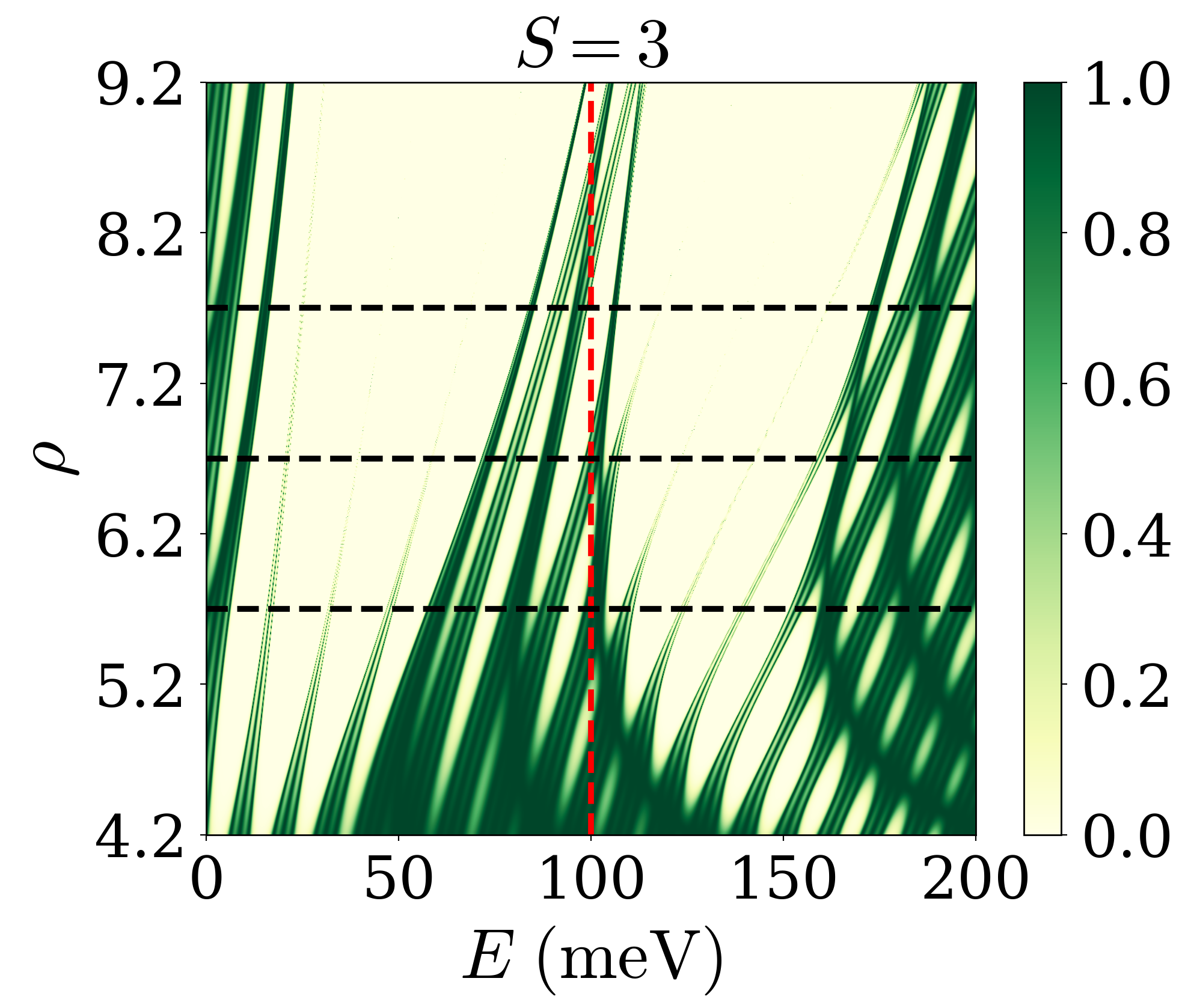}
\put(89,1.5){\bfseries (c)}
\end{overpic}%
\begin{overpic}[width=0.245\linewidth]{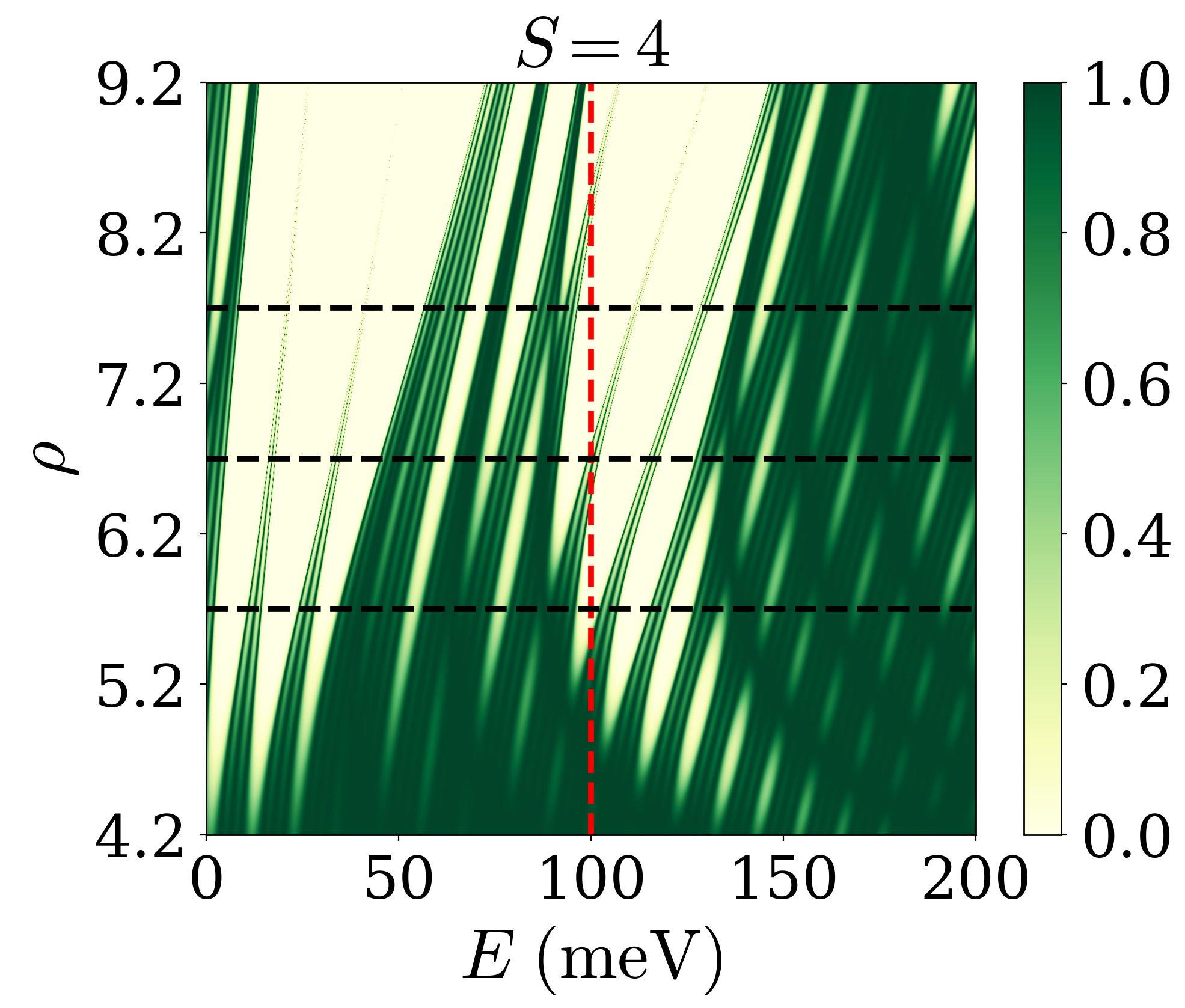}
\put(89,1.5){\bfseries (d)}
\end{overpic}
\begin{overpic}[width=0.245\linewidth]{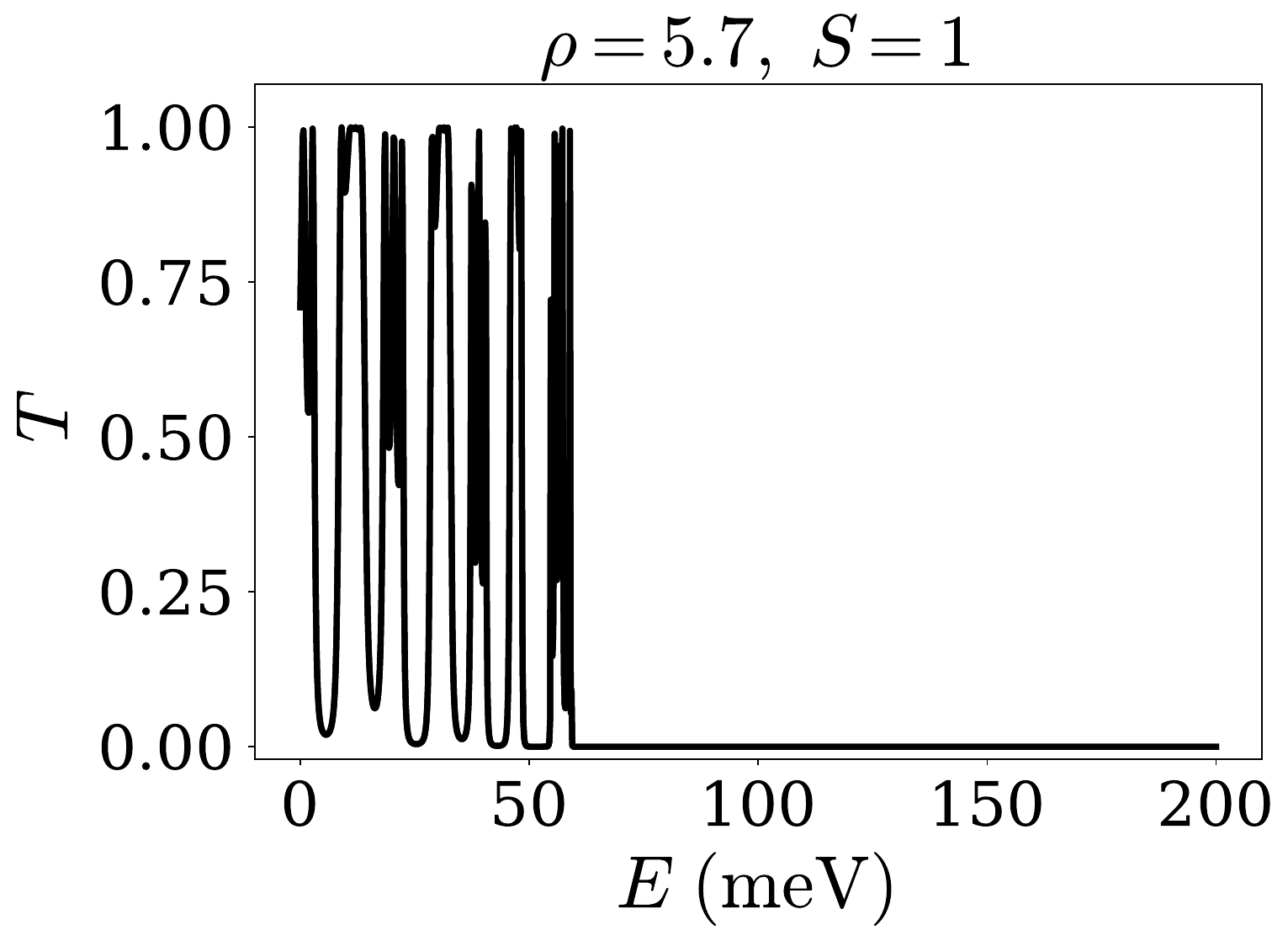}
\put(89, 1.5){\bfseries (e)}
\end{overpic}%
\begin{overpic}[width=0.245\linewidth]{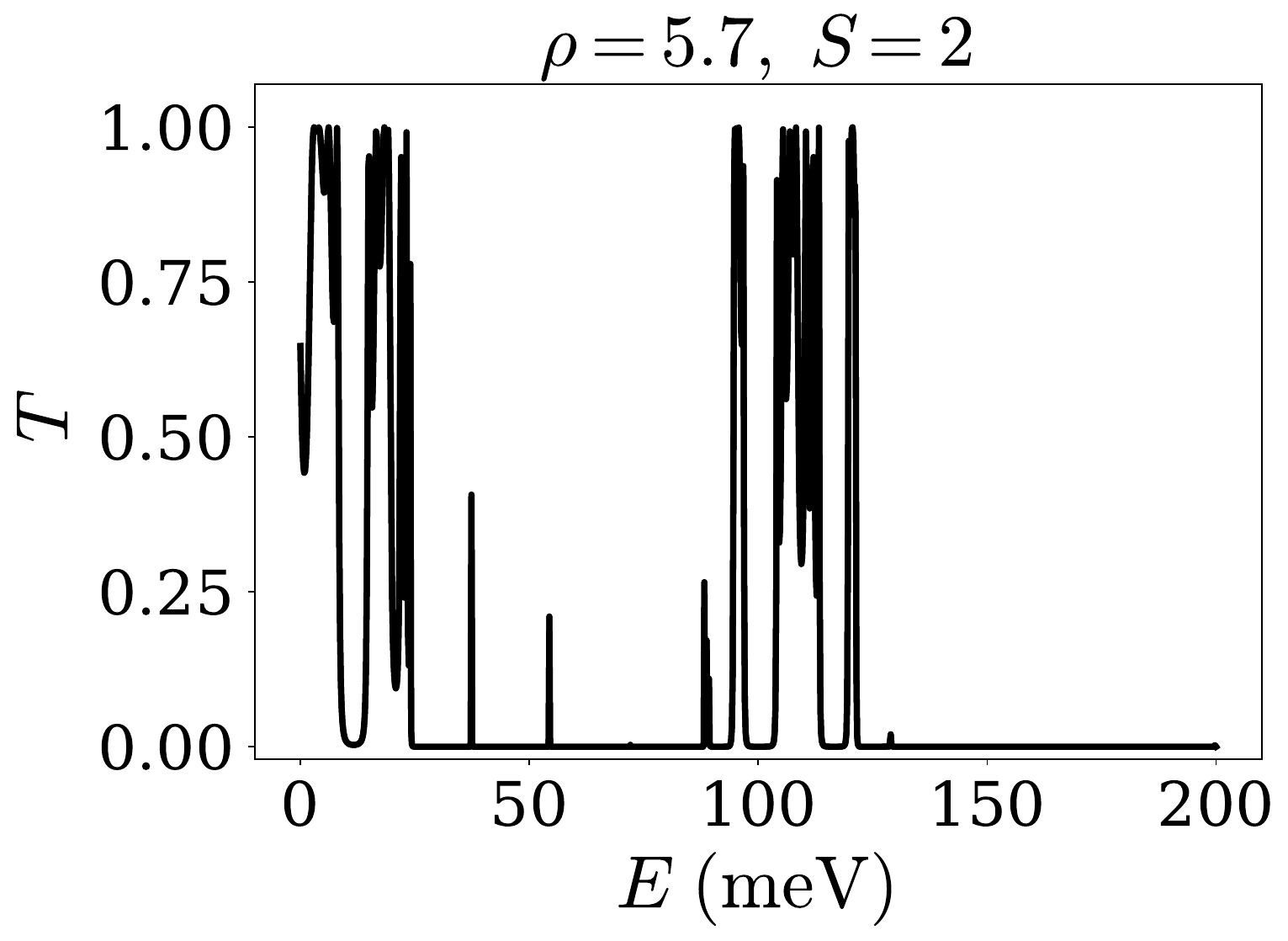}
\put(89, 1.5){\bfseries (f)}
\end{overpic}%
\begin{overpic}[width=0.245\linewidth]{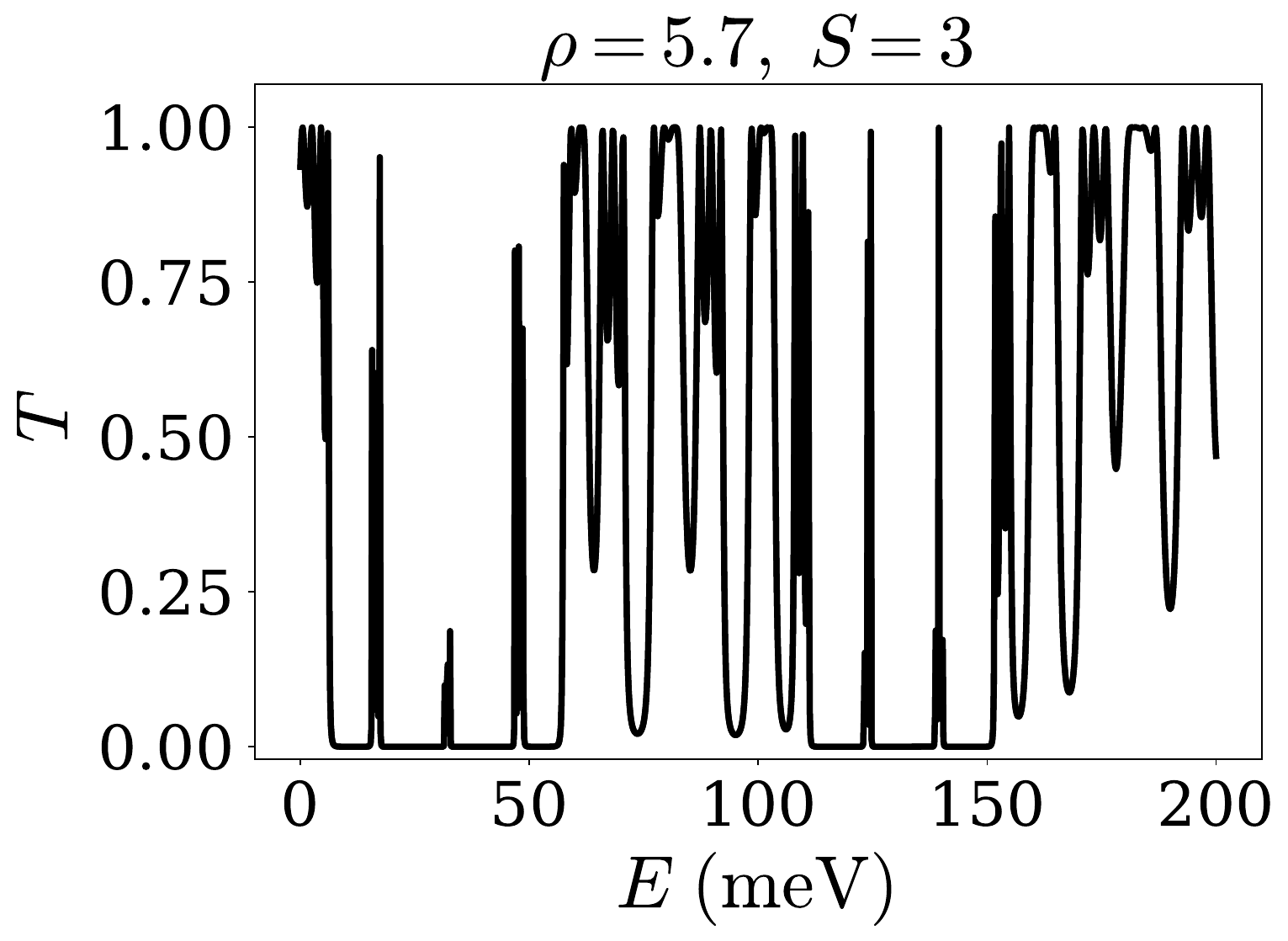}
\put(89, 1.5){\bfseries (g)}
\end{overpic}%
\begin{overpic}[width=0.245\linewidth]{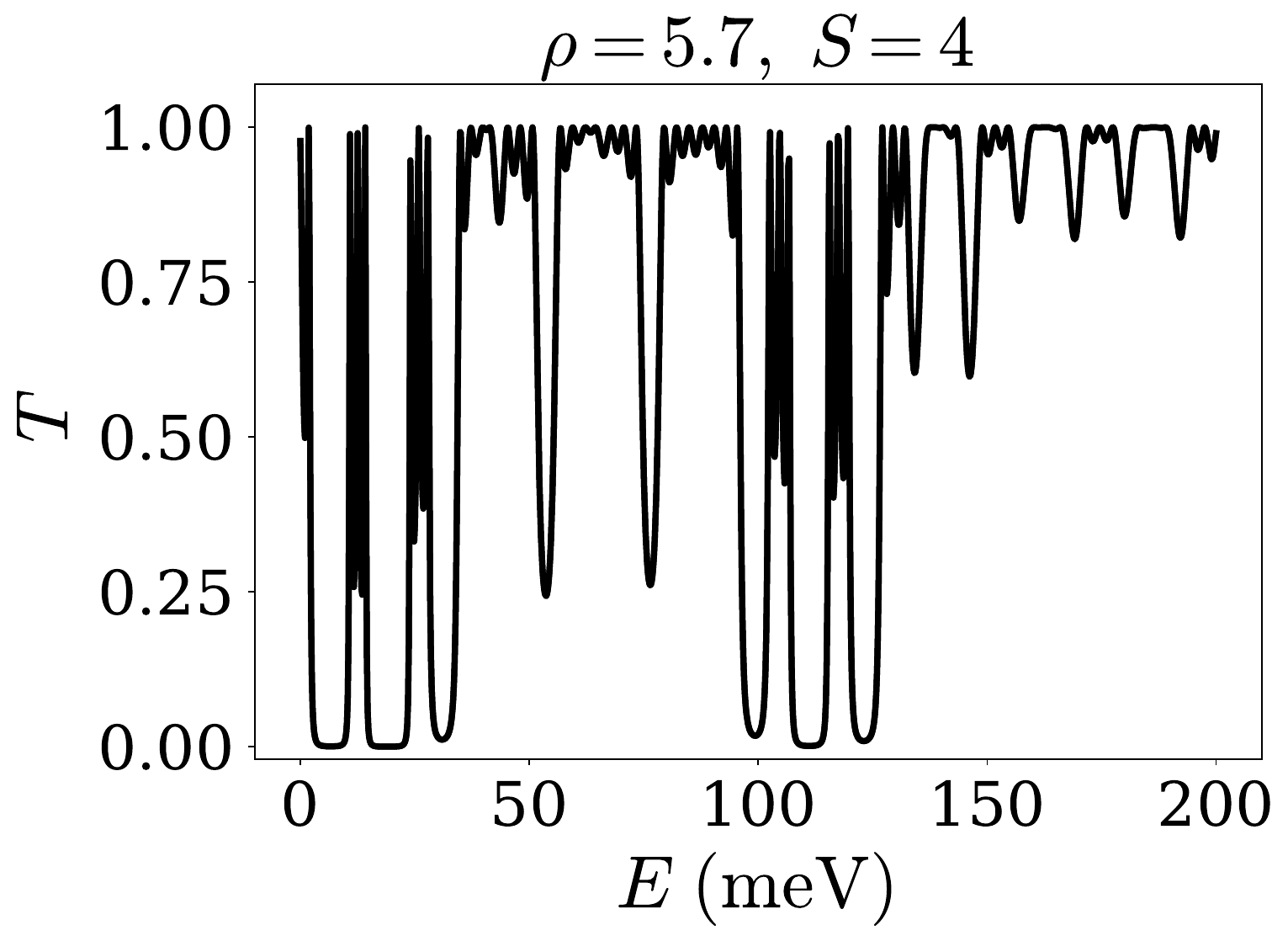}
\put(89, 1.5){\bfseries (h)}
\end{overpic}
\begin{overpic}[width=0.245\linewidth]{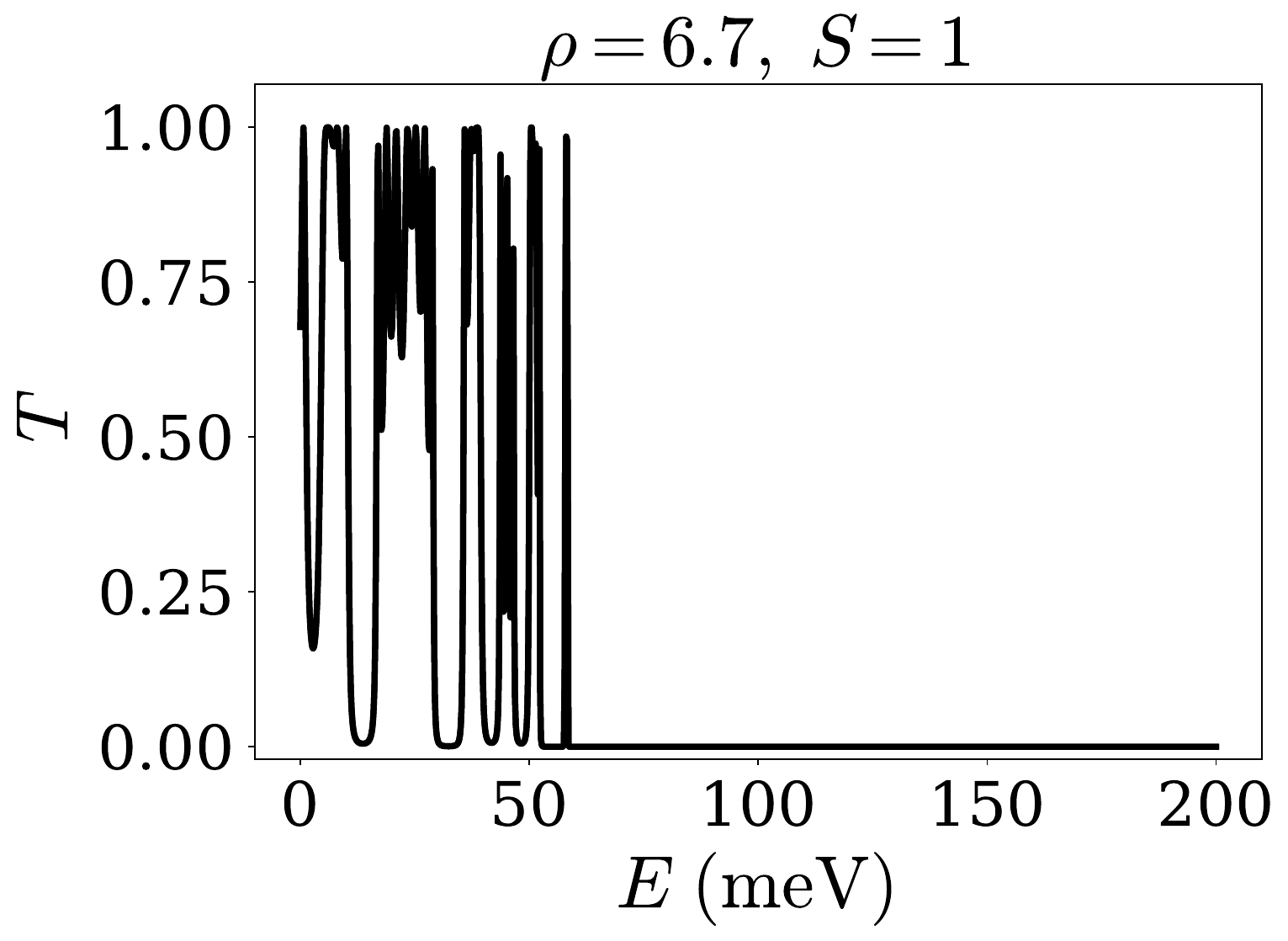}
\put(89, 1.5){\bfseries (i)}
\end{overpic}%
\begin{overpic}[width=0.245\linewidth]{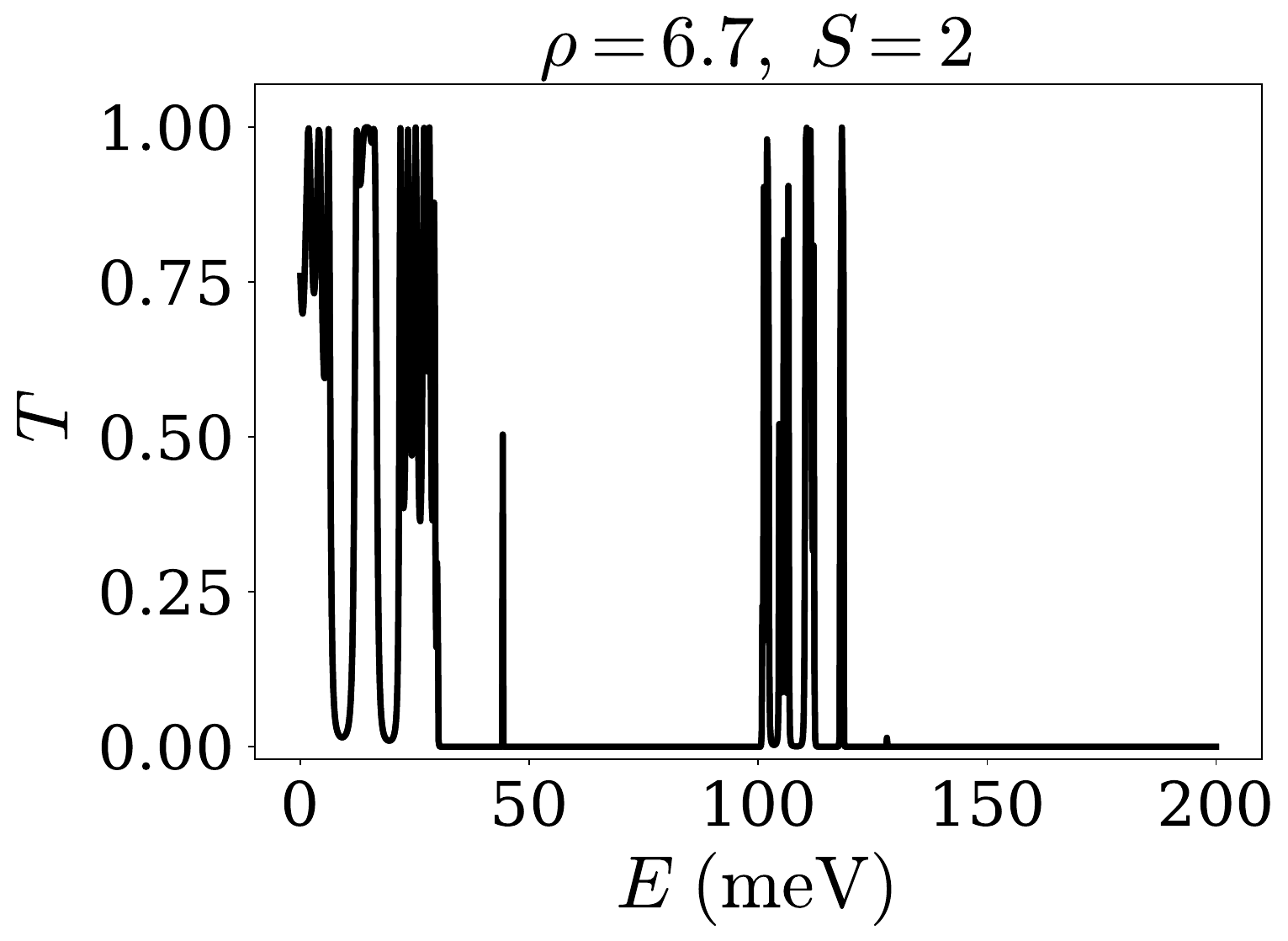}
\put(89, 1.5){\bfseries (j)}
\end{overpic}%
\begin{overpic}[width=0.245\linewidth]{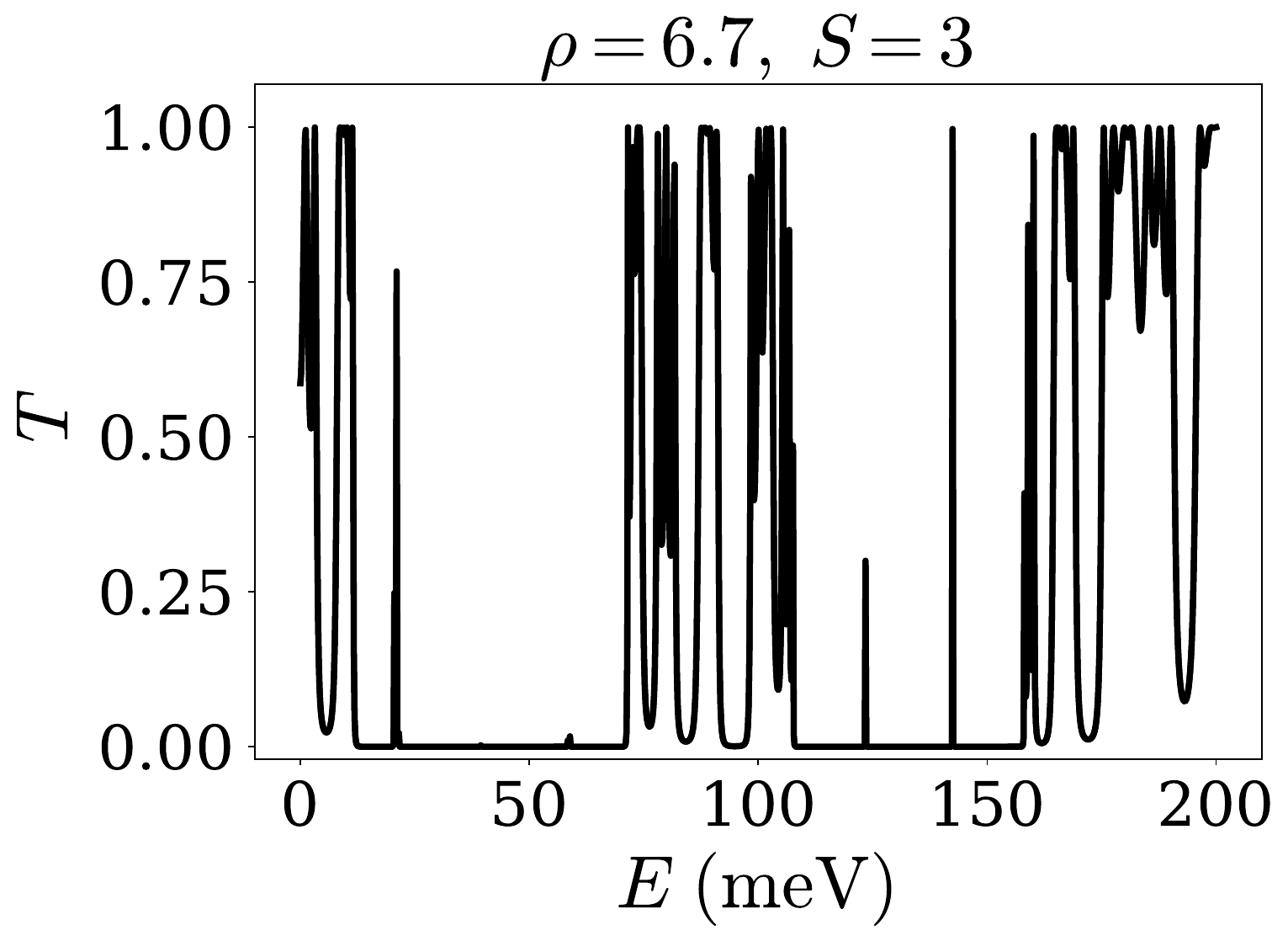}
\put(89, 1.5){\bfseries (k)}
\end{overpic}%
\begin{overpic}[width=0.245\linewidth]{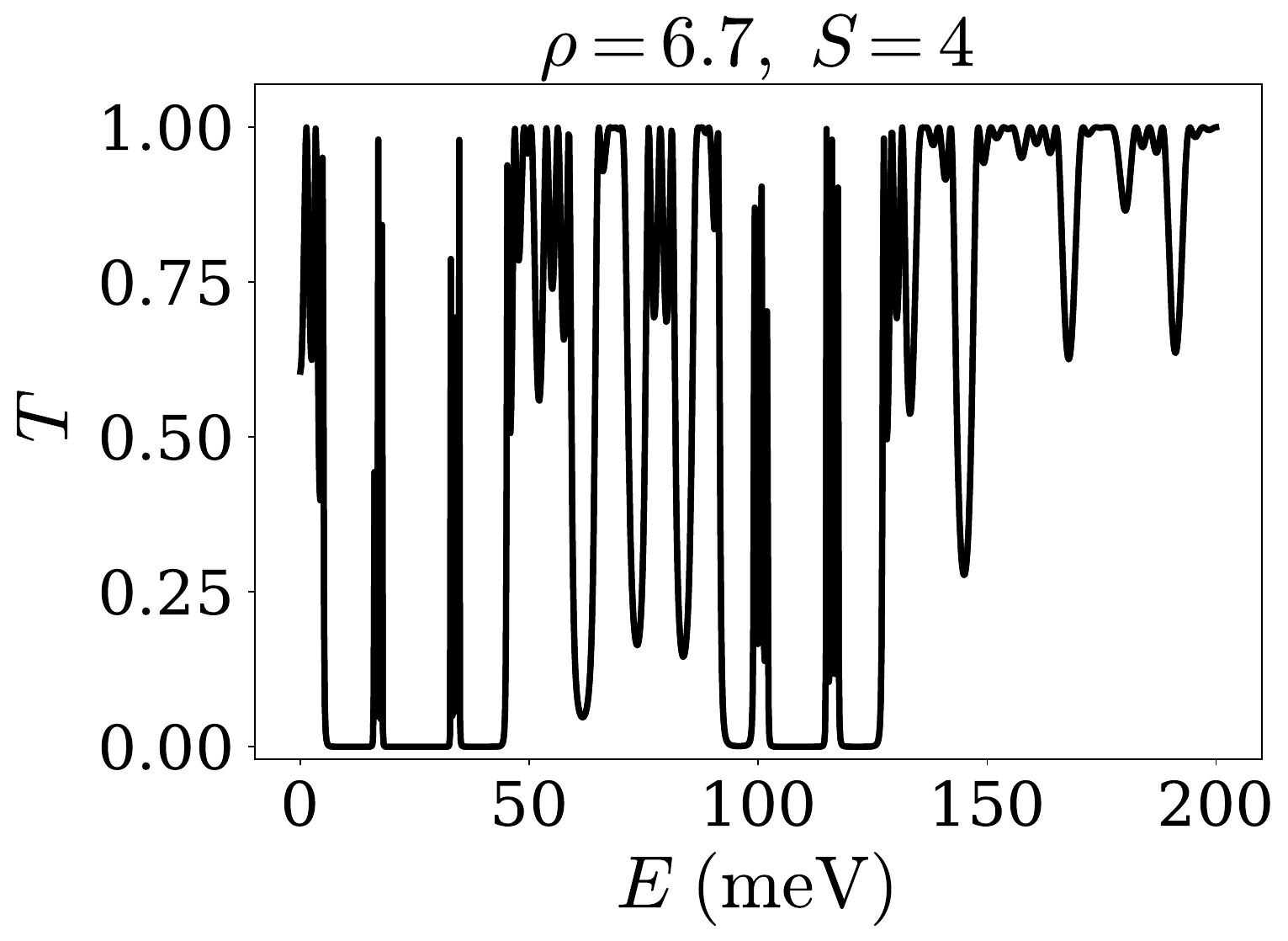}
\put(89, 1.5){\bfseries (l)}
\end{overpic}
\begin{overpic}[width=0.245\linewidth]{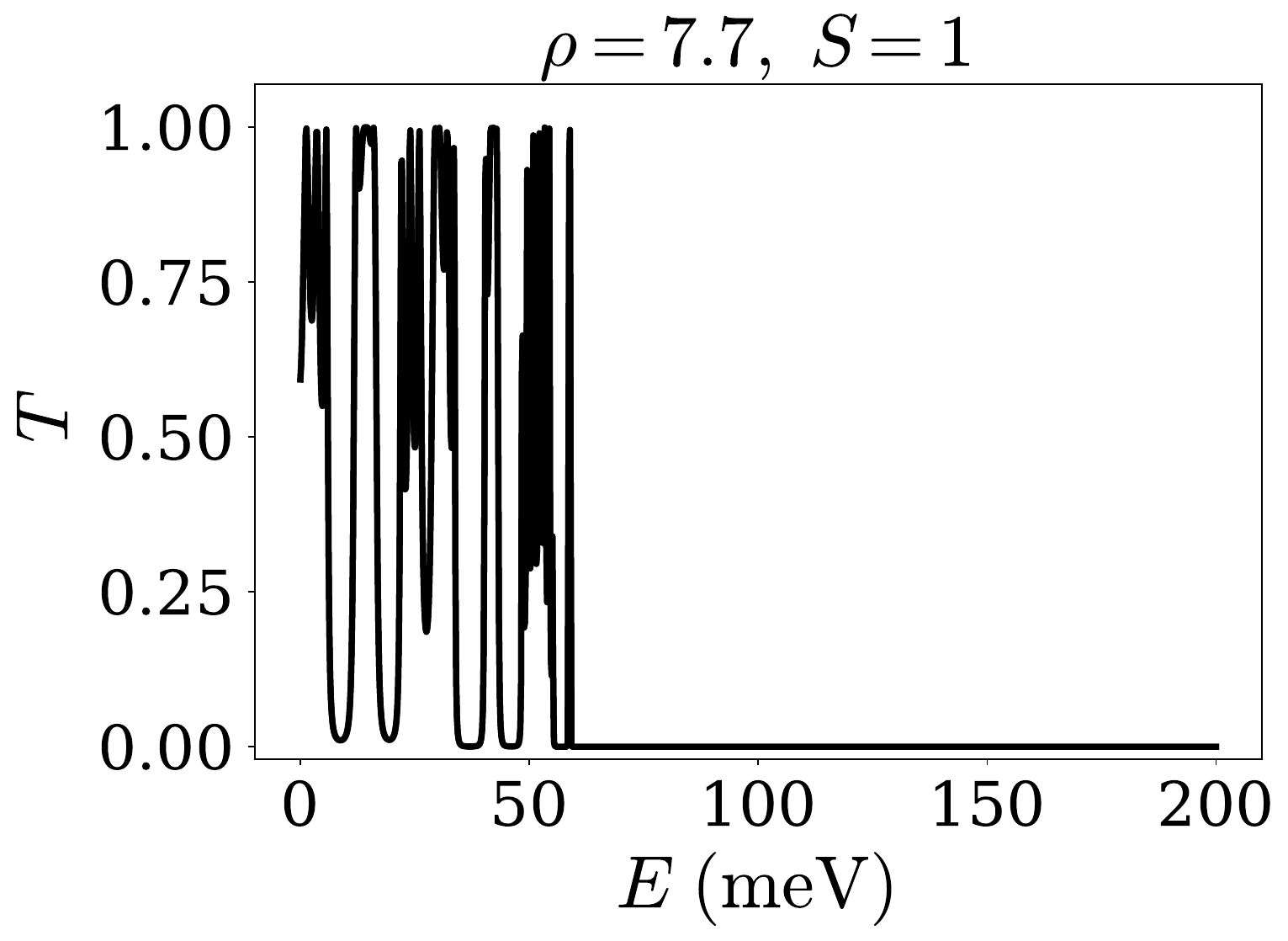}
\put(87, 1.5){\bfseries (m)}
\end{overpic}%
\begin{overpic}[width=0.245\linewidth]{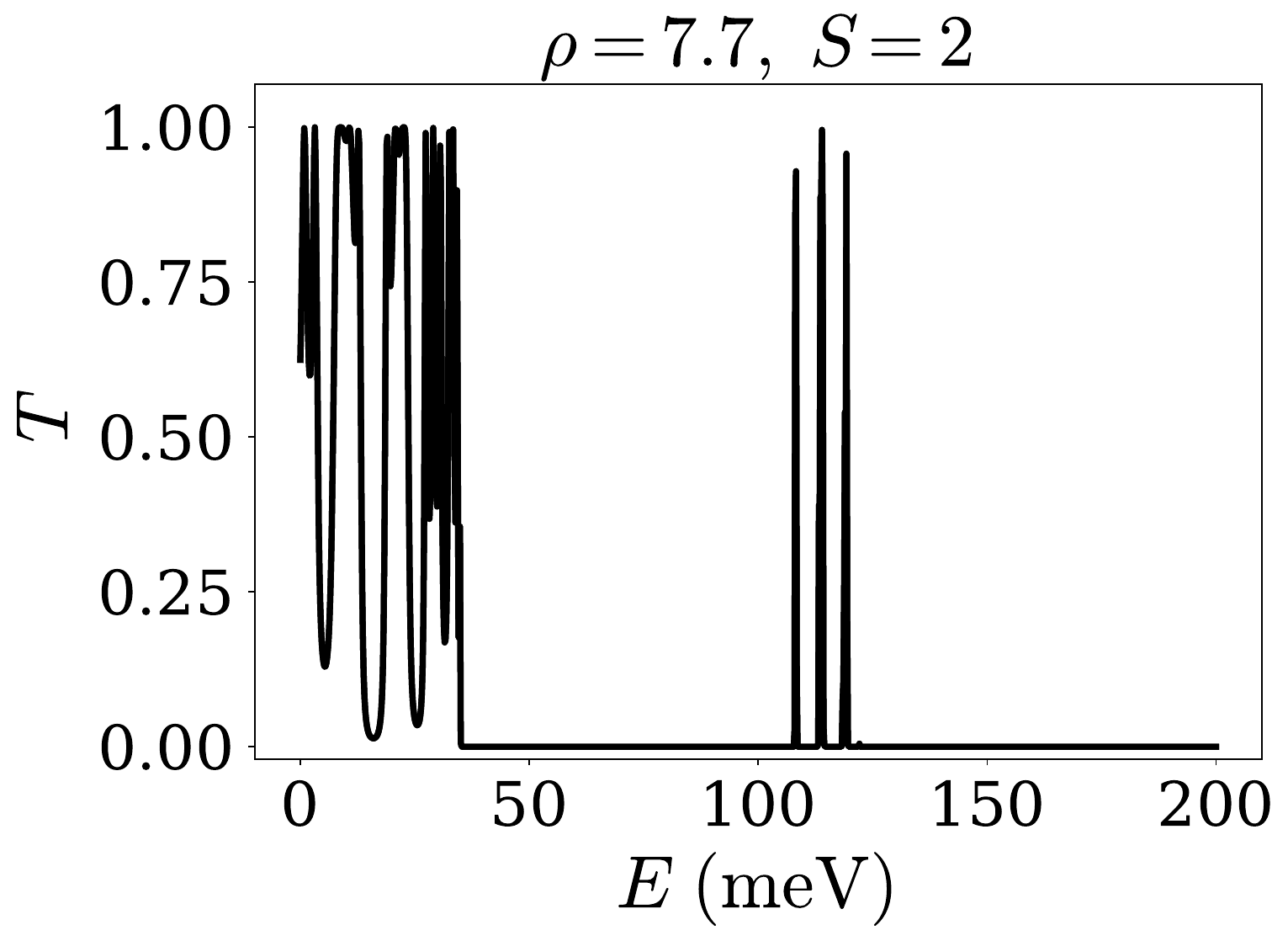}
\put(89, 1.5){\bfseries (n)}
\end{overpic}%
\begin{overpic}[width=0.245\linewidth]{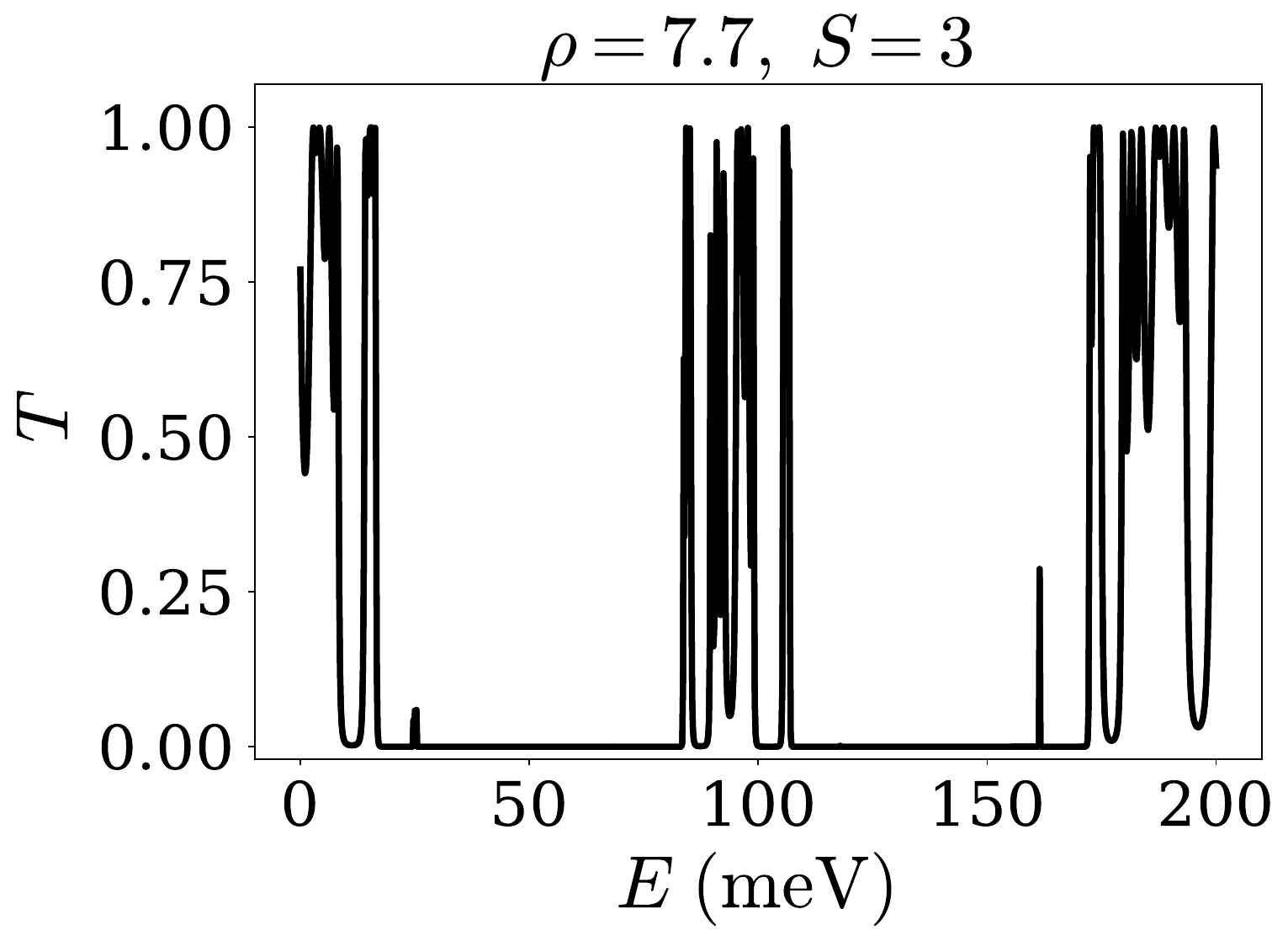}
\put(89, 1.5){\bfseries (o)}
\end{overpic}%
\begin{overpic}[width=0.245\linewidth]{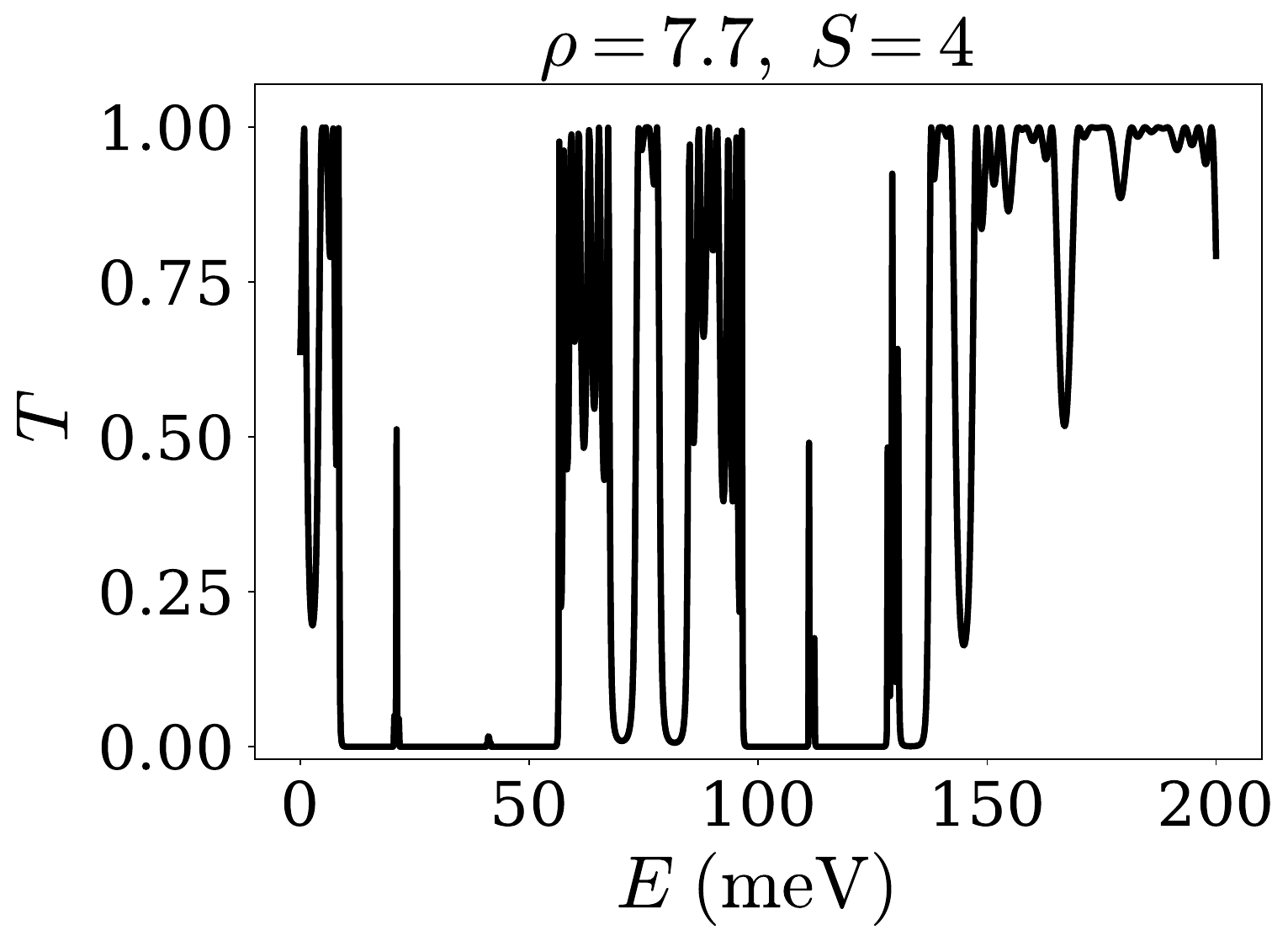}
\put(89, 1.5){\bfseries (p)}
\end{overpic}
\begin{overpic}[width=0.245\linewidth]{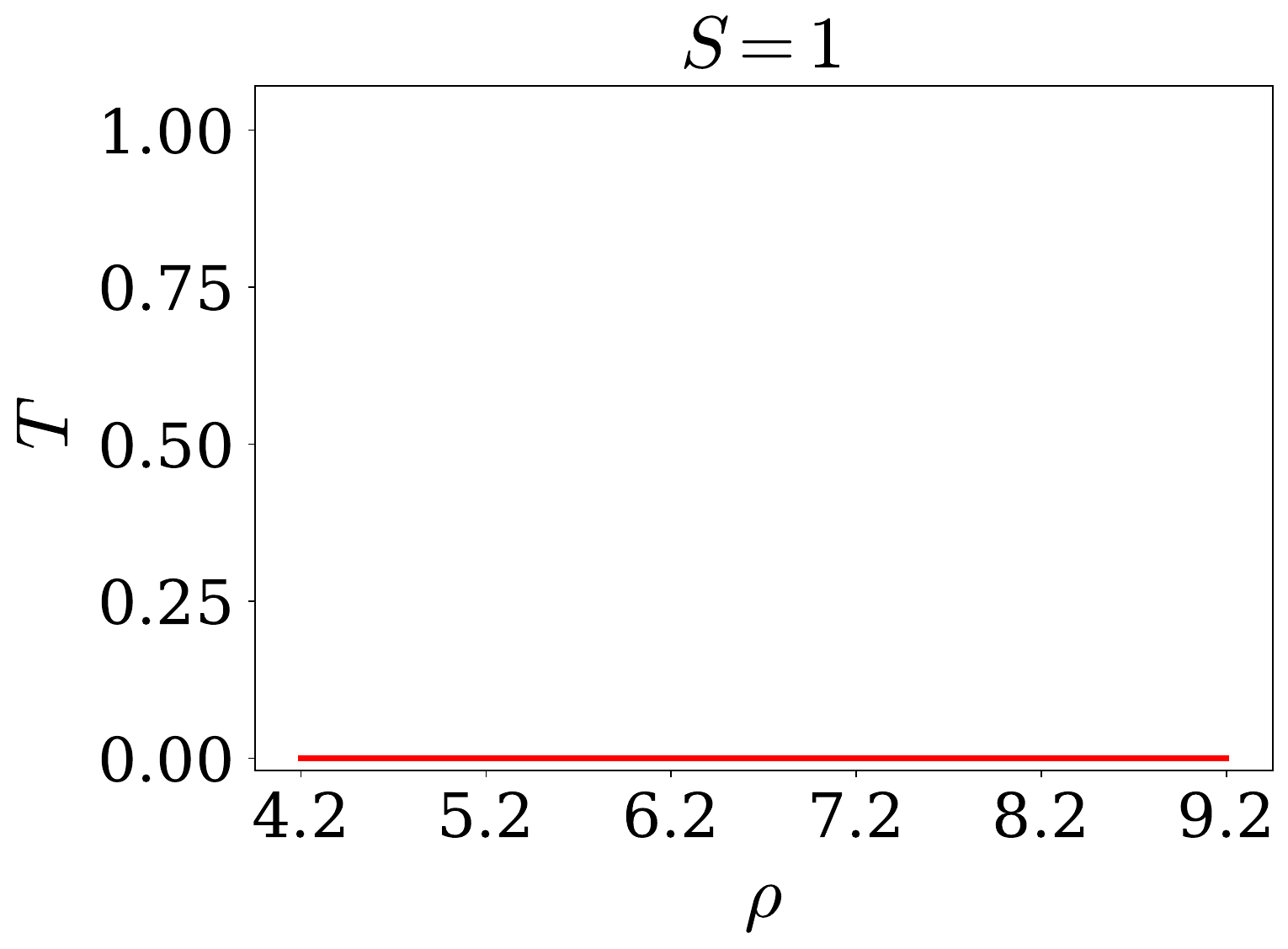}
\put(89, 1.5){\bfseries (q)}
\end{overpic}%
\begin{overpic}[width=0.245\linewidth]{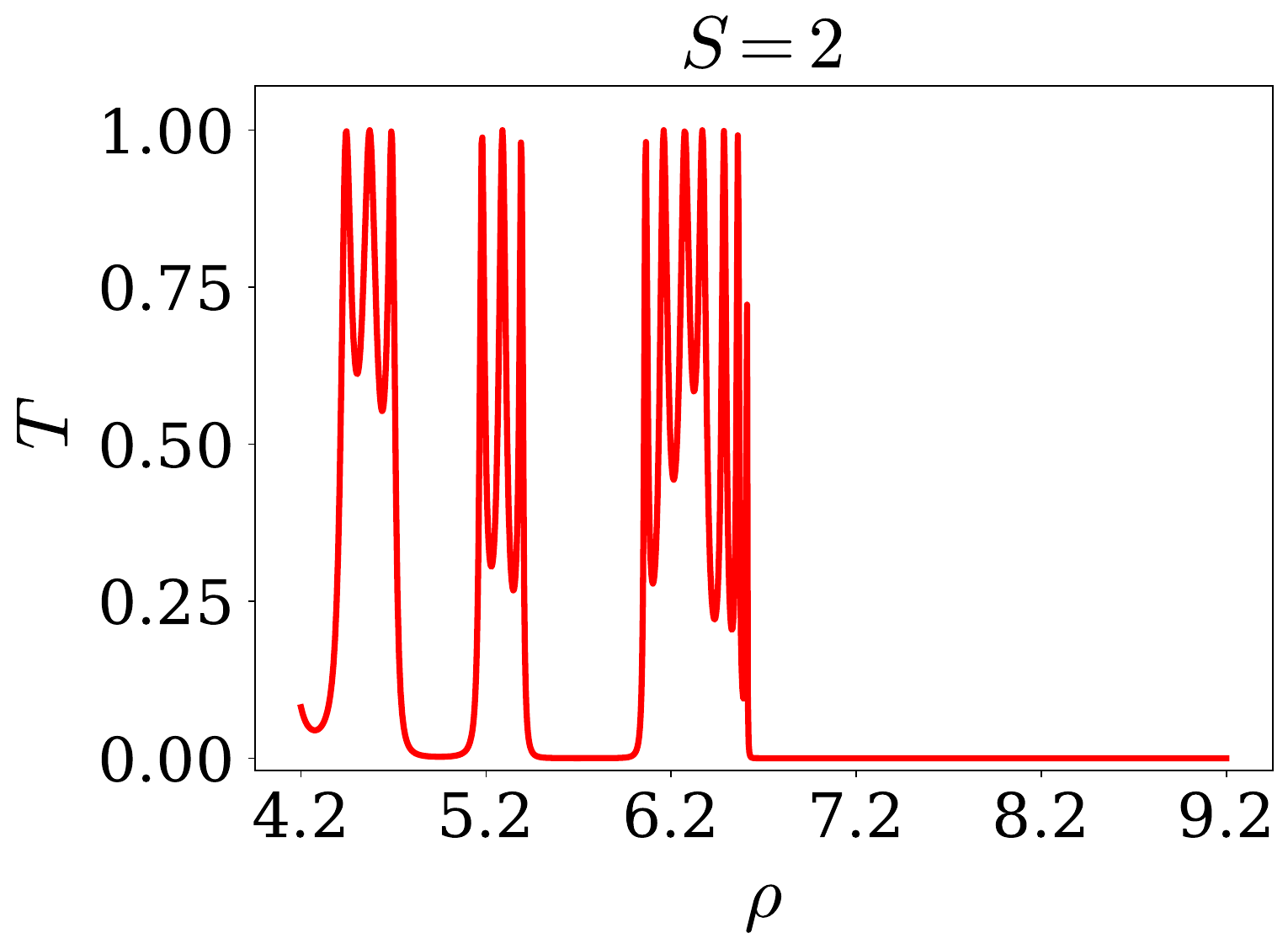}
\put(89, 1.5){\bfseries (r)}
\end{overpic}%
\begin{overpic}[width=0.245\linewidth]{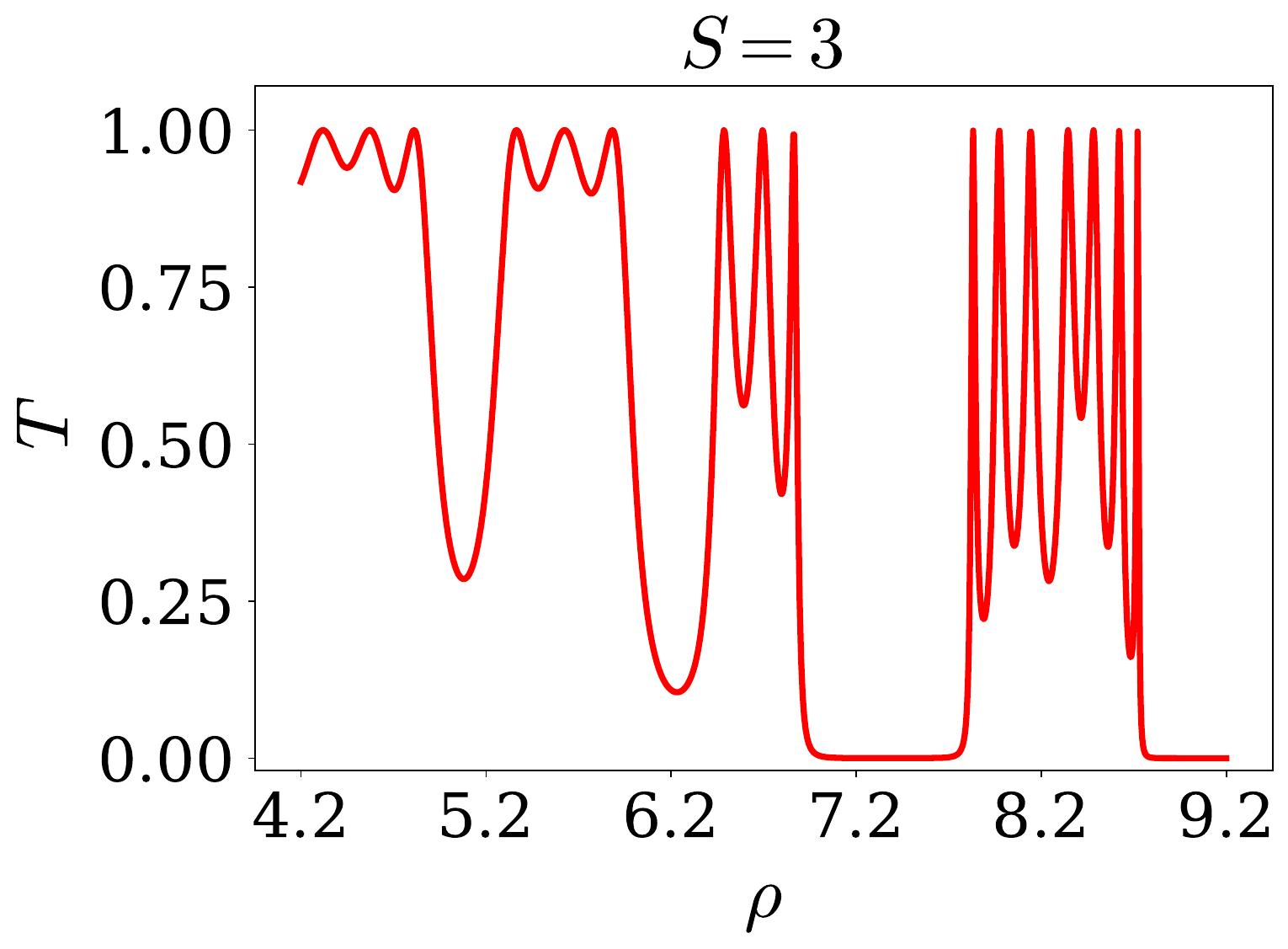}
\put(89, 1.5){\bfseries (s)}
\end{overpic}%
\begin{overpic}[width=0.245\linewidth]{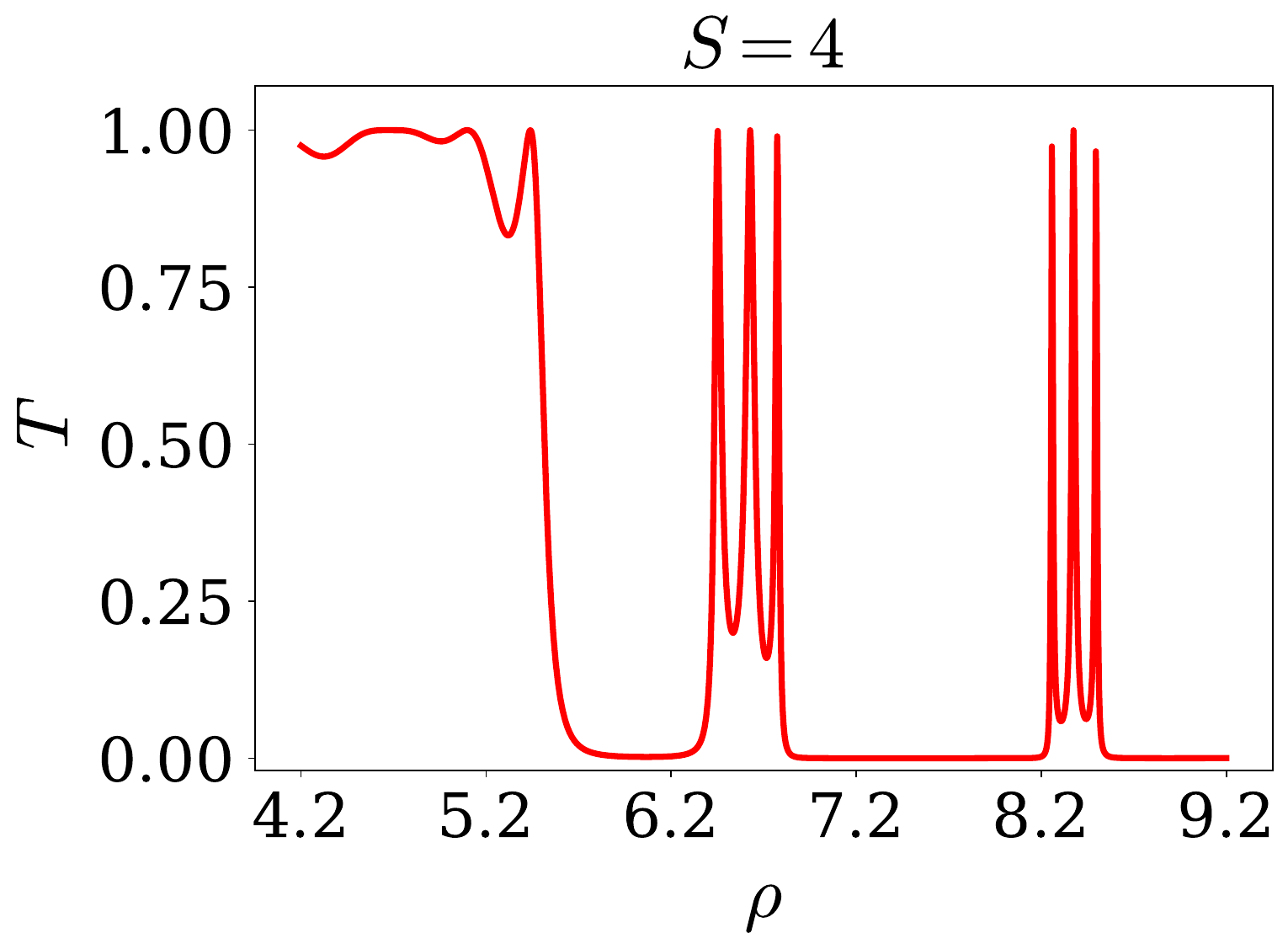}
\put(89, 1.5){\bfseries (t)}
\end{overpic}
\caption{Transmission characteristics of monolayer graphene subjected to a polyadic Cantor electrostatic barrier, with each column corresponding to a successive Cantor stage, $S=1$ to $4$. The top row [(a)--(d)] shows the transmission density maps $T(E,\rho)$ as functions of the Fermi energy $E$ and the scaling parameter $\rho$. The next three rows show sectional transmission profiles $T(E)$ at fixed values of the scaling parameter: $\rho=5.7$ [(e)--(h)], $\rho=6.7$ [(i)--(l)], and $\rho=7.7$ [(m)--(p)]. The bottom row [(q)--(t)] shows $T(\rho)$ evaluated at the Dirac point, $E=V_0$. In all panels $N=4$, $\phi=40^\circ$, $V_0=100$~meV, and $L=800$~nm.}
\label{fig:01}
\end{figure*}
\begin{figure*}[htbp]
\centering
\begin{overpic}[width=0.245\linewidth]{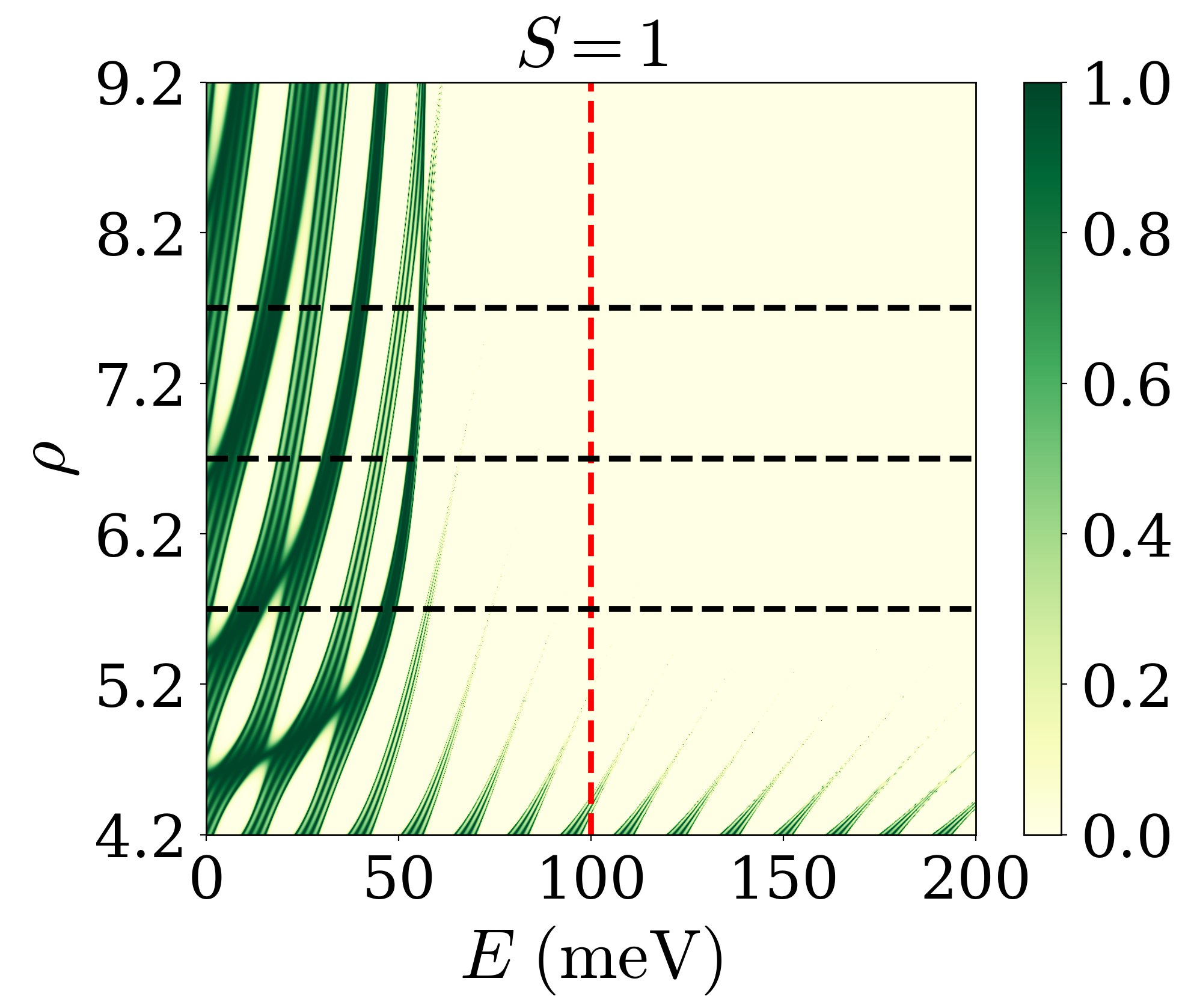}
\put(89,1.5){\bfseries (a)}
\end{overpic}%
\begin{overpic}[width=0.245\linewidth]{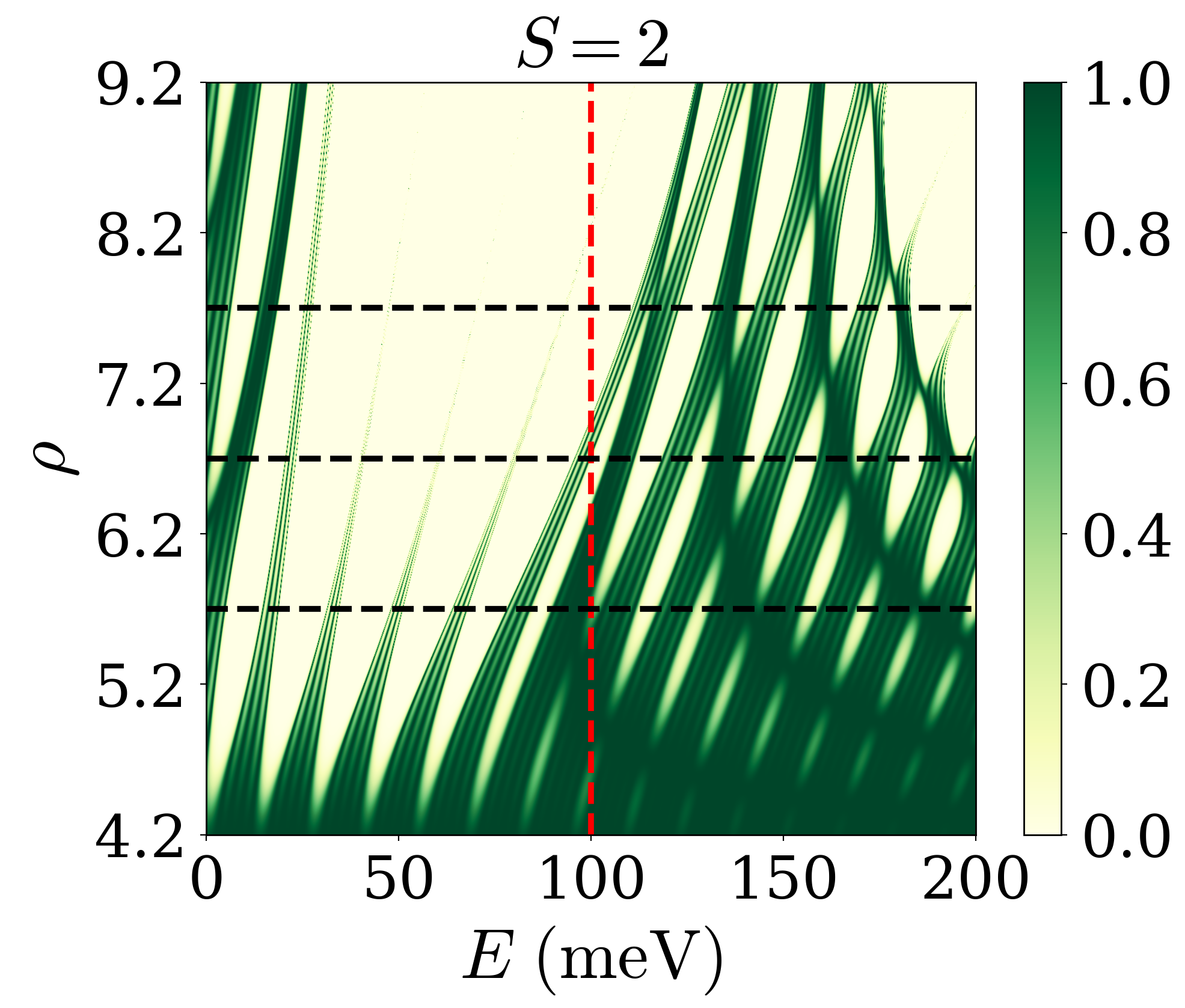}
\put(89,1.5){\bfseries (b)}
\end{overpic}%
\begin{overpic}[width=0.245\linewidth]{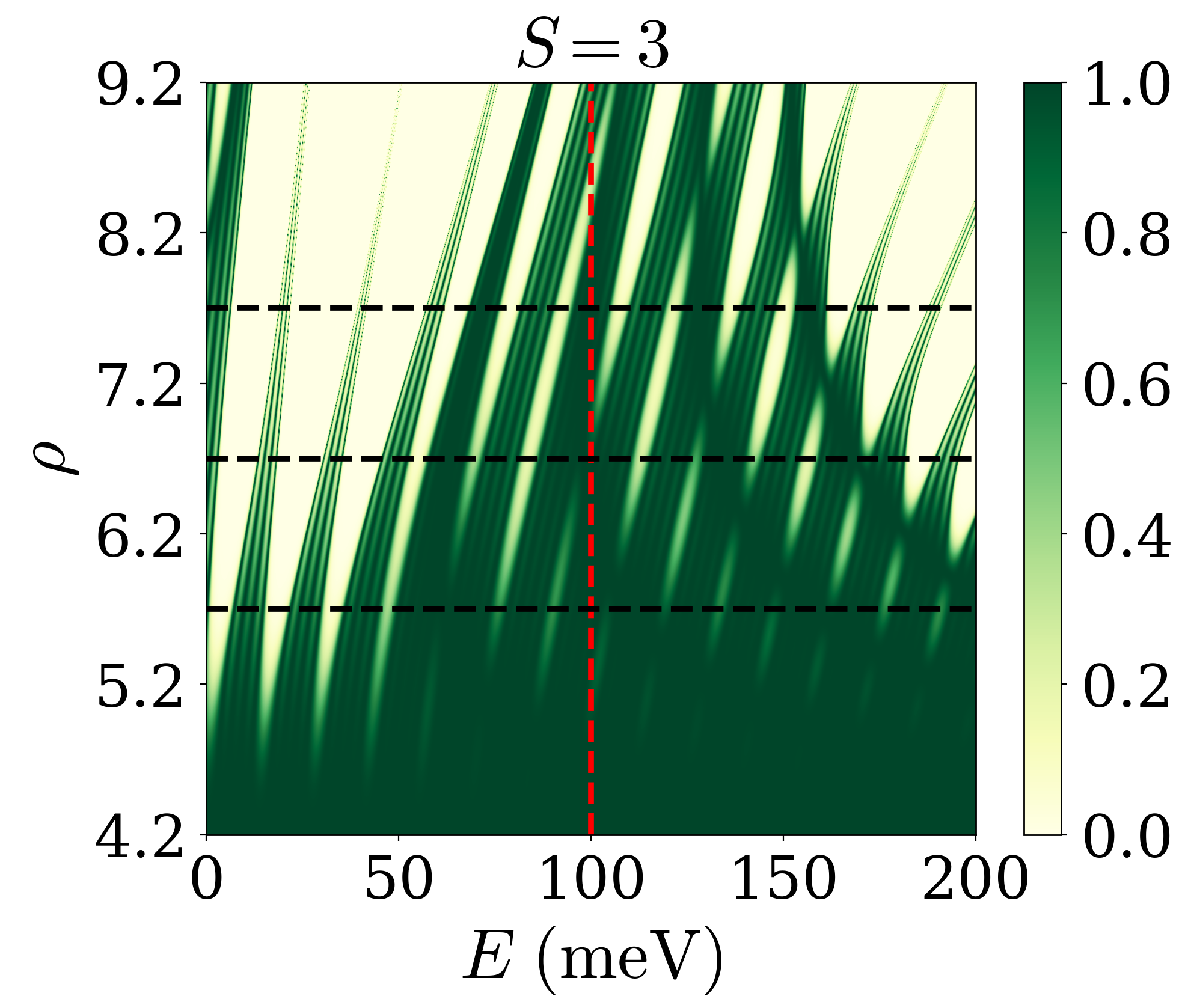}
\put(89,1.5){\bfseries (c)}
\end{overpic}%
\begin{overpic}[width=0.245\linewidth]{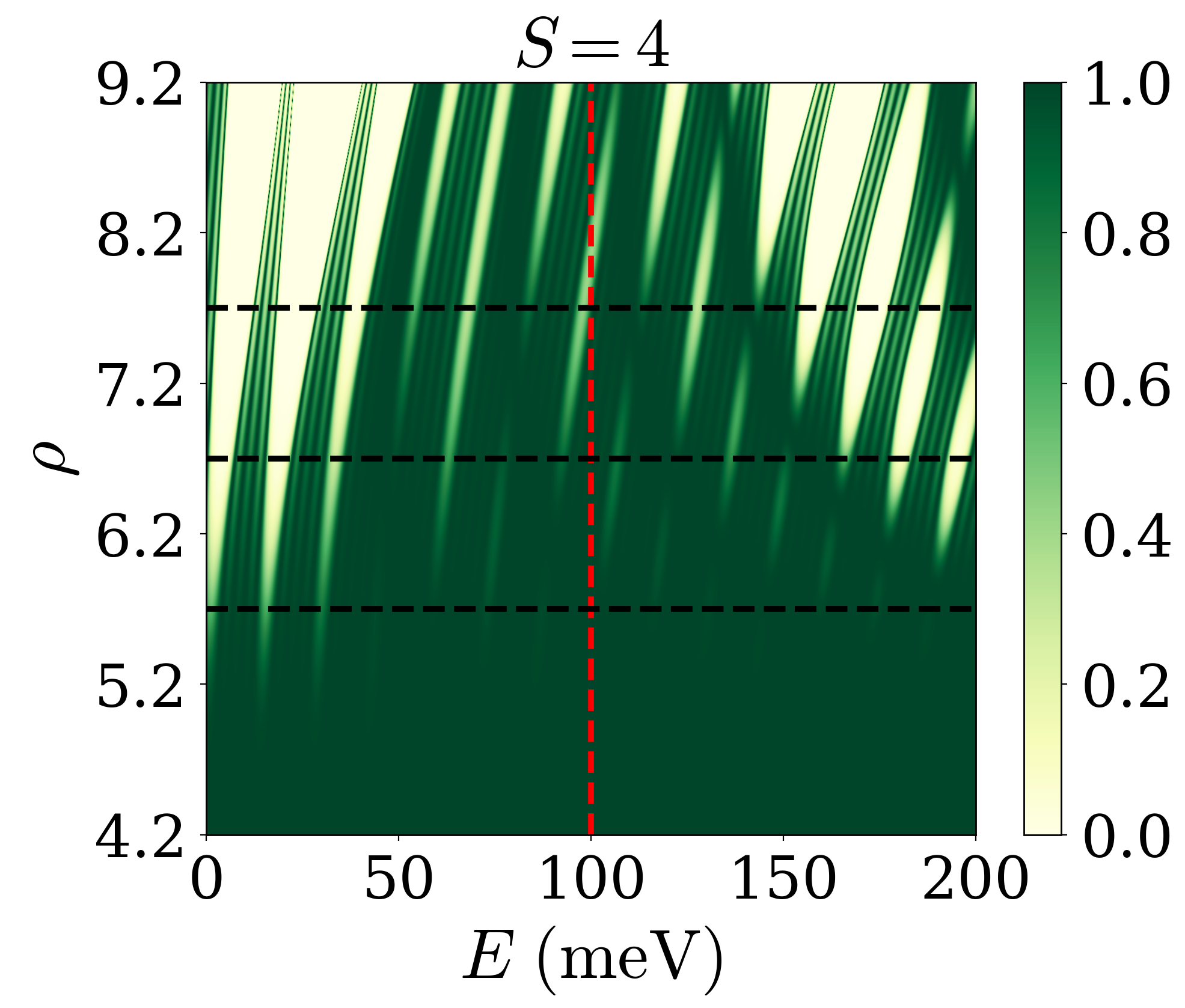}
\put(89,1.5){\bfseries (d)}
\end{overpic}
\begin{overpic}[width=0.245\linewidth]{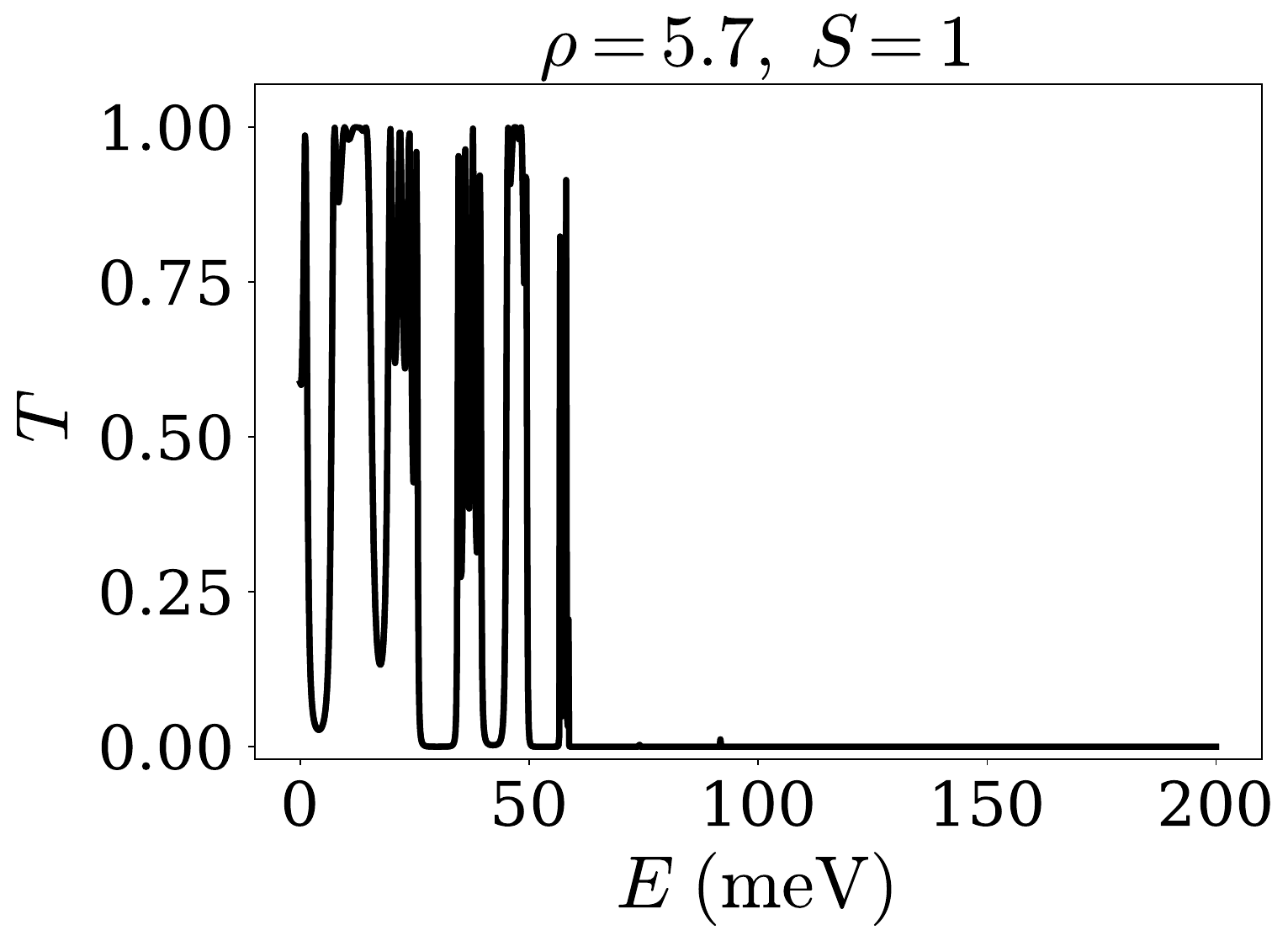}
\put(89,1.5){\bfseries (e)}
\end{overpic}%
\begin{overpic}[width=0.245\linewidth]{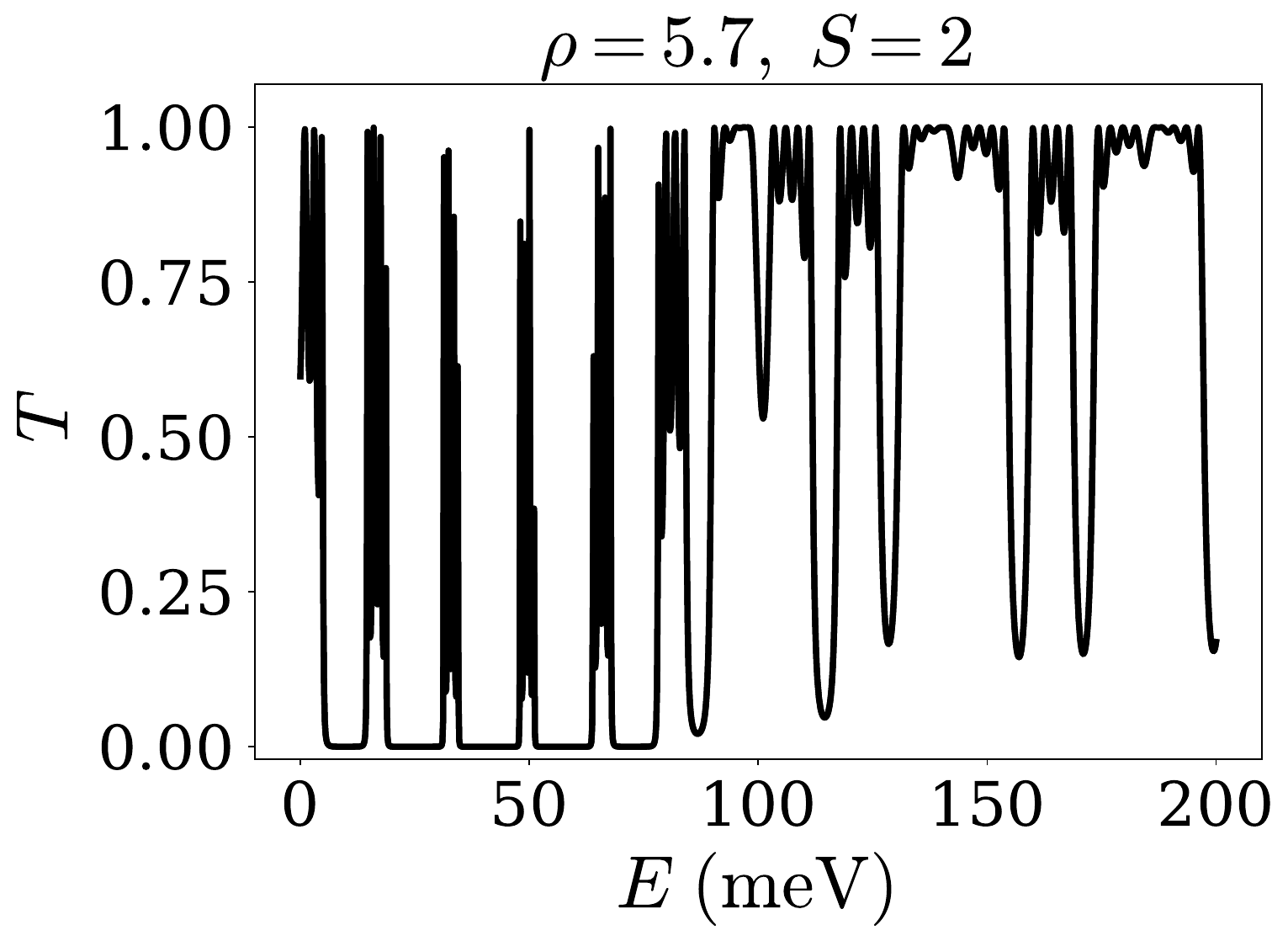}
\put(89,1.5){\bfseries (f)}
\end{overpic}%
\begin{overpic}[width=0.245\linewidth]{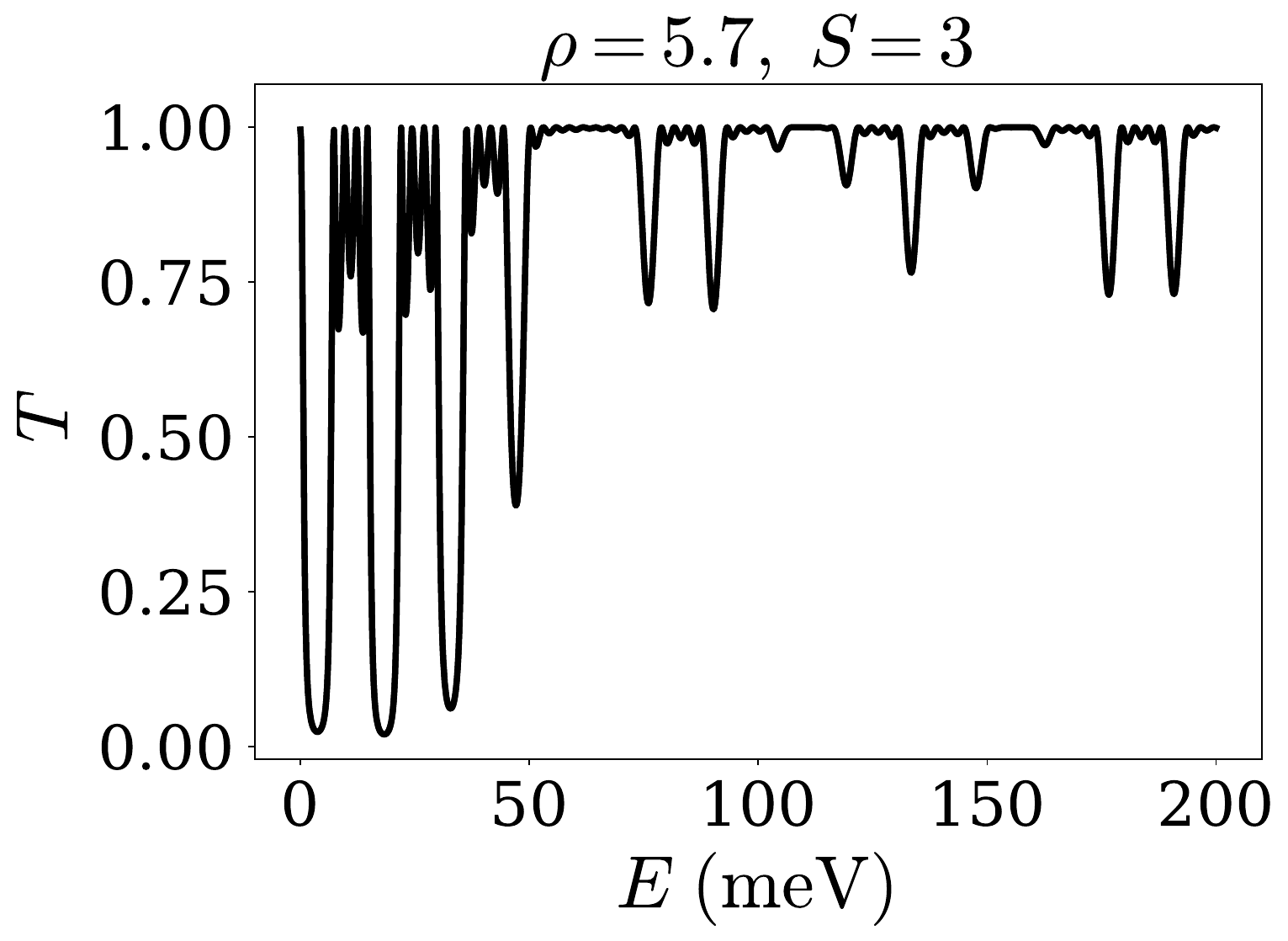}
\put(89,1.5){\bfseries (g)}
\end{overpic}%
\begin{overpic}[width=0.245\linewidth]{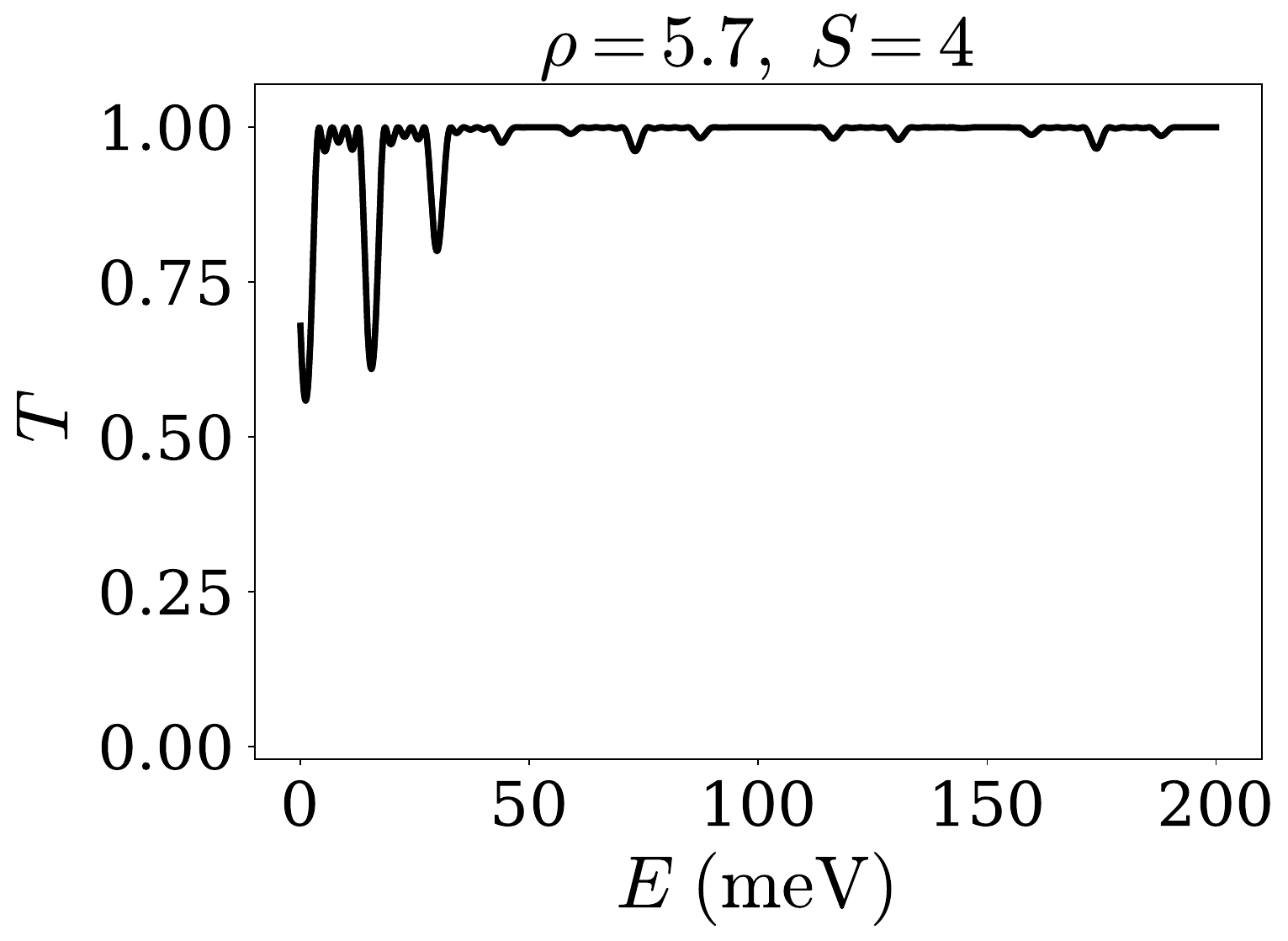}
\put(89,1.5){\bfseries (h)}
\end{overpic}
\begin{overpic}[width=0.245\linewidth]{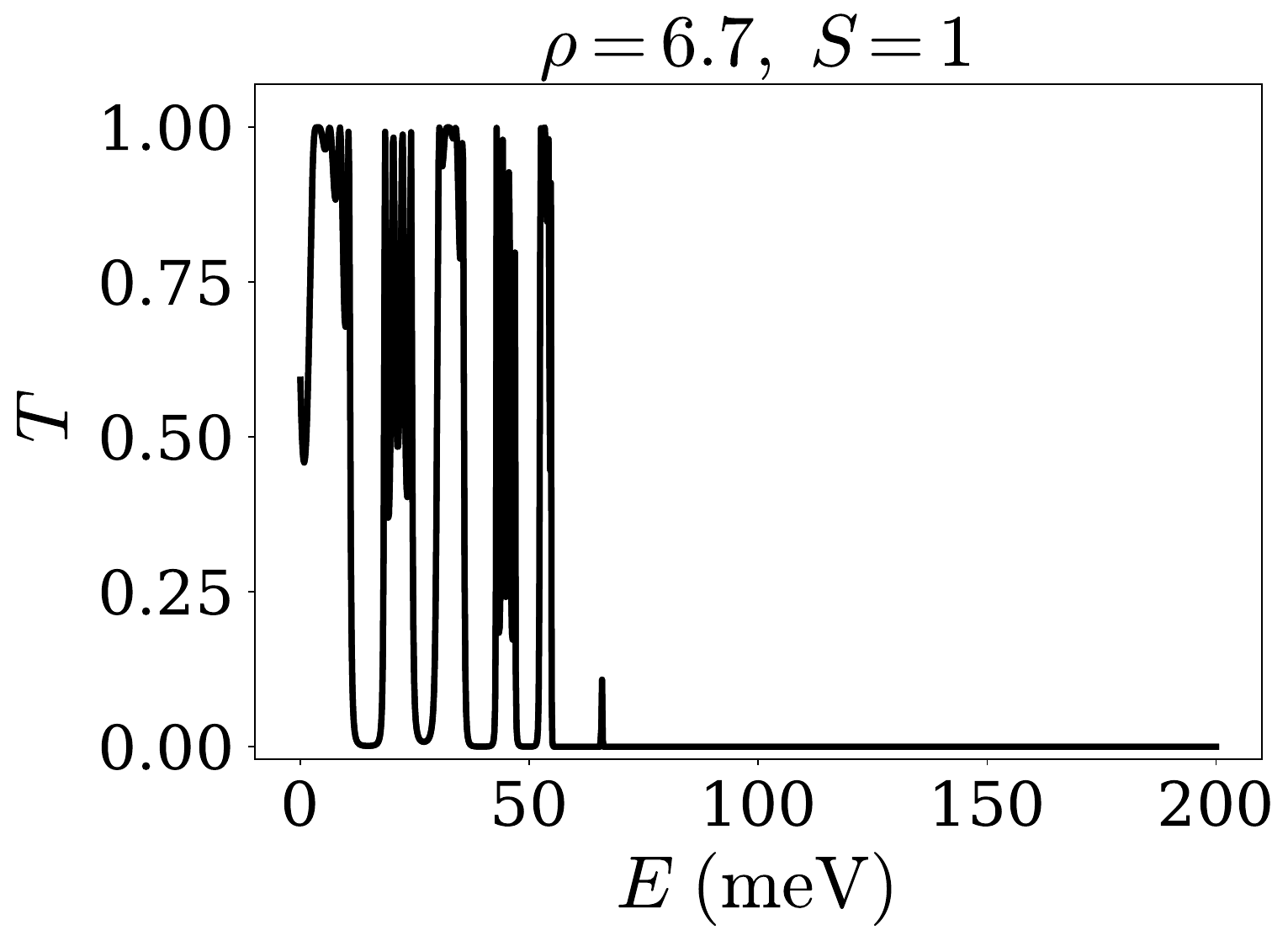}
\put(89,1.5){\bfseries (i)}
\end{overpic}%
\begin{overpic}[width=0.245\linewidth]{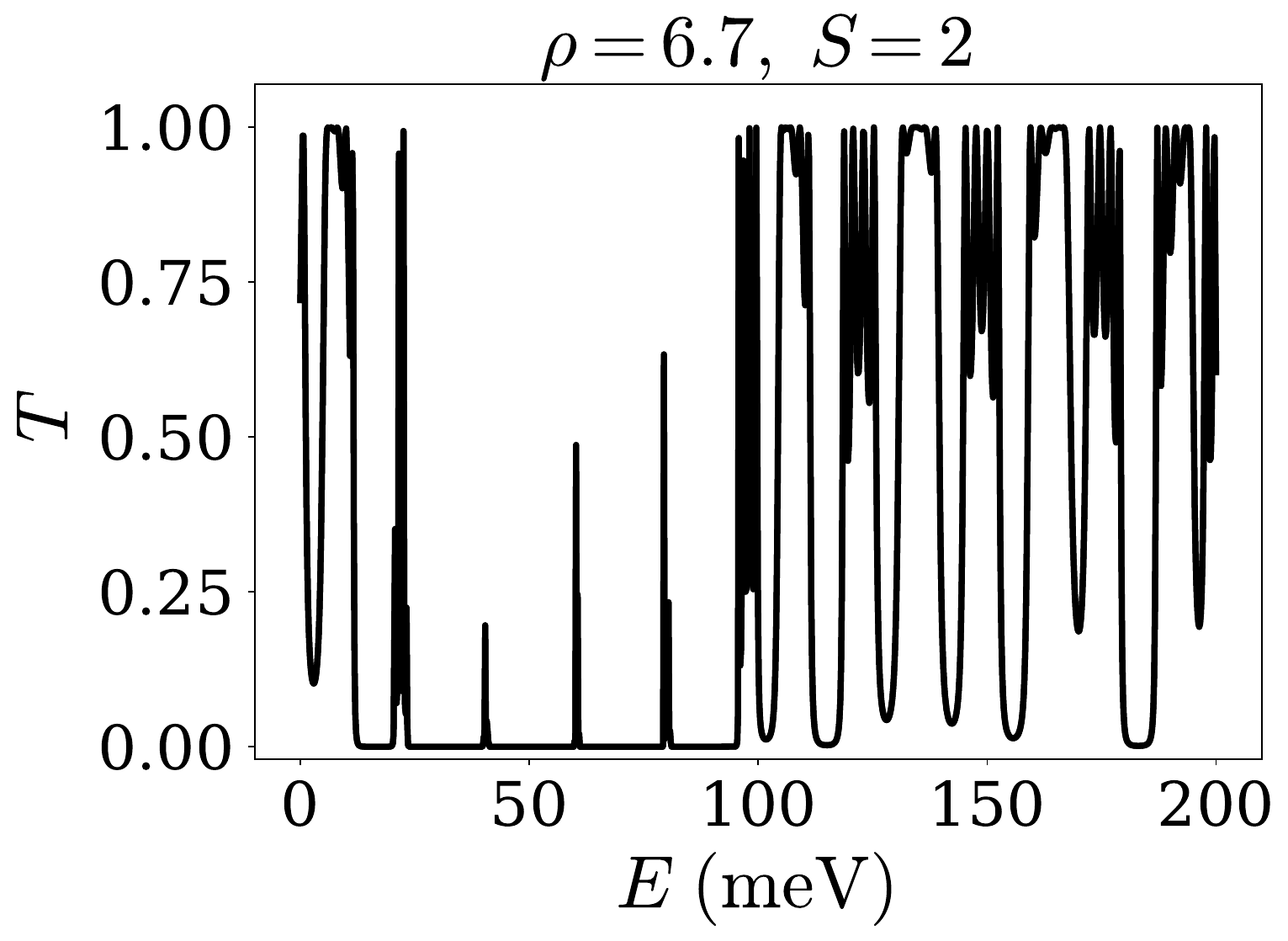}
\put(89,1.5){\bfseries (j)}
\end{overpic}%
\begin{overpic}[width=0.245\linewidth]{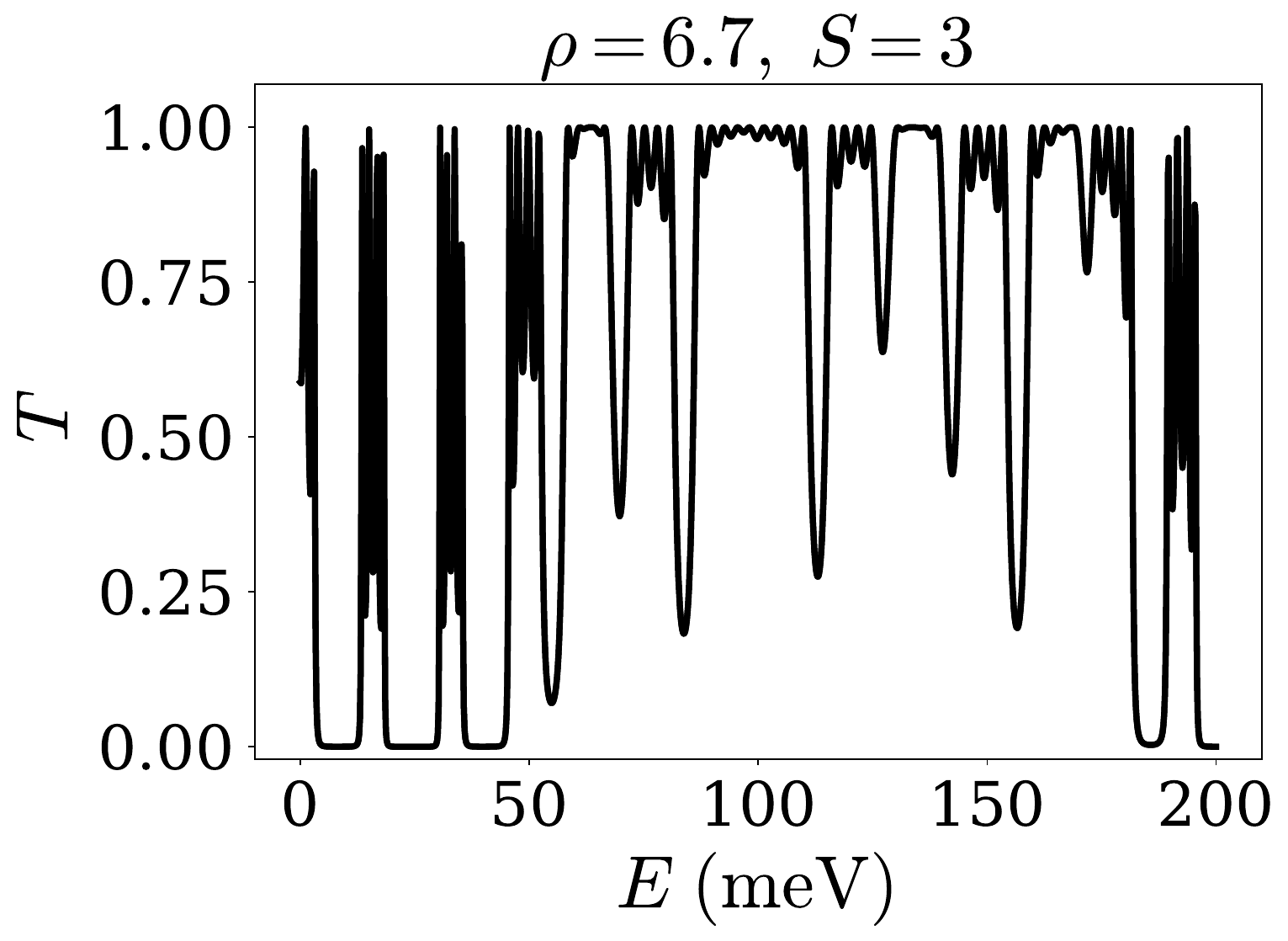}
\put(89,1.5){\bfseries (k)}
\end{overpic}%
\begin{overpic}[width=0.245\linewidth]{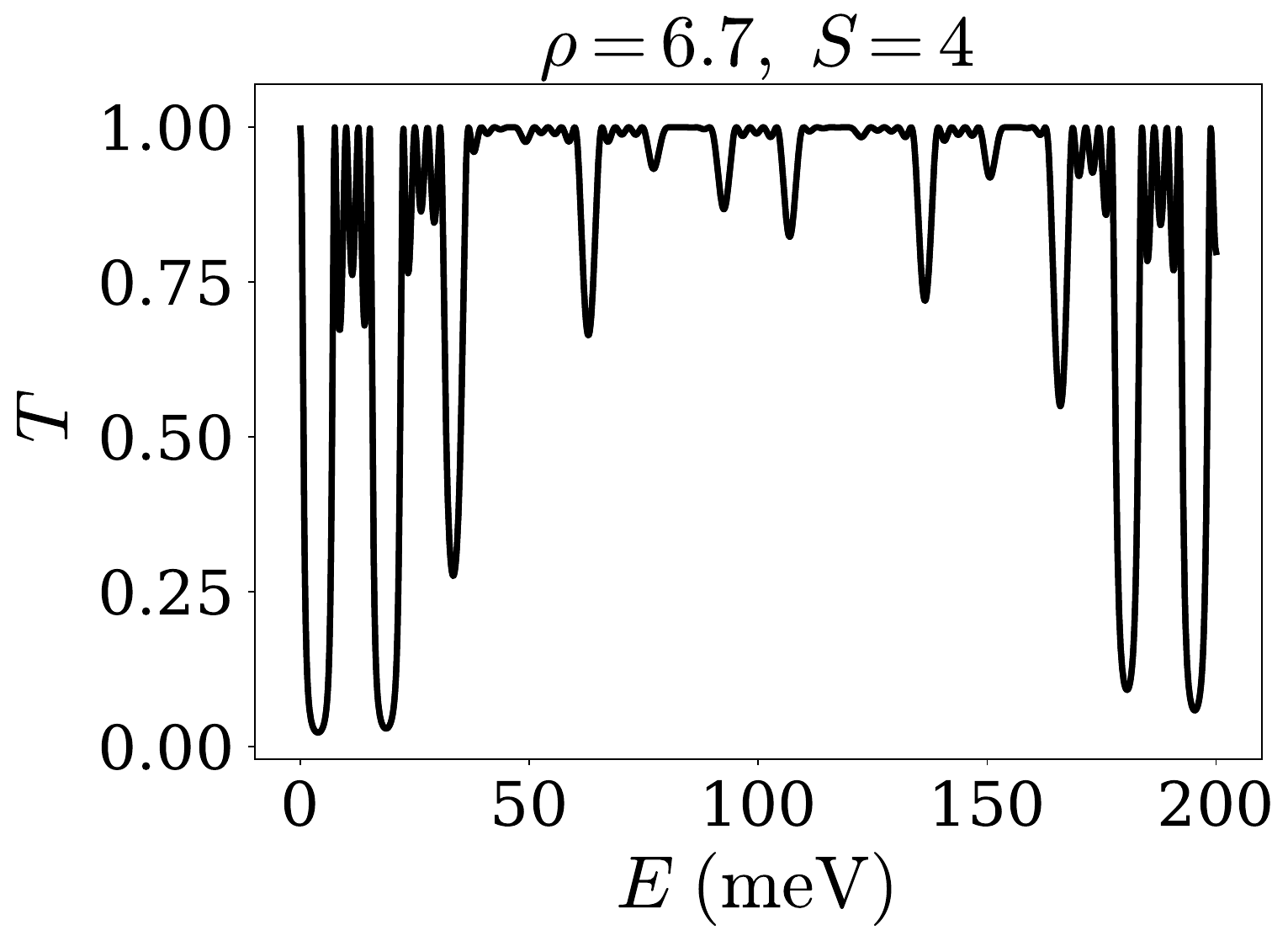}
\put(89,1.5){\bfseries (l)}
\end{overpic}
\begin{overpic}[width=0.245\linewidth]{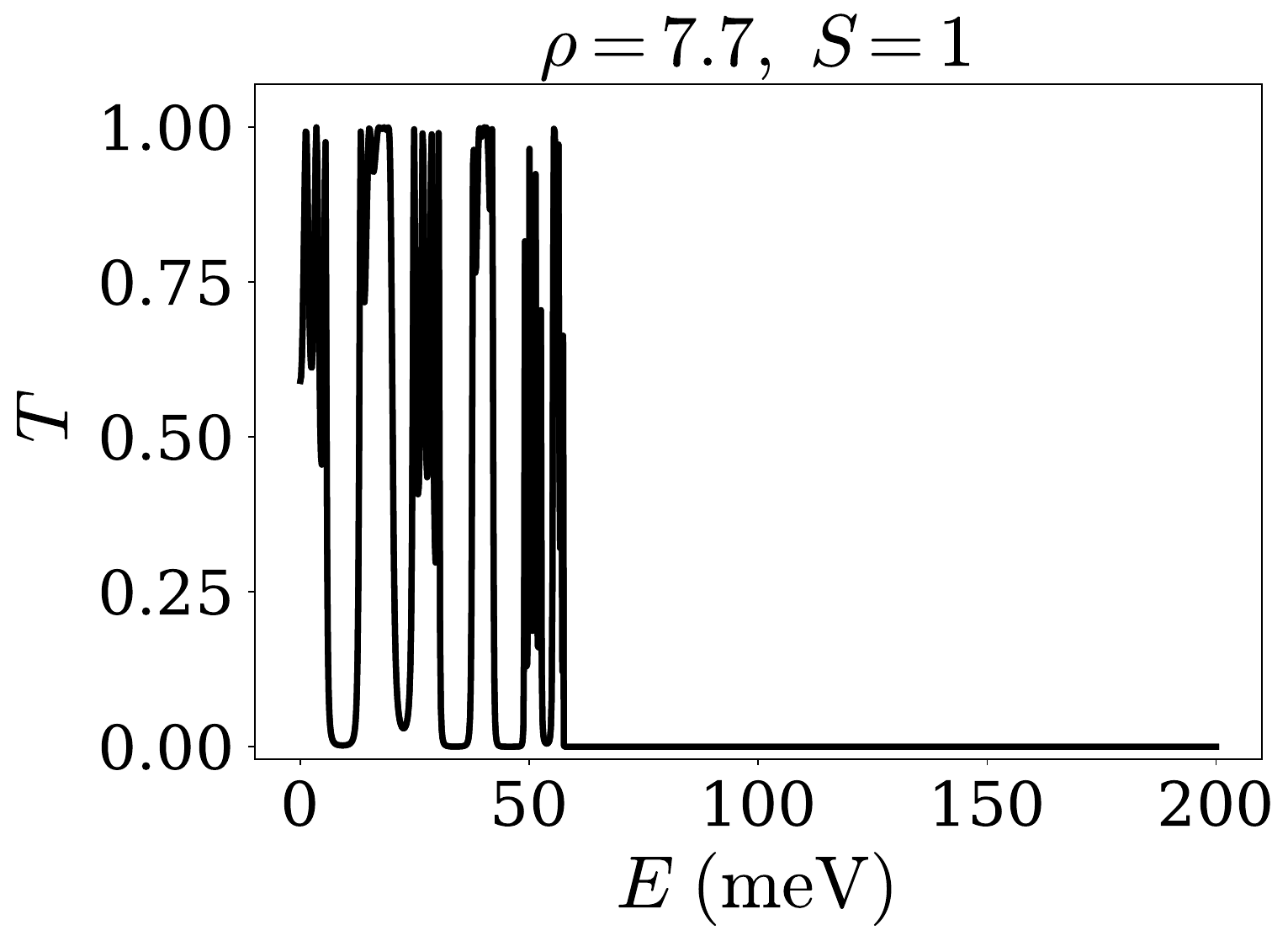}
\put(87,1.5){\bfseries (m)}
\end{overpic}%
\begin{overpic}[width=0.245\linewidth]{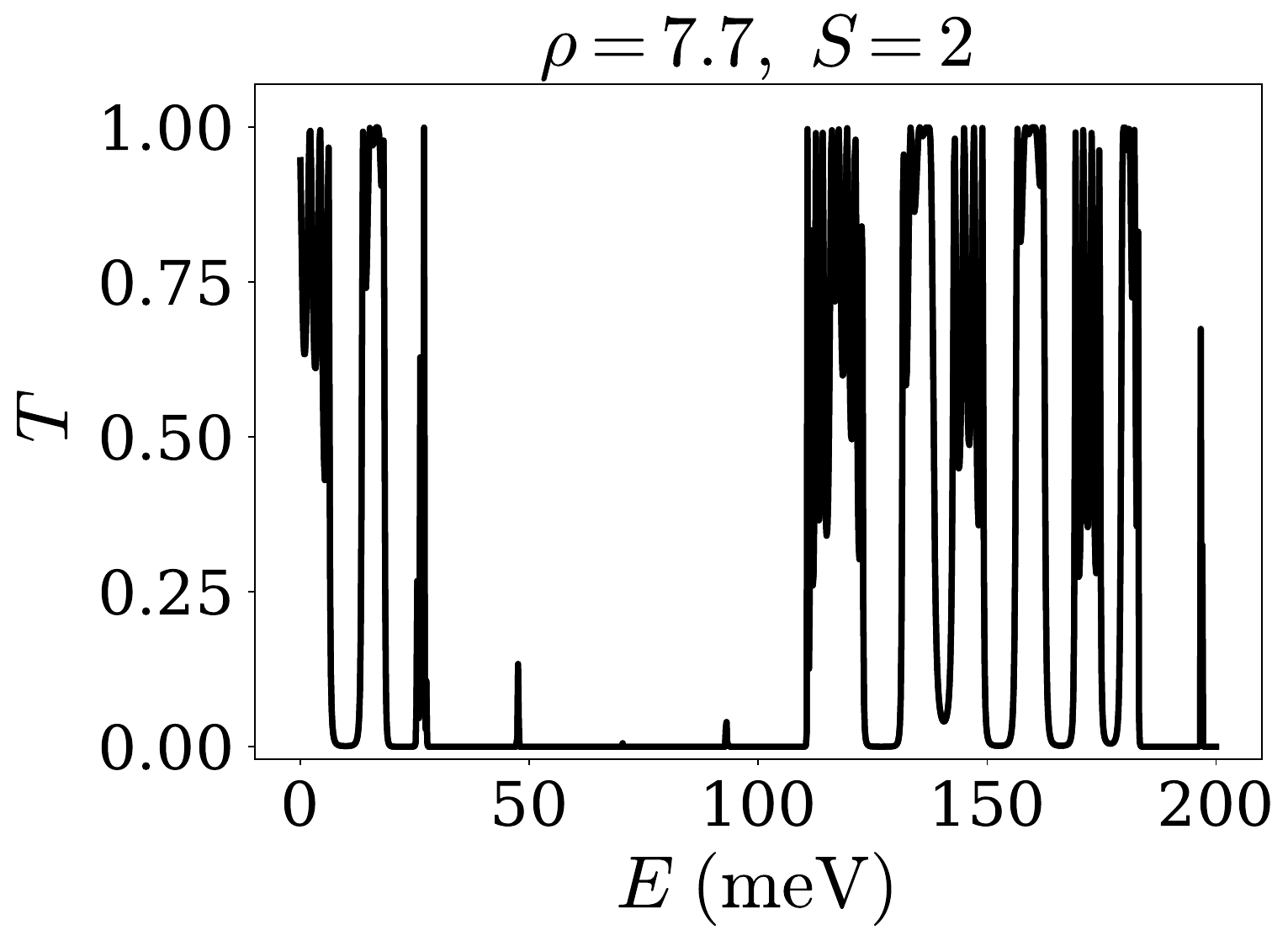}
\put(89,1.5){\bfseries (n)}
\end{overpic}%
\begin{overpic}[width=0.245\linewidth]{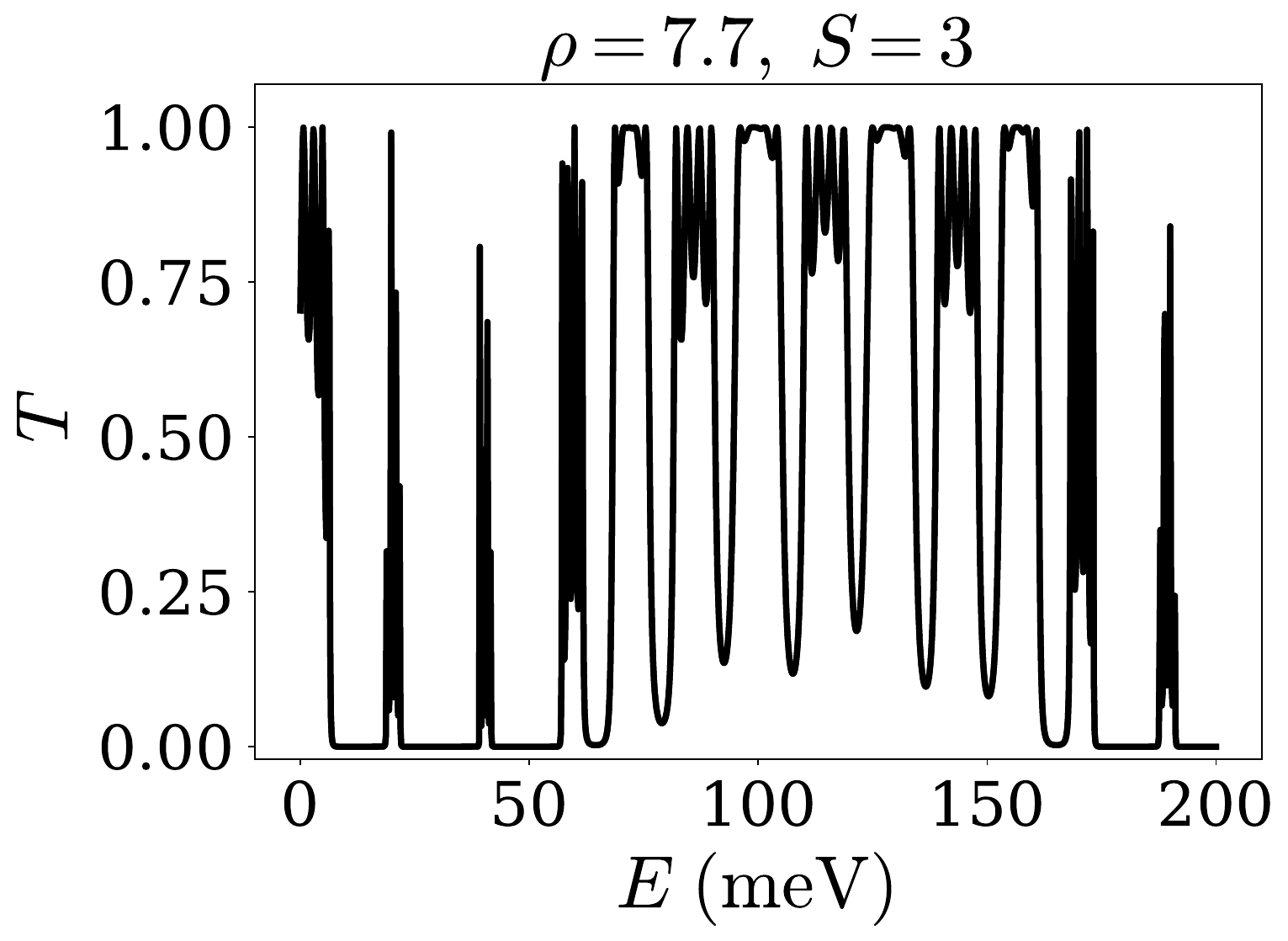}
\put(89,1.5){\bfseries (o)}
\end{overpic}%
\begin{overpic}[width=0.245\linewidth]{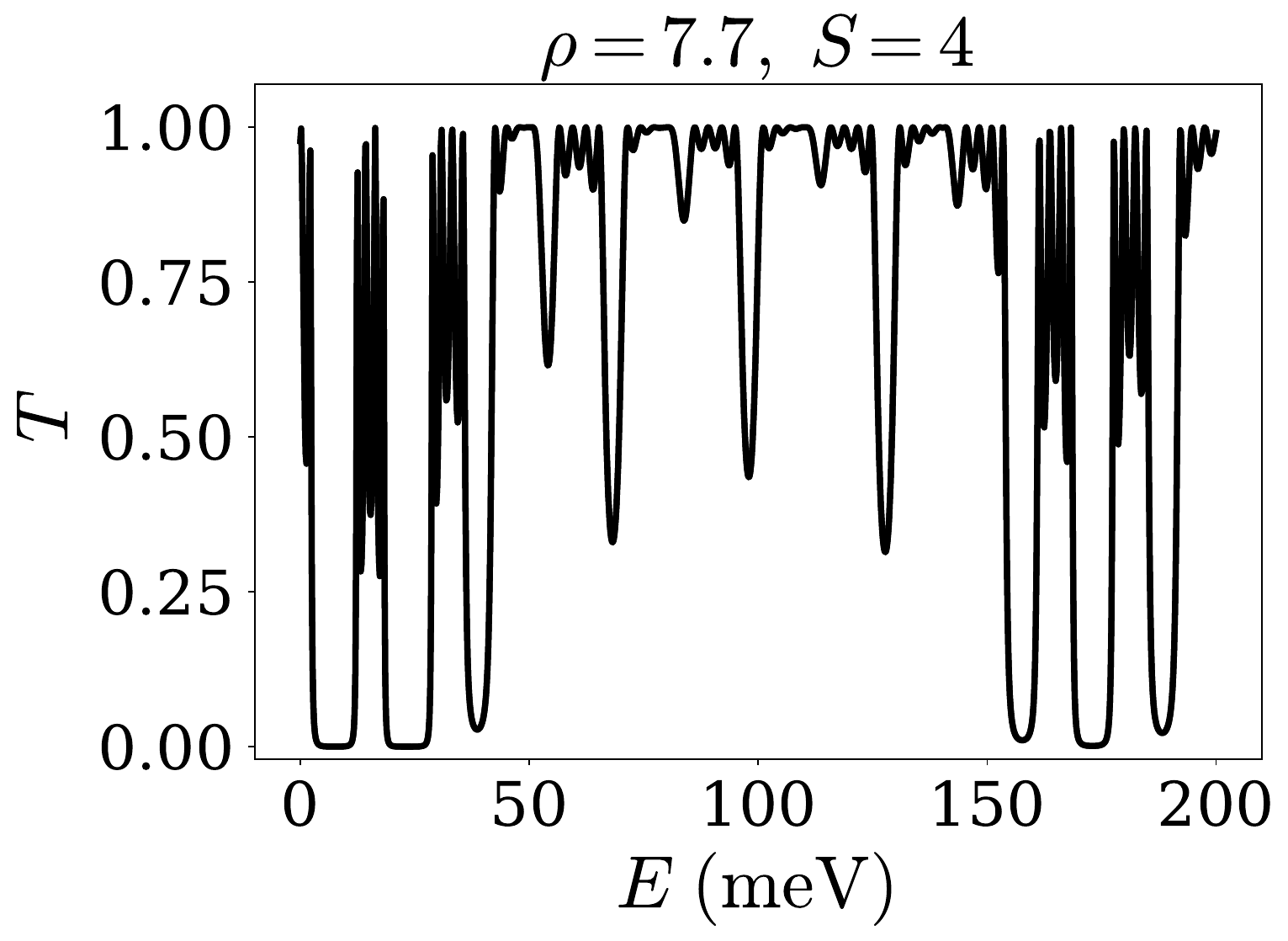}
\put(89,1.5){\bfseries (p)}
\end{overpic}
\begin{overpic}[width=0.245\linewidth]{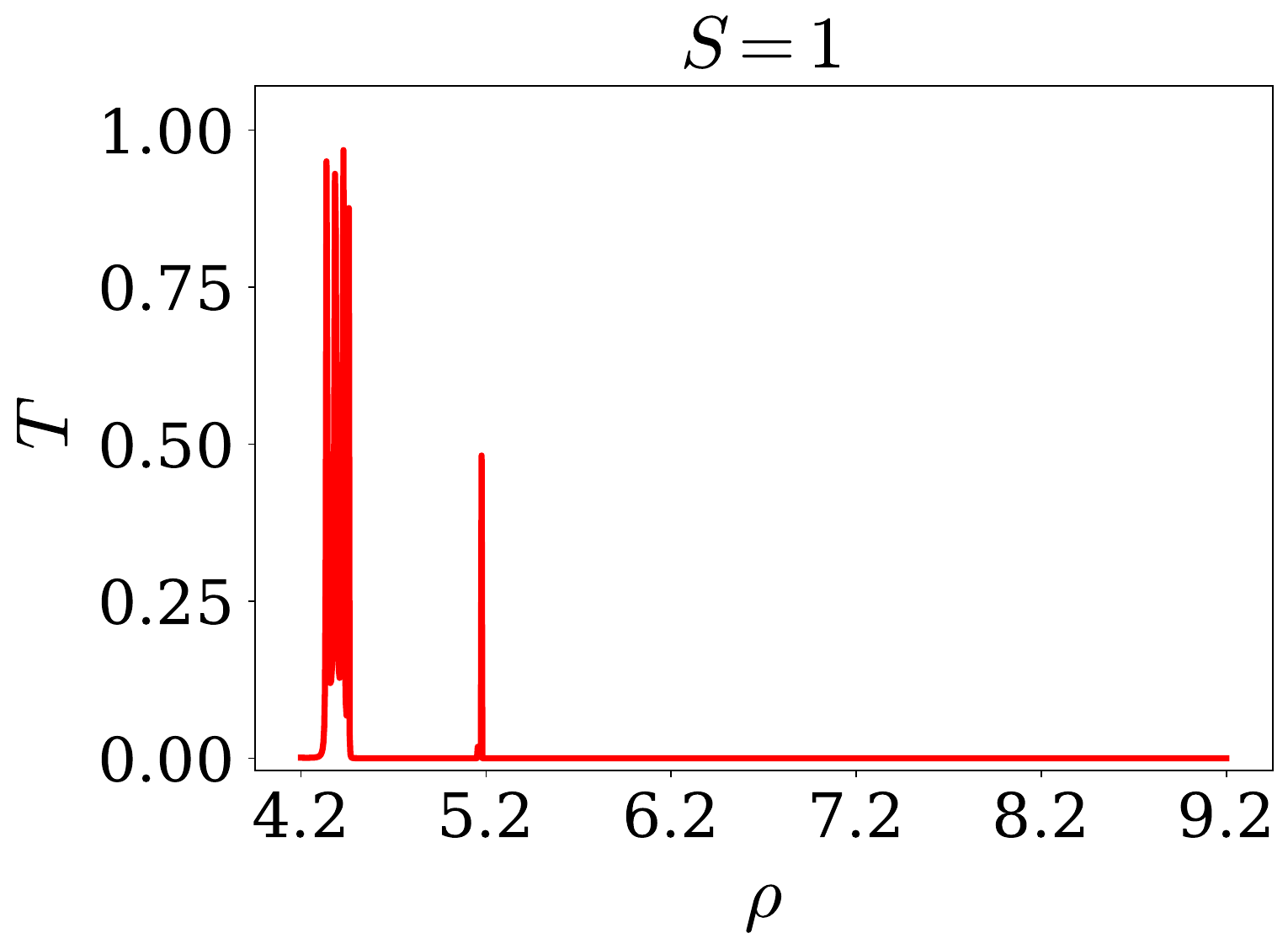}
\put(89,1.5){\bfseries (q)}
\end{overpic}%
\begin{overpic}[width=0.245\linewidth]{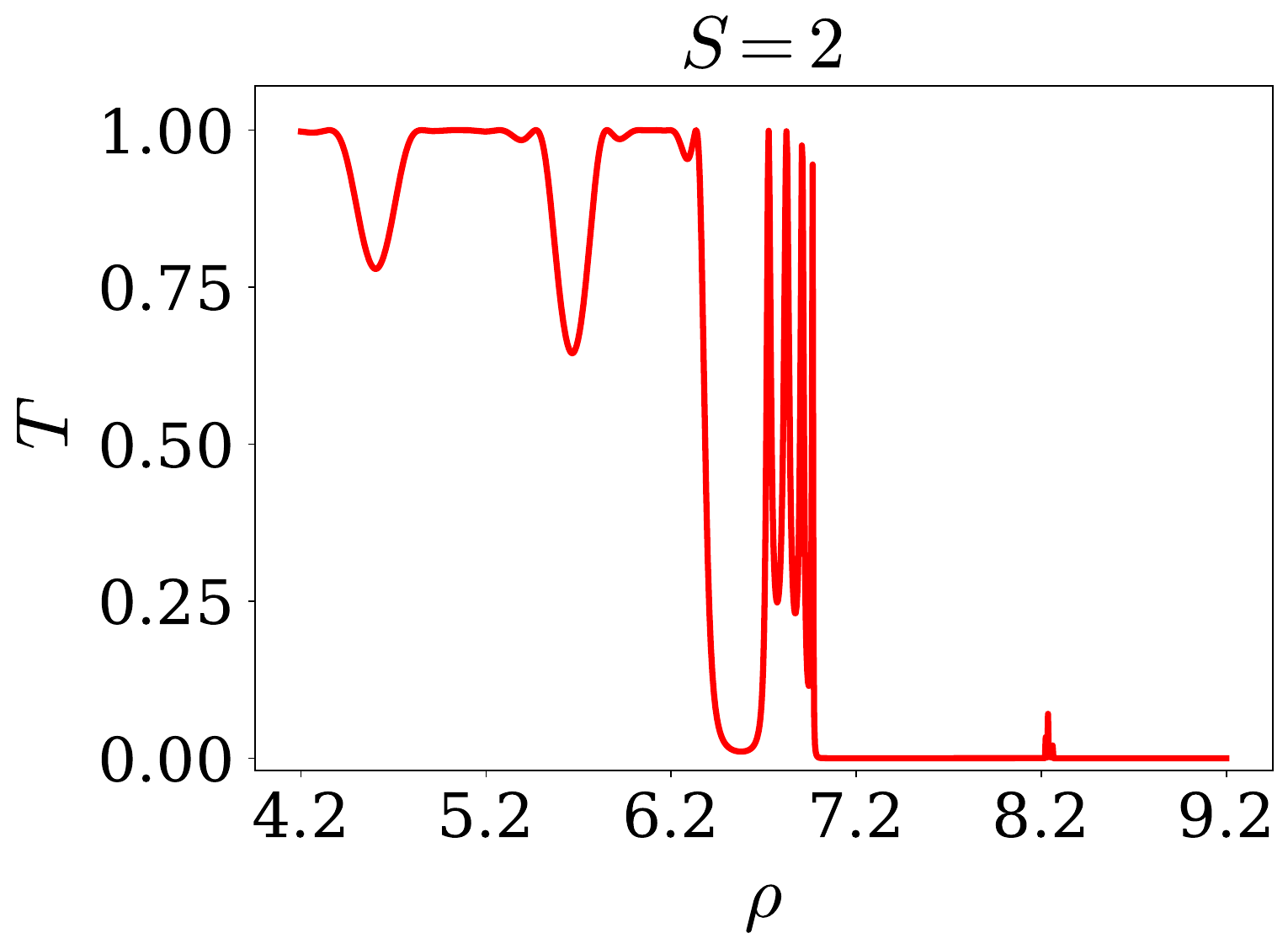}
\put(89,1.5){\bfseries (r)}
\end{overpic}%
\begin{overpic}[width=0.245\linewidth]{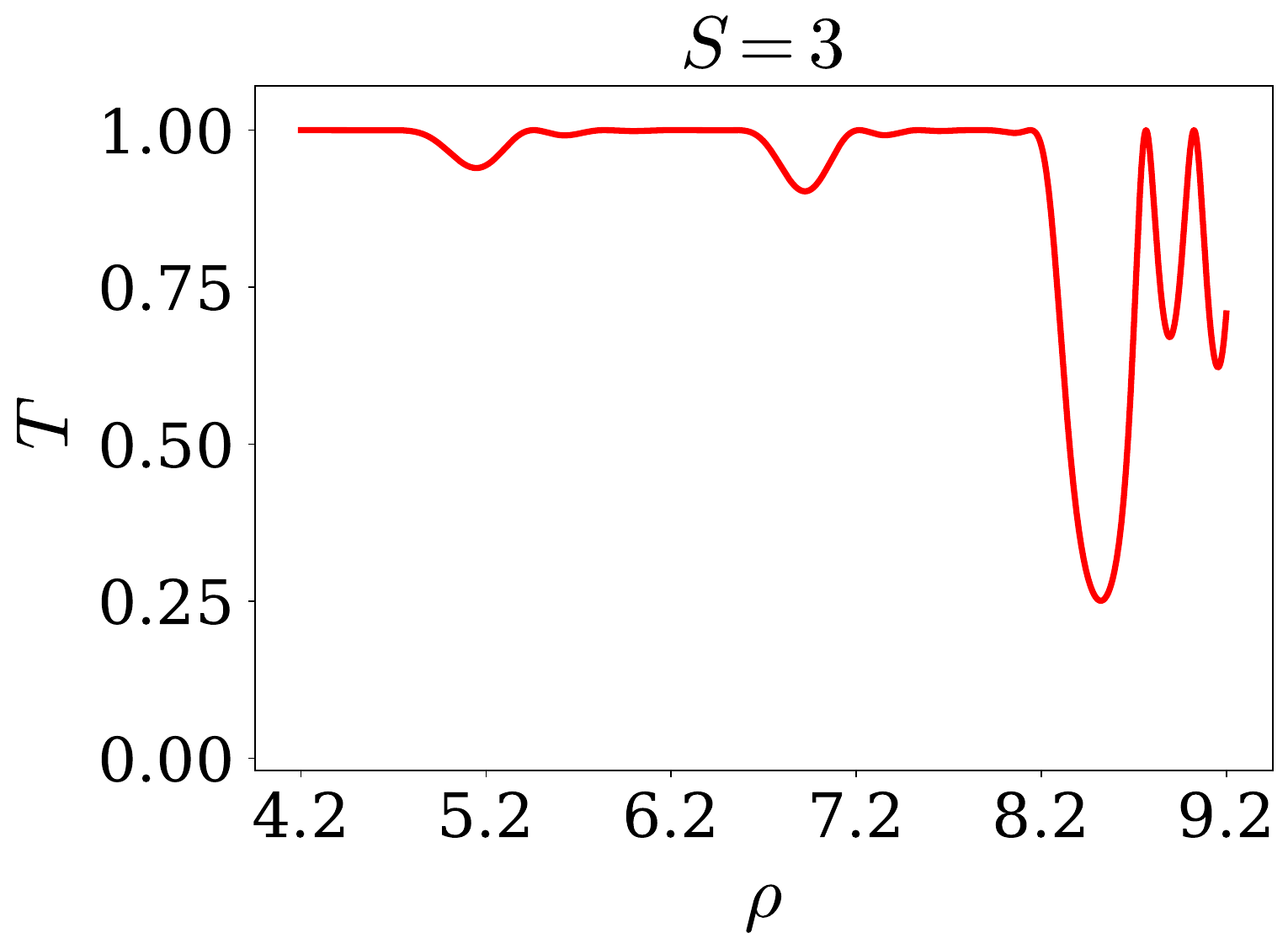}
\put(89,1.5){\bfseries (s)}
\end{overpic}%
\begin{overpic}[width=0.245\linewidth]{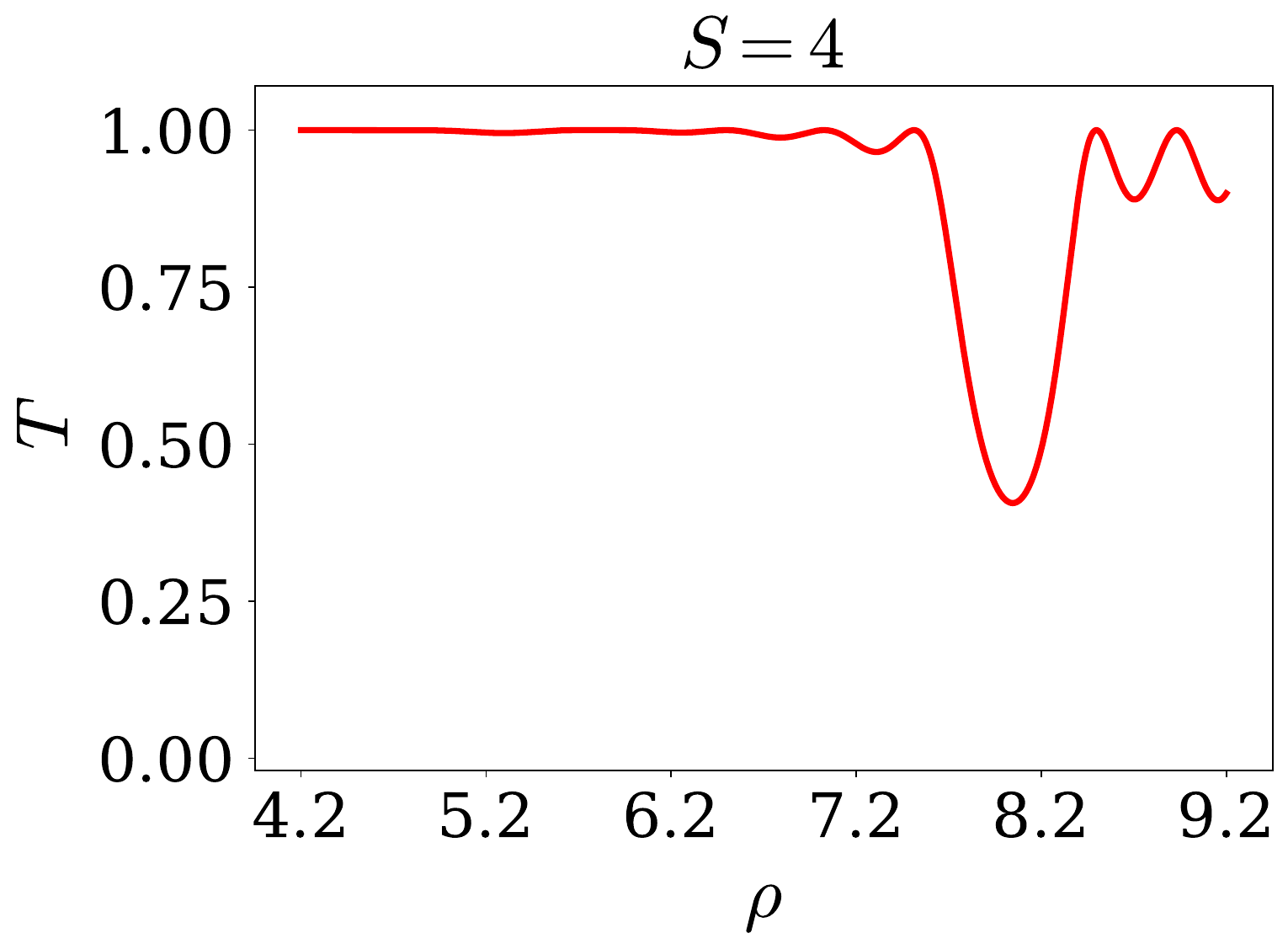}
\put(89,1.5){\bfseries (t)}
\end{overpic}
\caption{Same as Fig.~\ref{fig:01}, but for polyadic order $N=5$. The top row [(a)--(d)] shows the transmission density maps $T(E,\rho)$ for Cantor stages $S=1$ to $4$, and the remaining rows [(e)--(t)] show the corresponding sectional transmission profiles.}
\label{fig:01n_5}
\end{figure*}
\begin{figure*}[t]
\centering
\begin{minipage}[t]{0.48\textwidth}
\centering
\textbf{$N = 4$}\\[0.2cm]

\begin{overpic}[width=0.9\textwidth]{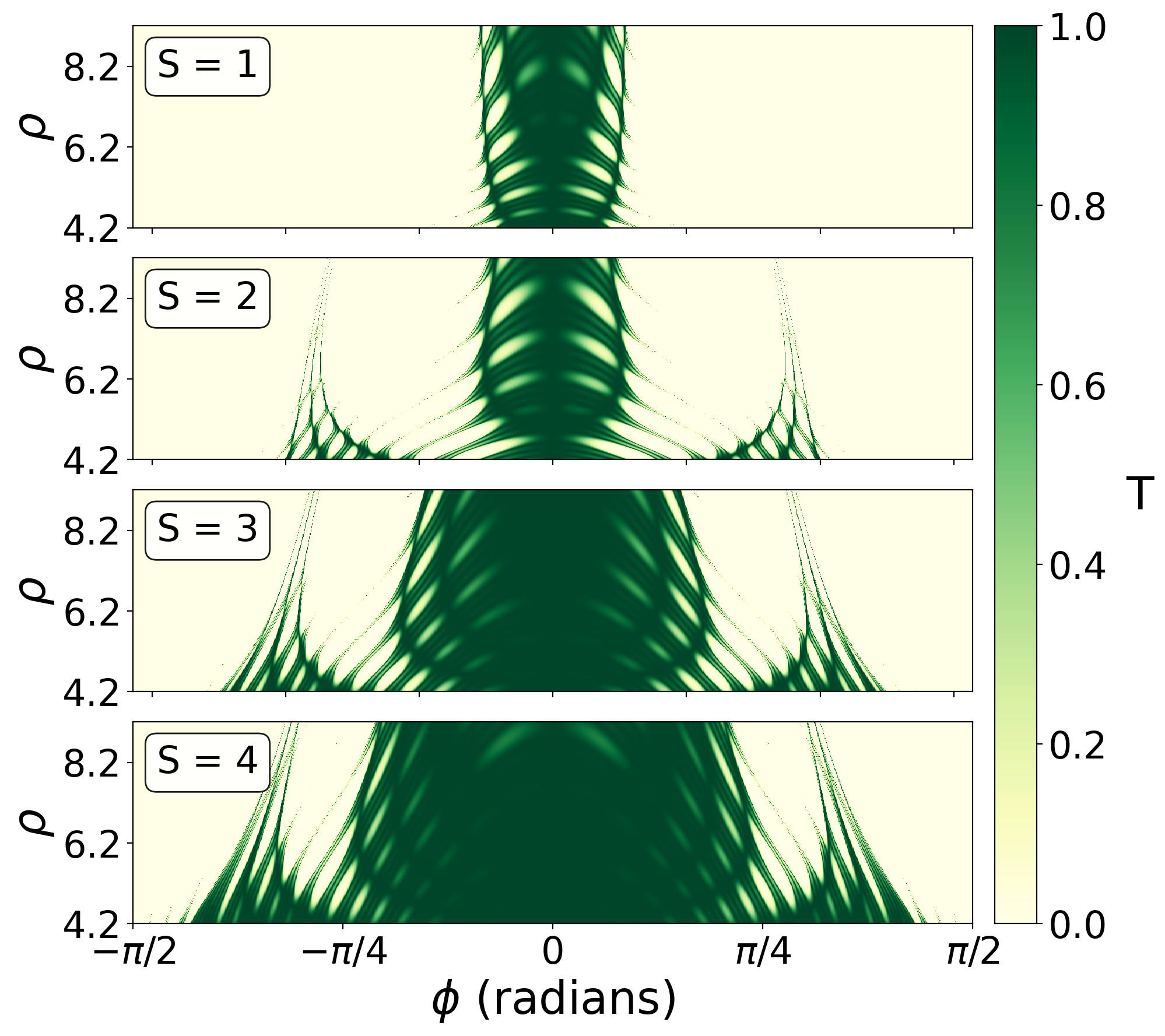}
\put(95,2){\bfseries (a)}
\end{overpic}\\[0.1cm]

\begin{overpic}[width=0.9\textwidth]{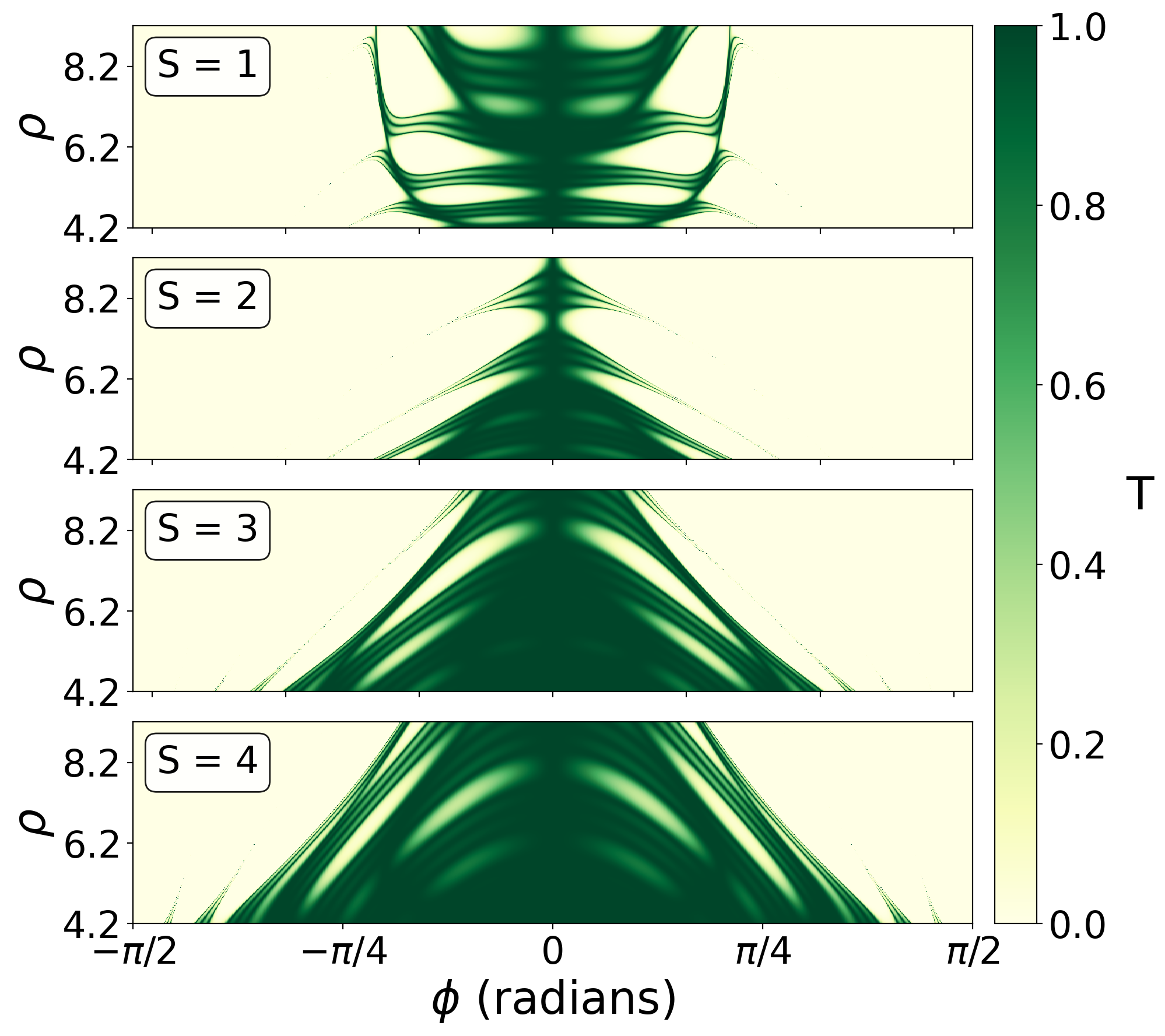}
\put(95,2){\bfseries (b)}
\end{overpic}\\[0.1cm]

\begin{overpic}[width=0.9\textwidth]{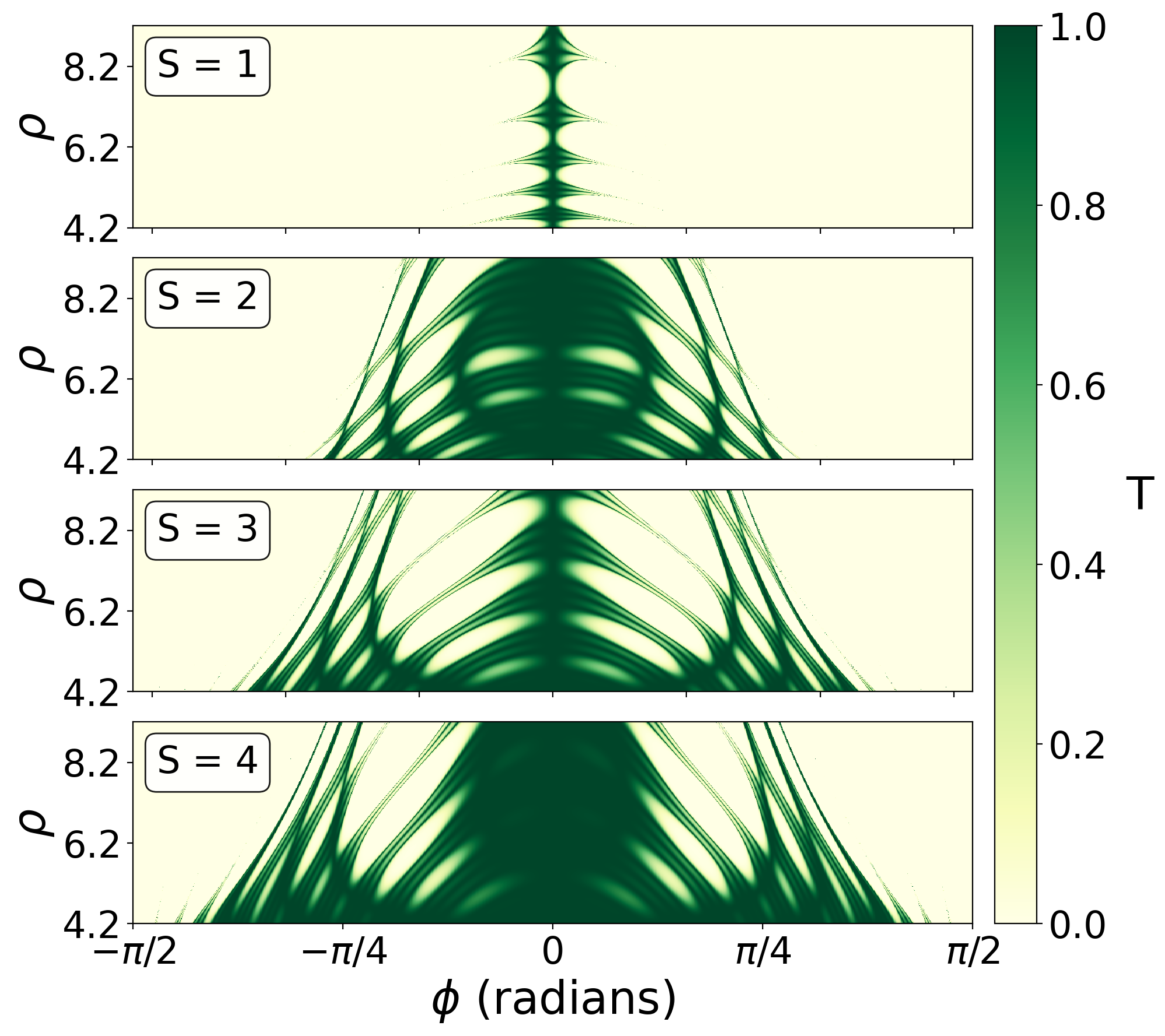}
\put(95,2){\bfseries (c)}
\end{overpic}
\end{minipage}%
\hfill
\begin{minipage}[t]{0.48\textwidth}
\centering
\textbf{$N = 5$}\\[0.2cm]

\begin{overpic}[width=0.9\textwidth]{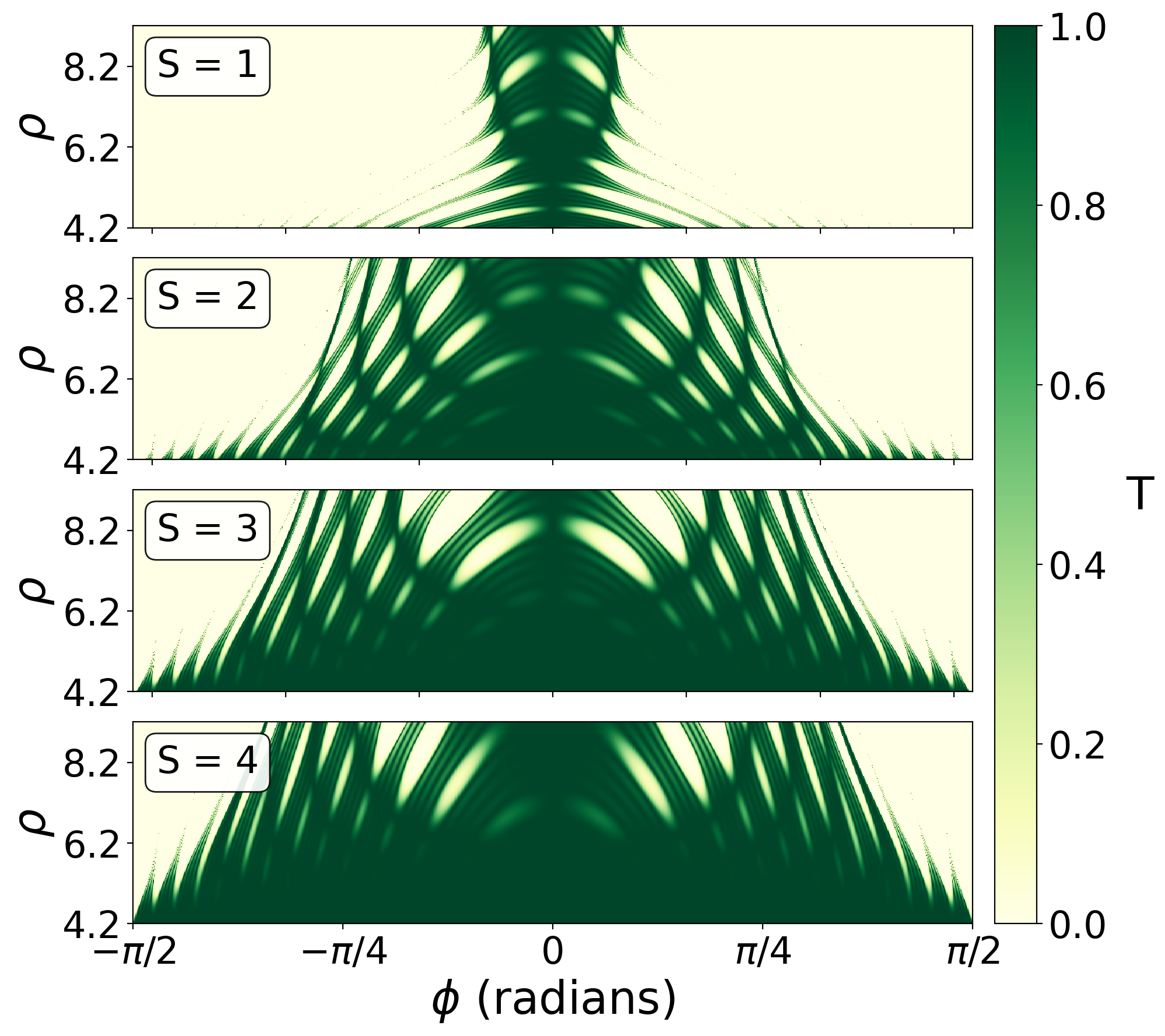}
\put(95,2){\bfseries (d)}
\end{overpic}\\[0.1cm]

\begin{overpic}[width=0.9\textwidth]{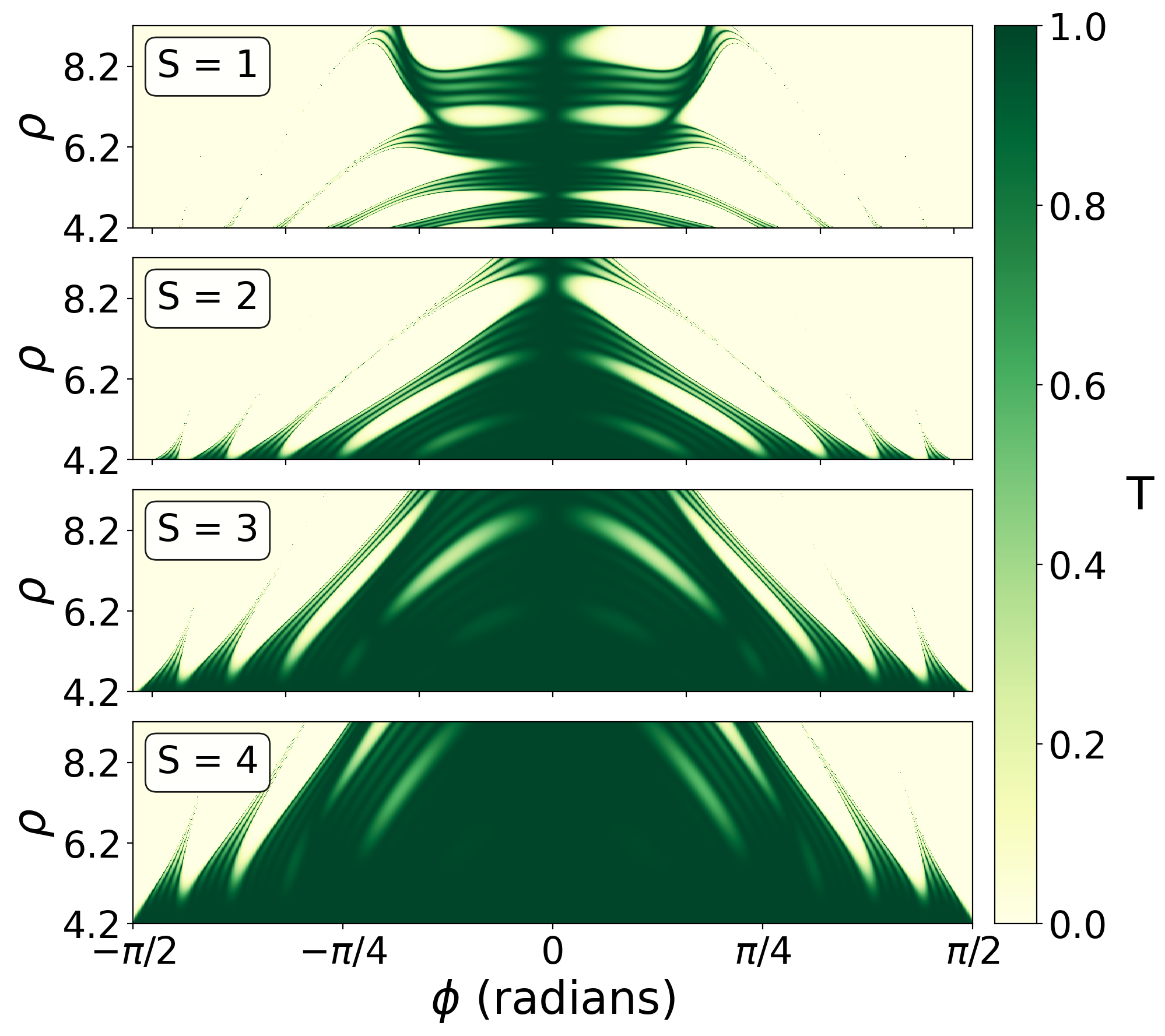}
\put(95,2){\bfseries (e)}
\end{overpic}\\[0.1cm]

\begin{overpic}[width=0.9\textwidth]{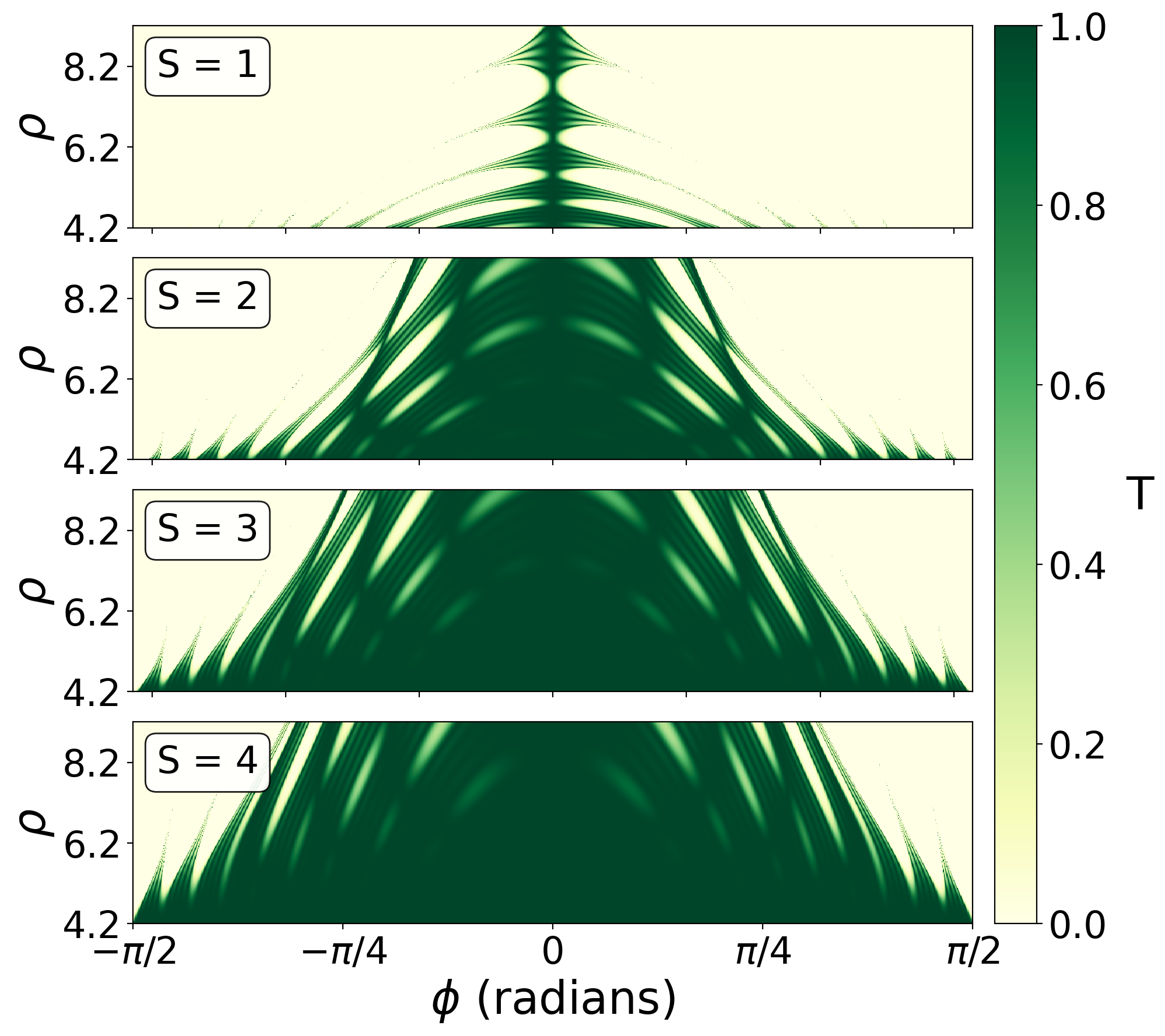}
\put(95,2){\bfseries (f)}
\end{overpic}
\end{minipage}

\caption{
Density maps of the transmission probability $T(\phi,\rho)$ as a function of the incident angle $\phi$ and the scaling parameter $\rho$, for successive Cantor stages $S=1$ to $4$. The left column [(a)--(c)] corresponds to $N=4$ and the right column [(d)--(f)] to $N=5$. The three rows correspond to the $n$--$n$--$n$ configuration, $E>V_0$ [(a) and (d)]; the $n$--$p$--$n$ configuration, $E<V_0$ [(b) and (e)]; and the Dirac point, $E=V_0$ [(c) and (f)]. In all panels $V_0=100$~meV, with $|E-V_0|=40$~meV for the $n$--$n$--$n$ and $n$--$p$--$n$ configurations; all remaining parameters are the same as in Fig.~\ref{fig:01}.
}
\label{fig:02}
\end{figure*}

We now analyze the transmission probability given by Eq.~\eqref{tn1n2ns}, using the geometric quantities $b_S$ and $r_q^{(S)}$ of Eqs.~\eqref{bs} and~\eqref{rq} together with the single-barrier transfer matrix derived above. Figures~\ref{fig:01} and~\ref{fig:01n_5} illustrate the transmission probability $T(E,\rho)$ of Dirac electrons through the polyadic Cantor electrostatic barrier for polyadic orders $N=4$ and $N=5$, respectively. In each figure, the columns correspond to successive Cantor stages $S=1$ to $S=4$. The top row displays the density maps $T(E,\rho)$ as a function of the Fermi energy $E$ and the scaling parameter $\rho$. The subsequent three rows present sectional transmission profiles $T(E)$ at fixed values of the scaling parameter ($\rho=5.7, 6.7, 7.7$), and the bottom row shows $T(\rho)$ evaluated at the Dirac point $E=V_0$. The angular response of the system is analyzed separately in Fig.~\ref{fig:02}, which displays the transmission probability $T(\phi,\rho)$ for the three doping configurations of physical interest: the $n$--$n$--$n$ regime ($E>V_0$), the $n$--$p$--$n$ regime ($E<V_0$), and the Dirac point ($E=V_0$).\\
\paragraph{Evolution from superlattice resonances to a near-transparent regime.} At the first Cantor stage, $S=1$, the PCP reduces to a periodic array of $N$ identical rectangular barriers, that is, a finite graphene superlattice~\cite{park2008anisotropic, barbier2009dirac, bai2007klein}. The density maps, Figs.~\ref{fig:01}(a) and~\ref{fig:01n_5}(a), are accordingly dominated by regions of low transmission, traversed by sharp bright bands. These features correspond to the superlattice resonances of the finite multi-barrier array, which are Fabry-P\'{e}rot-like in nature~\cite{shytov2008klein, young2009quantum}. These resonances arise from the constructive interference of partial waves scattered at successive interfaces. The corresponding line profiles, shown in Figs.~\ref{fig:01}(e, i, m) and~\ref{fig:01n_5}(e, i, m), exhibit dense Fabry-P\'{e}rot oscillations at low energies, where the pseudospin mismatch at the $n$--$p$ interfaces generates strong interference contrast, and collapse to zero above a threshold energy. From Eq.~\eqref{vec_inside}, propagation inside the barrier requires $|E-V_0| > E|\sin\phi|$, so for $\phi=40^\circ$ and $V_0=100$~meV the longitudinal wavevector $\widetilde{q}_x$ is imaginary throughout $61~\text{meV} \lesssim E \lesssim 280$~meV. This window covers the Dirac point and the entire $n$--$n$--$n$ range shown, and transport across it proceeds only through evanescent modes, whose strong decay across the wide barriers of stage $S=1$ suppresses the transmission almost completely.\\
\indent
As the Cantor stage increases, the initiator length $L$ is subdivided into a hierarchy of progressively thinner sub-barriers separated by superperiodic gaps, governed by Eqs.~\eqref{bs} and~\eqref{rq}. Each new hierarchical level introduces additional scattering interfaces and an expanding set of phase-coherent paths available to the Dirac electrons. Interference among these paths progressively populates the previously suppressed region of the $(E,\rho)$ plane with quasi-bound states~\cite{matulis2008quasibound, bai2017massless}, producing the densely structured maps of Figs.~\ref{fig:01}(b, c) and~\ref{fig:01n_5}(b, c).
The line profiles at $S=2$, Figs.~\ref{fig:01}(f, j, n) and~\ref{fig:01n_5}(f, j, n), capture the onset of this evolution: the subdivision of each stage-$1$ barrier into $N$ thinner segments weakens the evanescent decay across each of them, and transmission begins to appear in the window that was completely dark at $S=1$, while the background there remains well below unity. The filling sets in first at the smaller values of $\rho$, for which the barriers contract most rapidly with stage, and it is already far more advanced for $N=5$ than for $N=4$ at the same stage.
By the fourth Cantor stage, as shown in Figs.~\ref{fig:01}(d) and~\ref{fig:01n_5}(d), the increase in the number of interfaces is sufficient for the transmission background to approach unity over most of the $(E,\rho)$ plane, with the residual interference appearing only as narrow $\rho$-dependent dips. The corresponding line profiles at $S=3$, $4$, shown in Figs.~\ref{fig:01}(g, h, k, l, o, p) and~\ref{fig:01n_5}(g, h, k, l, o, p), confirm this evolution by displaying a high-transmission background close to unity punctuated by sharp oscillatory features whose positions depend sensitively on $\rho$. The wide suppressed region of the $S=1$ superlattice is thus progressively filled by hierarchical interference channels at higher stages opened by the recursive Cantor segmentation.\\
\paragraph{Transmission at the Dirac point.} The transmission $T(\rho)$ at the Dirac point $E=V_0$, displayed in the bottom row of each figure, provides a particularly direct probe of the hierarchical transport mechanism. At $E=V_0$ the longitudinal wavevector inside the barriers becomes purely imaginary, $\widetilde{q}_x = iq_y$, and transmission is carried entirely by evanescent modes. For $S=1$ the barriers are at their widest, $b_1 = (L/N)\left[1-(N-1)/\rho\right]$ from Eq.~\eqref{bs}, so the evanescent decay across each of them is strong and the transmission remains very small over the entire range of $\rho$, as shown in Figs.~\ref{fig:01}(q) and~\ref{fig:01n_5}(q). As $S$ increases, the recursive thinning of the sub-barriers reduces the evanescent decay across each segment, while the multiple interfaces enable efficient inter-segment tunneling of the evanescent modes. This enhancement is already apparent at $S=2$, Figs.~\ref{fig:01}(r) and~\ref{fig:01n_5}(r): for $N=4$ the transmission rises to near unity only over narrow intervals of $\rho$ and remains suppressed elsewhere, whereas for $N=5$ it is already high over a substantial part of the range. At $S=3$ and $S=4$ the $N=5$ system approaches unity over wide $\rho$ intervals, while for $N=4$ the high-transmission regions remain interspersed with suppressed ones, as seen in Figs.~\ref{fig:01}(s, t) and~\ref{fig:01n_5}(s, t). The near-transparent regime is thus more fully developed for $N=5$ than for $N=4$ at the same stage, since the higher polyadic order generates a larger number $N^S$ of thinner sub-barriers and a correspondingly larger number of tunneling channels for evanescent modes.\\
\paragraph{Role of the polyadic order.} A direct comparison of Figs.~\ref{fig:01} and~\ref{fig:01n_5} demonstrates that the polyadic order $N$ does not alter the overall character of the transmission: in both cases, the spectrum evolves from a sparse superlattice resonance structure at $S=1$ to a near-transparent background at $S=4$. The role of $N$ is to control the rate at which this evolution proceeds. Since the number of sub-barriers at stage $S$ scales as $N^S$, a higher polyadic order produces a denser hierarchy of scattering interfaces at any given stage, and consequently the near-transparent regime is reached at a lower Cantor stage for $N=5$ than for $N=4$. The recursive Cantor geometry, encoded in the stage-dependent quantities $b_S$ and $r_q^{(S)}$, thus dictates the overall structure of the transport spectra, while $N$ sets how rapidly the near-transparent regime is reached.\\
\paragraph{Angular response and the role of pseudospin.}
The angular dependence of the transmission, $T(\phi,\rho)$, displayed in Fig.~\ref{fig:02}, reveals how the chiral nature of the Dirac carriers in graphene interplays with the hierarchical geometry of the Cantor potential. The three rows of Fig.~\ref{fig:02} correspond to the $n$--$n$--$n$, $n$--$p$--$n$, and Dirac-point configurations. At the first Cantor stage, the three regimes exhibit distinct angular signatures, each governed by a different physical mechanism. For the $n$--$n$--$n$ configuration, shown in Figs.~\ref{fig:02}(a) and~\ref{fig:02}(d), the carriers remain electron-like throughout the structure, and reflection arises from Fabry-P\'{e}rot interference between successive interfaces; the transmission is concentrated in an angular window around $\phi=0$ and structured by superlattice resonances along $\rho$. For the $n$--$p$--$n$ configuration, shown in Figs.~\ref{fig:02}(b) and~\ref{fig:02}(e), the carriers undergo electron-to-hole conversion at each interface, producing a pseudospin mismatch at oblique angles. The single $n$--$p$--$n$ junction in graphene satisfies $T(\phi=0)=1$ irrespective of barrier height and width, a manifestation of the Klein tunneling effect~\cite{katsnelson2006chiral}, while at oblique angles the pseudospin mismatch modulates the transmission. The angular range follows from the condition $|E-V_0| > E|\sin\phi|$ for propagation inside the barrier: with $|E-V_0| = 40$~meV and $V_0 = 100$~meV this gives $|\phi| < 41.8^\circ$ in the $n$--$p$--$n$ case but only $|\phi| < 16.6^\circ$ in the $n$--$n$--$n$ case, beyond which transport is again carried by evanescent modes. The corresponding $S=1$ panels confirm these estimates, with perfect transmission preserved at $\phi=0$ by Klein tunneling. At the Dirac point, as shown in Figs.~\ref{fig:02}(c) and~\ref{fig:02}(f), the evanescent modes inside the barrier decay as $e^{-|\widetilde{q}_x|x}$ with $|\widetilde{q}_x| = q_f|\sin\phi|$, which grows monotonically with $|\phi|$ away from normal incidence, sharply confining the $S=1$ transmission to a narrow strip around normal incidence, the narrowest angular selectivity of the three regimes.\\
\indent
A central observation of Fig.~\ref{fig:02} is that these three distinct angular characteristics progressively disappear as the Cantor stage increases. For $S \geq 2$, the high-transmission region of each configuration extends to substantially larger angles, and by $S=4$ it covers most of the angular range in all three regimes, approaching $|\phi| \to \pi/2$ for $N=5$. The origin of this broadening is twofold. First, the hierarchical thinning of the sub-barriers reduces both the pseudospin mismatch accumulated across each $n$--$p$ interface and the evanescent decay across each segment, weakening the angular selectivity that governs the $S=1$ transmission. Second, the multiple superperiodic length scales open additional phase-coherent transmission paths at oblique angles that are suppressed at $S=1$. Klein tunneling still protects $T=1$ at $\phi=0$, but the additional features at oblique angles arise from hierarchical interference rather than from Klein tunneling itself. The polyadic order $N$ again controls the rate of this evolution; for $N=5$, the high-transmission window is broader and shows more $\rho$-dependent resonance features than for $N=4$ at the same Cantor stage, as seen by comparing Figs.~\ref{fig:02}(a--c) and~\ref{fig:02}(d--f).\\
\indent
The transmission characteristics described above are governed by two main effects. Increasing the Cantor stage $S$ takes the system from a sparse superlattice spectrum at $S=1$, with Fabry-P\'{e}rot resonances and a suppressed region near $E=V_0$, to a near-transparent regime at $S=4$ populated by narrow interference dips. This evolution is visible both in the $(E,\rho)$ plane and in the angular response. It originates from the increase in the number of sub-barriers at higher Cantor stages, which generates a larger set of phase-coherent paths for the Dirac electrons. On the other hand, increasing the polyadic order $N$ controls the rate at which the near-transparent regime is reached, without altering the overall shape of the transmission spectrum. The interplay between the hierarchical geometry and the chirality of the Dirac carriers is most clearly seen in the angular response. At $S=1$, the three regimes show distinct angular signatures from single-barrier mechanisms (Fabry-P\'{e}rot interference, Klein tunneling, and evanescent-mode narrowing); these progressively disappear at higher $S$, where the recursive Cantor geometry dominates. The near-transparent regime emerging at higher $S$ is the natural setting in which to identify scaling laws governing the transmission. We develop these scaling laws in the following subsections for the Cantor stage $S$, the potential height $V_0$, the initiator length $L$, and the angle of incidence $\phi$.
%
%
\subsection{Scaling laws for transmission}
\label{Sec:ScalingRule}
The polyadic Cantor construction distributes the electrostatic barrier in a self-similar and hierarchical manner, producing scale-dependent and self-similar transport properties. In this subsection, we identify scaling laws governing the transmission in the near-transparent regime, with respect to the Cantor stage $S$, the potential height $V_0$, the initiator length $L$, and the angle of incidence $\phi$. These scaling laws are displayed in log--log representations throughout the following analysis.\\
\paragraph{Scaling with Cantor stages.}
To analyze the scaling behavior of $T_S(E)$ with respect to the Cantor stage, as given in Eq.~\eqref{tn1n2ns}, it is convenient to introduce the deviation from perfect transmission
\begin{equation}
X_S(E) \equiv \frac{1 - T_S(E)}{T_S(E)}
= |M_{12}^{(S)}|^{2}\,
\prod_{q=1}^{S} U_{N-1}^{2}(\Gamma_q^{(S)}).
\label{eq:XSdef}
\end{equation}
This equation shows that the scaling of $X_S$ with the Cantor stage is determined by two independent factors: the squared single-barrier transfer matrix element $|M_{12}^{(S)}|^{2}$ which depends on $S$ through $b_S$, and the product of the square of the CPs over all $S$ hierarchical levels, whose arguments $\Gamma_q^{(S)}$ depend on $S$. Equation~\eqref{bs} can also be written as
\begin{equation}
b_S = \frac{L}{(N\lambda)^{S}},
\quad \text{with} \quad
\lambda \equiv \frac{\rho}{\rho - (N-1)},
\label{eq:bS_lambda}
\end{equation}
where $\lambda > 1$ is the \emph{geometric contraction factor}: at each Cantor stage, the single-barrier width shrinks by an additional factor of $N\lambda$. For sufficiently large $S$, the phase $\widetilde{q}_x b_S$ accumulated across a single barrier becomes small, and the off-diagonal element of the single-barrier transfer matrix as expressed in Eq.~\eqref{m12} can be approximated as $|M_{12}^{(S)}| \propto \widetilde{q}_x b_S$. Therefore,
\begin{equation}
|M_{12}^{(S)}|^{2} \propto b_S^{2} 
= \frac{L^{2}}{(N\lambda)^{2S}},
\label{eq:M12scaling}
\end{equation}
with a proportionality constant that depends on $V_0$, $E$, and $\phi$ but is independent of $S$. For the Chebyshev product, the superperiodic distance $r_q^{(S)}$ together with the longitudinal wavevector $q_x = E\cos\phi/(\hbar v_f)$ outside the barrier, determine the phase accumulated over the superperiodic distance $r_q^{(S)}$, given by
\begin{equation}
q_x\,r_q^{(S)} = \frac{E\,\mathcal{R}}{(N\lambda)^{\,S+1-q}},
\quad
\mathcal{R} \equiv \frac{L(1 + 1/\rho)\,\lambda\cos\phi}{\hbar v_f}.
\label{eq:phase_q}
\end{equation}
The accumulated phase $q_x\,r_q^{(S)}$ depends on $S$ and $q$ only through the combination $S + 1 - q$. Consequently, for any two stages $S_j > S_i$ with $m = S_j - S_i$, the phases at level $q$ of the stage $S_i$ system match those at level $q' = m + q$ of the stage $S_j$ system (for $q = 1, \ldots, S_i$):
\begin{equation}
q_x\,r_{q'}^{(S_j)} = q_x\,r_{q}^{(S_i)}.
\label{eq:phase_selfsim}
\end{equation}
The outermost $S_i$ hierarchical levels of any stage $S_j$ system share identical Bloch phases with the corresponding levels of a stage $S_i$ system. This phase matching allows us to relate the Chebyshev products of stages $S_i$ and $S_j$. Splitting the stage $S_j$ product into inner ($q = 1, \ldots, m$) and outer ($q = m+1, \ldots, S_j$) contributions, and using Eq.~\eqref{eq:phase_selfsim} for the outer levels, gives
\begin{equation}
\prod_{q=1}^{S_j} U_{N-1}^{2}(\Gamma_q^{(S_j)})
= \left[\prod_{q=1}^{m} U_{N-1}^{2}(\Gamma_q^{(S_j)})\right]
\prod_{q=1}^{S_i} U_{N-1}^{2}(\Gamma_q^{(S_i)}).
\label{eq:prodsplit}
\end{equation}
\begin{figure}[thb]
\centering
\begin{overpic}[width=0.495\linewidth]{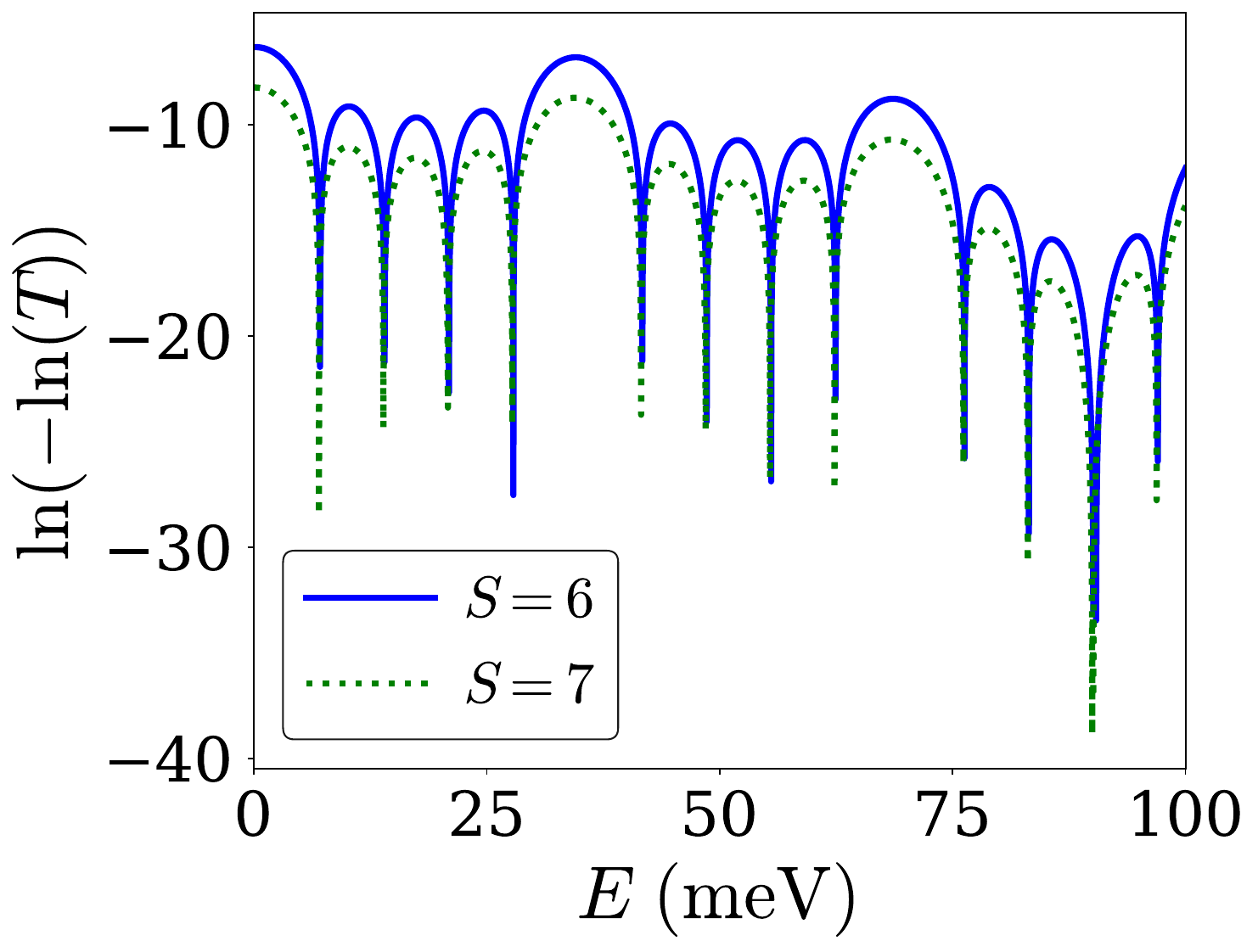}
\put(90,1.5){\bfseries (a)}
\end{overpic}%
\begin{overpic}[width=0.495\linewidth]{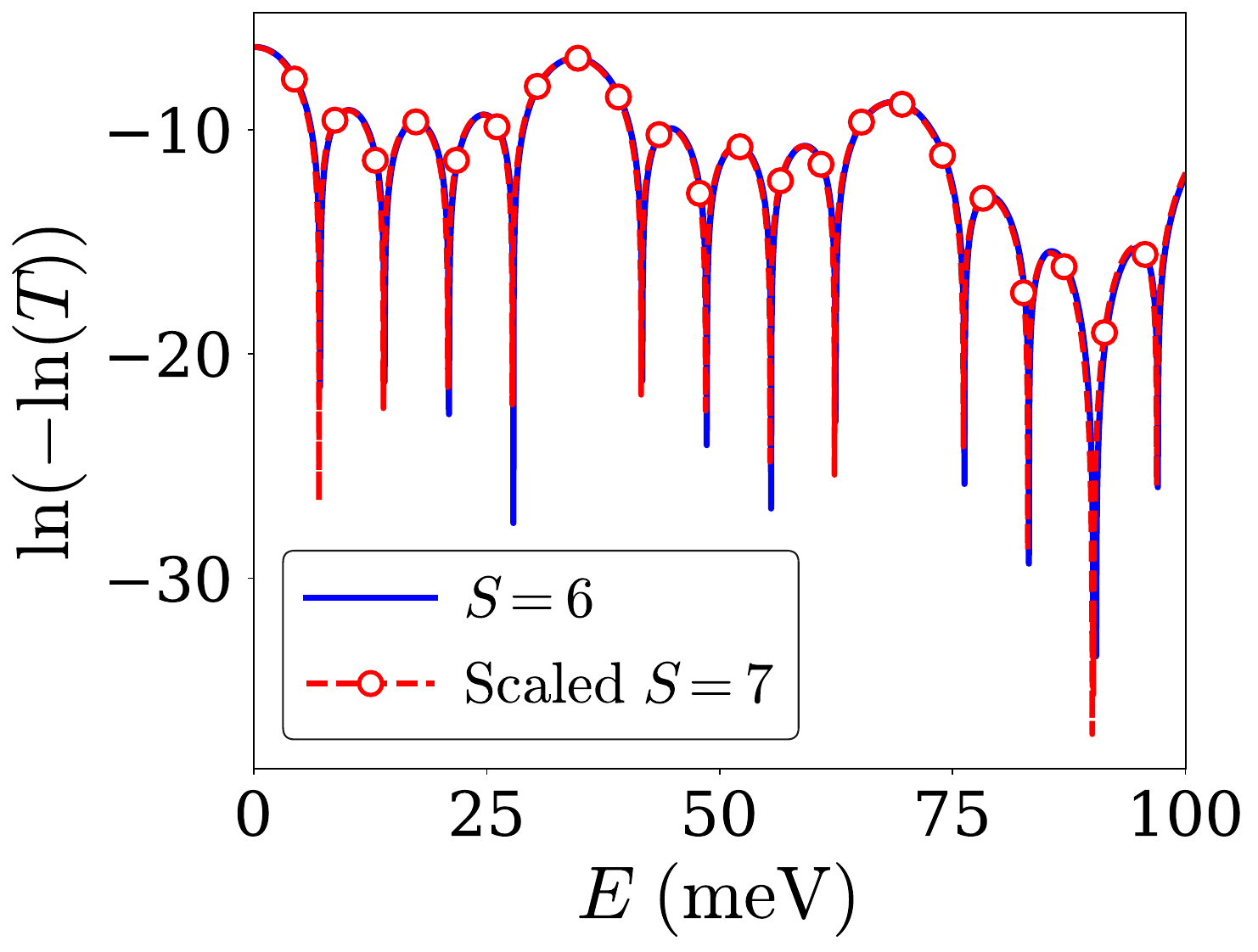}
\put(90,1.5){\bfseries (b)}
\end{overpic}\\[0.15cm]

\begin{overpic}[width=0.495\linewidth]{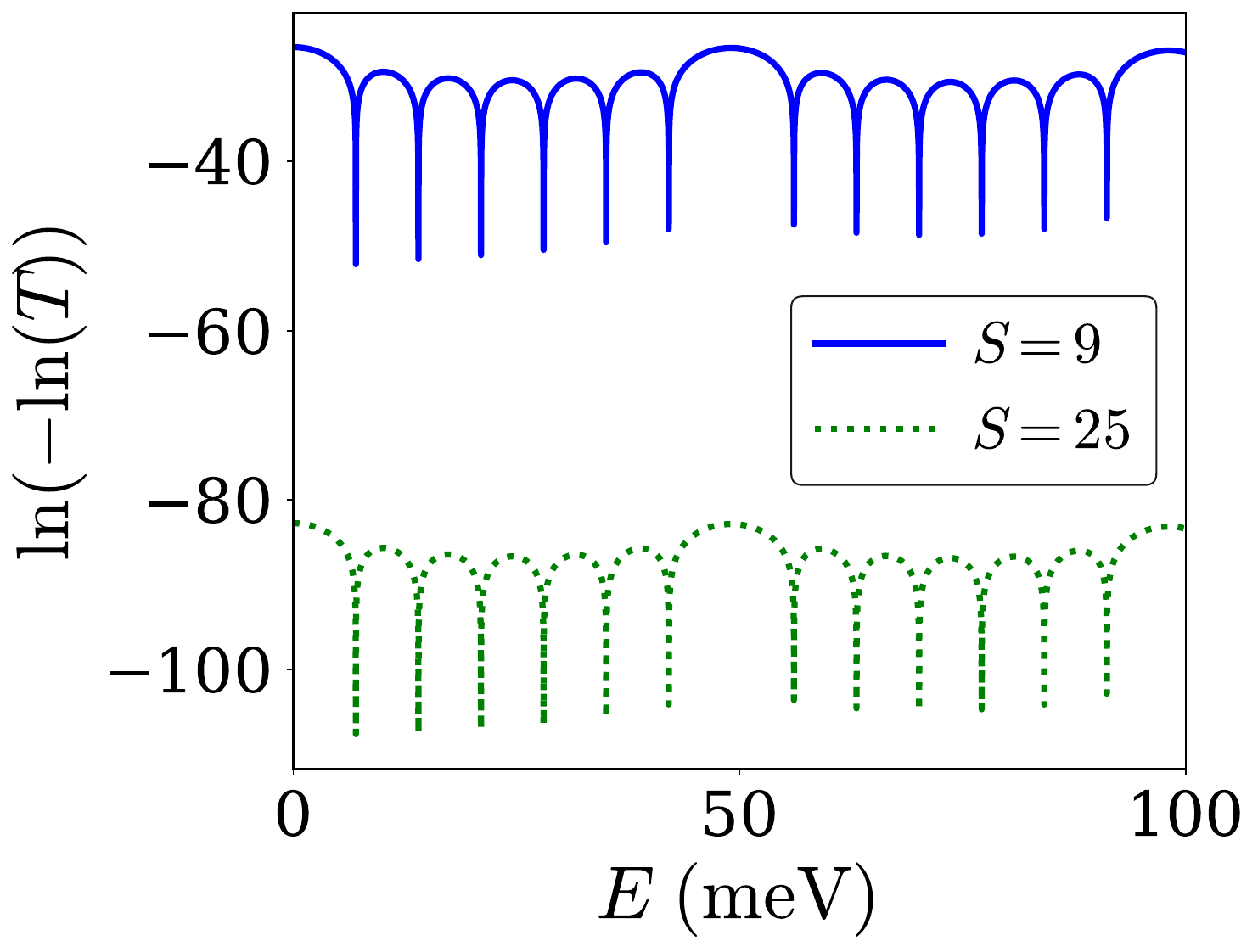}
\put(90,1.5){\bfseries (c)}
\end{overpic}%
\begin{overpic}[width=0.495\linewidth]{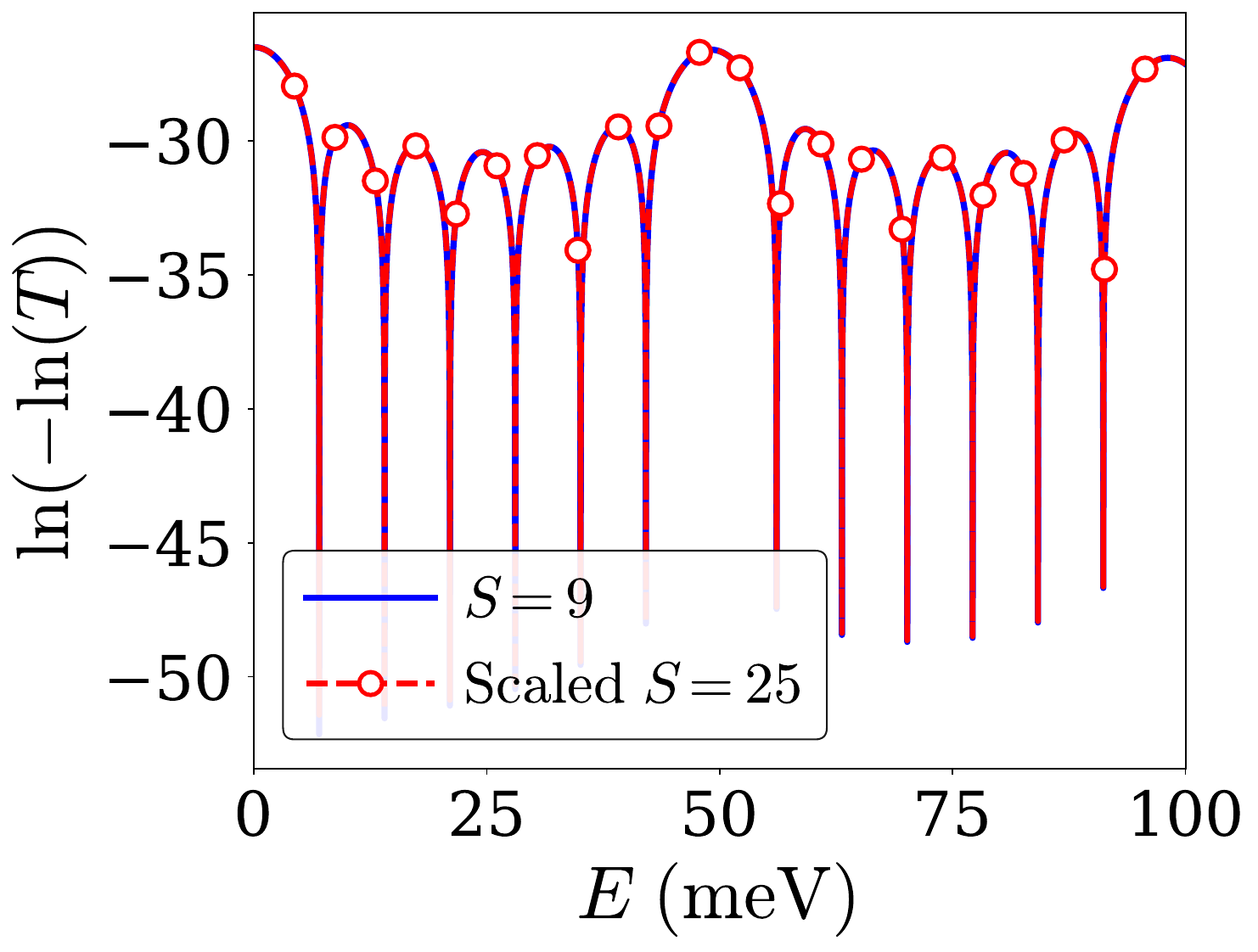}
\put(90,1.5){\bfseries (d)}
\end{overpic}

\caption{Double-logarithmic scaling of the transmission with the Cantor stage $S$. Plots (a) and (b) correspond to $\phi=30^\circ$, $V_0=50$~meV, $L=300$~nm, $N=5$, and $\rho=6.5$; plots (c) and (d) to $\phi=30^\circ$, $V_0=50$~meV, $L=300$~nm, $N=7$, and $\rho=7.25$. Plots (a) and (c) show the unscaled quantity $\ln(-\ln T_S)$ as a function of the Fermi energy $E$ for two Cantor stages, for which the curves do not coincide [$S_i=6$, $S_j=7$ in (a); $S_i=9$, $S_j=25$ in (c)]. Plots (b) and (d) show the same data after rescaling according to Eq.~\eqref{eq:scaling_law}, which produces a collapse of the curves onto a common envelope.}
\label{fig:scalingcollapse}
\end{figure}

\noindent
To evaluate their contribution, consider the level $q$ unit cell transfer matrix $M_q$, formed from $N$ identical copies of the level ($q-1$) unit cell separated by superperiodic distance $r_q^{(S_j)}$:
\begin{equation}
M_q = \bigl[P_q^{(S_j)}\bigr]^{N}\,\begin{pmatrix} e^{i N q_x r_q^{(S_j)}} & 0 \\ 0 & e^{-i N q_x r_q^{(S_j)}} \end{pmatrix}\,,
\label{eq:Mq_def}
\end{equation}
where
\begin{equation}
P_q^{(S_j)} \equiv M_{q-1}\,\begin{pmatrix} e^{-i q_x r_q^{(S_j)}} & 0 \\ 0 & e^{i q_x r_q^{(S_j)}} \end{pmatrix}\,,
\label{eq:Pmatrix}
\end{equation}
is the \emph{shifted} transfer matrix~\cite{griffiths2001waves}. At the inner levels where the accumulated phases $q_x\,r_q^{(S_j)}$ become very small for sufficiently large stage $S_i \gg 1$, $P_q^{(S_j)} \to M_{q-1}$, so $M_q \approx M_{q-1}^{\,N}$. In the regime of nearly perfect transmission, the level $(q-1)$ unit cell matrix is close to the identity, $M_{q-1} = \indicator + \epsilon_{q-1}$ with $\|\epsilon_{q-1}\| \ll 1$ (where $\|\cdot\|$ denotes any submultiplicative matrix norm), giving $M_{q-1}^{\,N} \approx \indicator + N\,\epsilon_{q-1}$, so that the off-diagonal element is amplified by a factor of $N$:
\begin{equation}
\bigl[M_{q-1}^{\,N}\bigr]_{12} \approx N\,[M_{q-1}]_{12}.
\label{eq:M12_amp}
\end{equation}
The Chebyshev identity for the $N$th power of a $2\times 2$ unimodular matrix gives~\cite{griffiths2001waves}
\begin{equation}
\bigl[(P_q^{(S_j)})^N\bigr]_{12}
= [P_q^{(S_j)}]_{12}\;U_{N-1}(\Gamma_q^{(S_j)}),
\label{eq:chebyshev_id}
\end{equation}
where $\Gamma_q^{(S_j)} = \tfrac{1}{2}\,\mathrm{tr}\bigl[P_q^{(S_j)}\bigr]$ is the Bloch-phase argument given by Eq.~\eqref{eq17}. Taking the limit $P_q^{(S_j)} \to M_{q-1}$ on both sides reduces Eq.~\eqref{eq:chebyshev_id} to
\begin{equation}
\bigl[M_{q-1}^{\,N}\bigr]_{12}
= [M_{q-1}]_{12}\;U_{N-1}(\Gamma_q^{(S_j)}).
\end{equation}
Comparing with Eq.~\eqref{eq:M12_amp} and noting that $[M_{q-1}]_{12} \ne 0$ in general, gives
\begin{equation}
U_{N-1}(\Gamma_q^{(S_j)}) \approx N,
\label{eq:UN_inner}
\end{equation}
and hence $\prod_{q=1}^{m} U_{N-1}^{2}(\Gamma_q^{(S_j)}) \approx N^{2m}$. Combining this with Eq.~\eqref{eq:prodsplit}, the Chebyshev product satisfies
\begin{equation}
\frac{\displaystyle\prod_{q=1}^{S_j}
U_{N-1}^{2}(\Gamma_q^{(S_j)})}
{\displaystyle\prod_{q=1}^{S_i}
U_{N-1}^{2}(\Gamma_q^{(S_i)})}
= N^{2(S_j - S_i)}.
\label{eq:prodratio}
\end{equation}
Each additional Cantor stage thus contributes a multiplicative factor $N^{2}$ to the product of CPs. Substituting Eqs.~\eqref{eq:M12scaling} and~\eqref{eq:prodratio} into Eq.~\eqref{eq:XSdef}, the ratio of $X_S$ values at the two stages reads
\begin{equation}
\frac{X_{S_j}(E)}{X_{S_i}(E)}
= \frac{1}{(N\lambda)^{2(S_j - S_i)}}
\cdot N^{2(S_j - S_i)}
= \frac{1}{\lambda^{2(S_j - S_i)}}.
\label{eq:Xratio}
\end{equation}
The factors $1/(N\lambda)^{2(S_j-S_i)}$ and $N^{2(S_j-S_i)}$ represent the shrinking of the individual barriers and the amplification of the off-diagonal transfer matrix element across the additional inner Cantor levels, respectively. The barriers thin faster than they multiply, leaving a net suppression of $X_S$ by $1/\lambda^{2}$ per Cantor stage. From Eq.~\eqref{eq:XSdef}, $T_S = (1+X_S)^{-1}$, so that $-\ln T_S = \ln(1+X_S) \approx X_S$ in the near-transparent regime $X_S \ll 1$. Taking the logarithm of Eq.~\eqref{eq:Xratio} yields the double-logarithmic scaling relation
\begin{equation}
\ln(-\ln T_{S_i}(E))
\approx \ln(-\ln T_{S_j}(E)) + \ln\alpha,
\label{eq:scaling_law}
\end{equation}
where
\begin{align}
\alpha &= \exp \bigl(\beta\,(S_j - S_i)\bigr),
\label{eq:alpha}\\
\beta &= 2\ln\lambda
= 2\ln\!\left(\frac{\rho}{\rho - (N-1)}\right).
\label{eq:beta} 
\end{align}
Equation~\eqref{eq:scaling_law} predicts that the double-logarithmic transmission curves at different Cantor stages collapse onto a single curve when shifted vertically by $\beta$ per stage. This scaling law shows that transport through polyadic Cantor electrostatic barriers exhibits a universal self-similar behavior, controlled by the scaling parameter $\rho$ and the polyadic order $N$.\\
\indent
Next, Fig.~\ref{fig:scalingcollapse} validates this analytical prediction. Plots~\ref{fig:scalingcollapse}(a) and~\ref{fig:scalingcollapse}(c) show $\ln(-\ln T_S)$ vs. Fermi energy $E$ in unscaled form for two Cantor stages $S_i$ and $S_j$: the curves exhibit similar qualitative features but remain clearly separated, reflecting the explicit stage dependence. Plots~\ref{fig:scalingcollapse}(b) and~\ref{fig:scalingcollapse}(d) display the same data after rescaling according to Eq.~\eqref{eq:scaling_law}, and the spectra collapse onto a common envelope over a wide energy range, confirming the predicted double-logarithmic scaling. The collapse is clearly observed once the system reaches the near-transparent regime, in which the leading-order scaling law holds. This regime is attained at lower Cantor stages for larger $N$, and at correspondingly higher stages for smaller $N$.\\
\indent
The scaling relation established above is the first of four scaling laws governing transport through graphene with polyadic Cantor barriers. The transmission probability depends not only on the Cantor stage but also on the potential height $V_0$, the initiator length $L$, and the angle of incidence $\phi$. The form of the transmission probability suggests that these parameters may enter the scaling behavior through additional rescaling factors. We therefore investigate, in the following subsections, the scaling relations associated with $V_0$, $L$, and $\phi$, and finally combine all four into a unified multi-parameter scaling law.\\
%
\paragraph{Scaling with potential heights.}
\begin{figure}[t]
\centering
\begin{overpic}[width=0.495\linewidth]{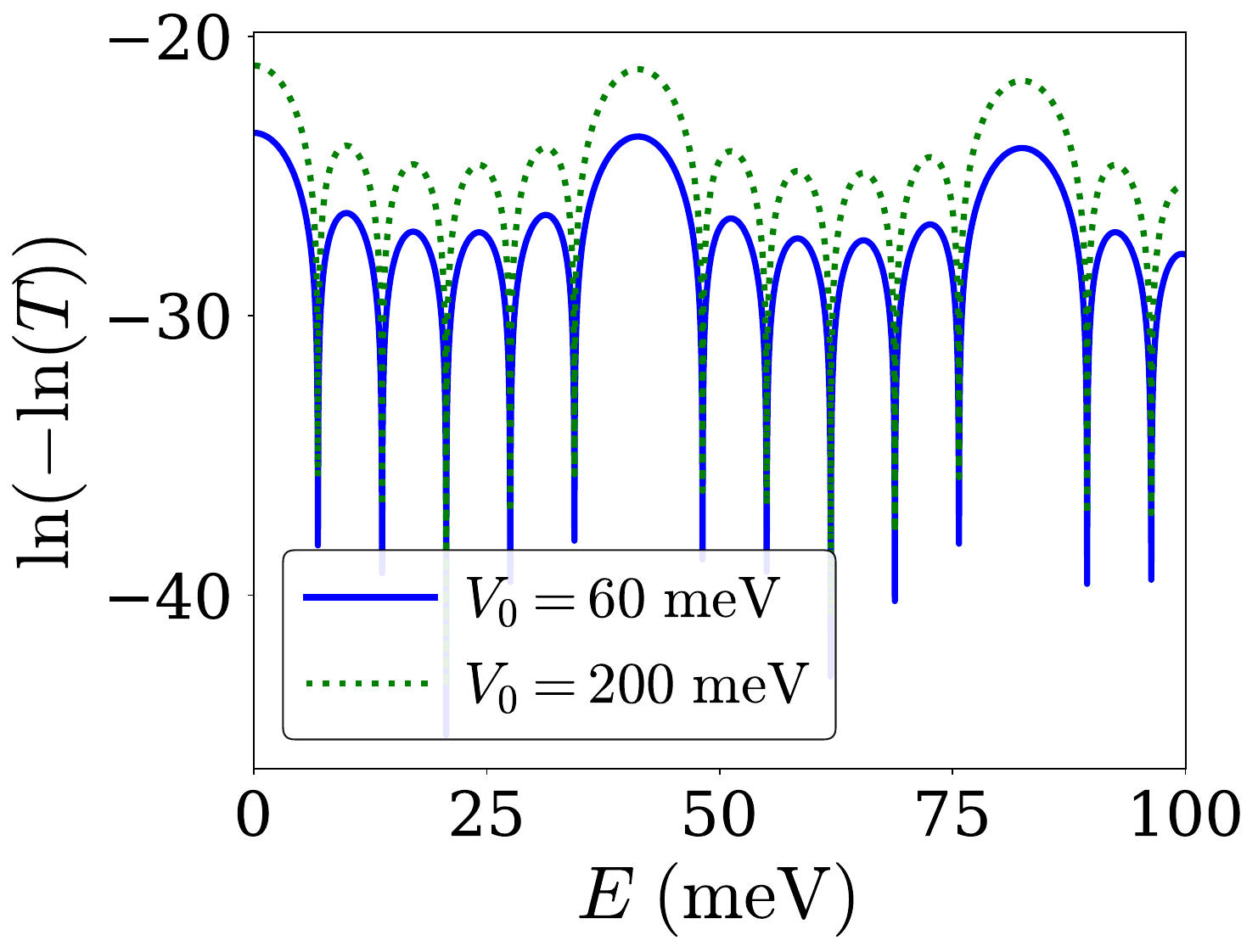}
\put(90,1.5){\bfseries (a)}
\end{overpic}%
\begin{overpic}[width=0.495\linewidth]{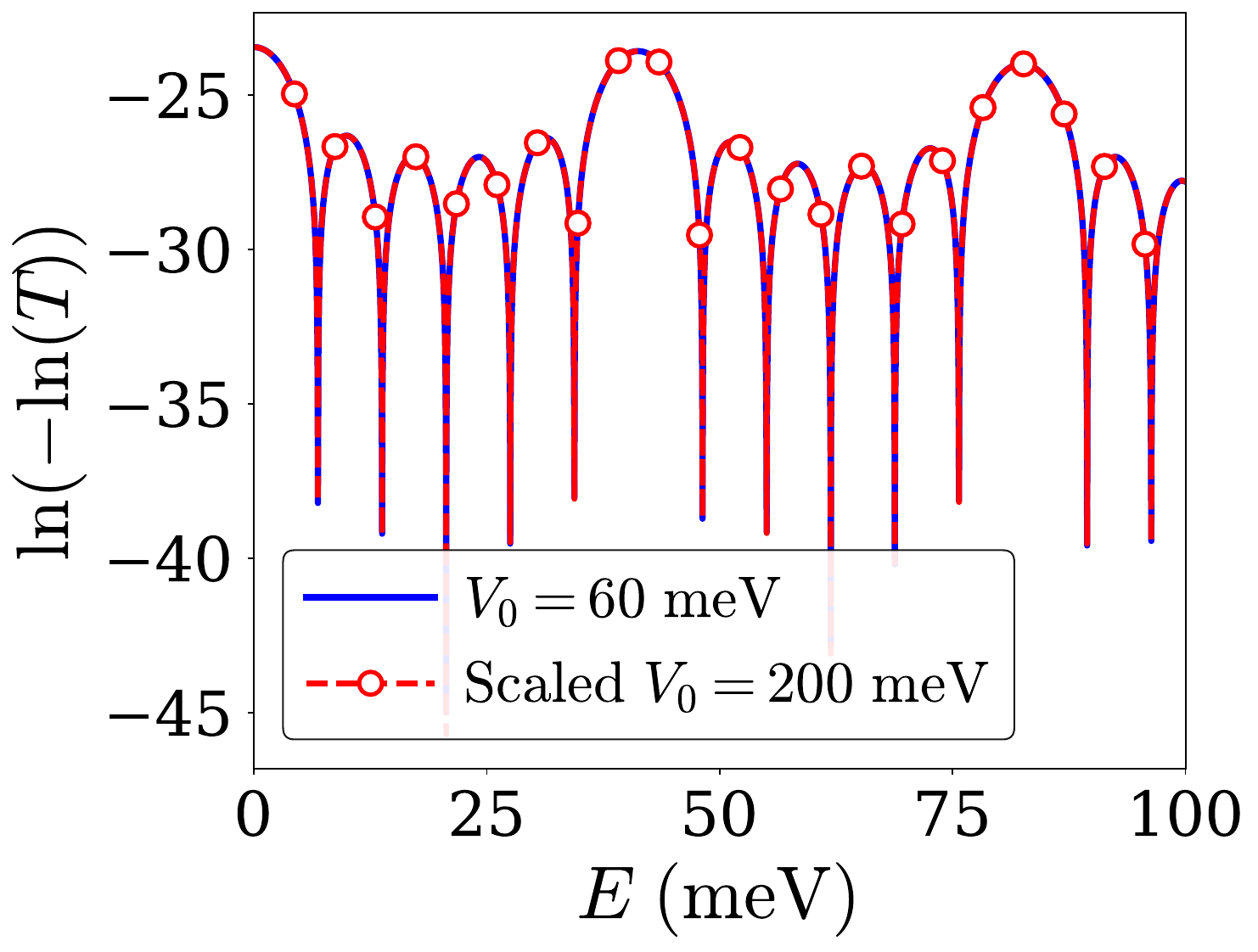}
\put(90,1.5){\bfseries (b)}
\end{overpic}
\caption{Double-logarithmic scaling of the transmission with the potential height $V_0$. Plots (a) and (b) correspond to $\phi=30^\circ$, $N=6$, $L=300$~nm, $S=9$, and $\rho=6.25$. Plot (a) shows the unscaled quantity $\ln(-\ln T_{V_0})$ as a function of the Fermi energy $E$ for two potential heights, $V_0=60$~meV and $V_0=200$~meV, for which the curves do not coincide. Plot (b) shows the same data after rescaling according to Eq.~\eqref{eq:Vscaling}, which produces a collapse of the curves onto a common envelope.}
\label{fig:Vscaling}
\end{figure}
We now derive the scaling behavior of the transmission probability with respect to the potential height $V_0$. Throughout the following scaling analysis, the Cantor stage $S$ and the polyadic order $N$ are held fixed at values for which the system is in the near-transparent regime. Also, the Fermi energy $E$, the initiator length $L$, and the angle of incidence $\phi$ are held fixed, while the potential height $V_0$ is varied.
The deviation from perfect transmission, $X_S = (1-T_S)/T_S$ as given in Eq.~\eqref{eq:XSdef}, depends on the potential height $V_0$ through the single-barrier element $|M_{12}^{(S)}|^{2}$ and the Chebyshev arguments $\Gamma_q^{(S)}$.\\
\indent
The Chebyshev argument $\Gamma_q^{(S)}$ at hierarchical level $q$ is the half-trace of the shifted transfer matrix, which combines the level $(q-1)$ unit cell with the accumulated phase $q_x r_q^{(S)}$ accumulated over the superperiodic distance. The wavevector $q_x$ outside the barrier and the superperiodic distance $r_q^{(S)}$ are both independent of the potential height. In the near-transparent regime the unit cell matrices are close to the identity, so the Chebyshev arguments reduce to $\Gamma_q^{(S)} \approx \cos(q_x r_q^{(S)})$ and are independent of $V_0$ to leading order. The product of CPs therefore does not contribute to the scaling with the potential height, and we write
\begin{equation}
\prod_{q=1}^{S} U_{N-1}^2(\Gamma_q^{(S)}) \equiv C_S,
\label{eq:CS_def}
\end{equation}
where $C_S$ is independent of $V_0$ at fixed Cantor stage. The entire dependence of $X_S$ on the potential height is thus carried by the single-barrier transfer matrix element and is given as (see Eq.~\eqref{m12})
\begin{equation}
|M_{12}^{(S)}|^{2}
= \tfrac{1}{4}\,\mathcal{B}^{2}\sec^{2}\theta\,
\sin^{2}(\widetilde{q}_x b_S).
\label{eq:M12_full}
\end{equation}
In the near-transparent regime the Cantor stage is large and the single-barrier width $b_S$ is small, so the phase $\widetilde{q}_x b_S$ is small and $\sin(\widetilde{q}_x b_S) \approx \widetilde{q}_x b_S$. Eq.~\eqref{eq:M12_full} then reduces to
\begin{equation}
|M_{12}^{(S)}|^{2} \approx
\frac{b_S^{2}}{4}\,
\bigl[\,\mathcal{B}\sec\theta\;\widetilde{q}_x\,\bigr]^{2}.
\label{eq:M12_smallphase}
\end{equation}
To evaluate the bracketed quantity, we introduce $\kappa$, the magnitude of the  Fermi wavevector inside the barrier, $\kappa \equiv \sqrt{q_y^{2} + \widetilde{q}_x^{2}}$. From Eq.~\eqref{vec_inside},
\begin{equation}
\kappa = \frac{|E - V_0|}{\hbar v_f},
\label{eq:kappa}
\end{equation}
so that $\sec\theta = \kappa/\widetilde{q}_x$ and $\sin\theta = q_y/\kappa$. Substituting $\sin\theta$ into $\mathcal{B} = 2\tan\phi - 2s\widetilde{s}\sin\theta\sec\phi$ and using $q_y = E\sin\phi/\hbar v_f$ gives
\begin{equation}
\mathcal{B} = 2\tan\phi
\left(1 - s\widetilde{s}\,\frac{E}{|E - V_0|}\right).
\label{eq:Bmid}
\end{equation}
For electron-like incidence, $E>0$ and using the sign identity $|E-V_0| = \widetilde{s}\,(E-V_0)$, the above equation reduces to
\begin{equation}
\mathcal{B} = -\,\widetilde{s}\,
\frac{2\,V_0\tan\phi}{|E - V_0|}.
\label{eq:Bclosed}
\end{equation}
The bracketed quantity in Eq.~\eqref{eq:M12_smallphase} then simplifies, since 
$\sec\theta\,\widetilde{q}_x = \kappa$, to
\begin{equation}
\mathcal{B}\sec\theta\;\widetilde{q}_x
= \mathcal{B}\,\kappa
= -\,\widetilde{s}\,\frac{2\tan\phi}{\hbar v_f}\;V_0\,.
\label{eq:Bqx_linear}
\end{equation}
The single-barrier scattering amplitude is therefore exactly linear in the potential height. Substituting Eq.~\eqref{eq:Bqx_linear} into Eq.~\eqref{eq:M12_smallphase} gives
\begin{equation}
|M_{12}^{(S)}|^{2} \approx
\frac{b_S^{2}\tan^{2}\phi}{\hbar^{2} v_f^{2}}\;V_0^{2}.
\label{eq:M12_V0sq}
\end{equation}
By combining the $V_0$-independent product of CPs, Eq.~\eqref{eq:CS_def}, with the quadratic dependence of the single-barrier element, Eq.~\eqref{eq:M12_V0sq}, the deviation from perfect transmission is found to obey
\begin{equation}
X_S(V_0) = C_S\,|M_{12}^{(S)}|^{2} \propto V_0^{2},
\label{eq:XS_V0sq}
\end{equation}
so that, for two potential heights $V_0$ and $V_0'$,
\begin{equation}
\frac{X_S(V_0)}{X_S(V_0')} = \frac{V_0^{2}}{V_0'^{\,2}}.
\label{eq:Xratio_V0}
\end{equation}
Since $-\ln T_S \approx X_S$ in the near-transparent regime, taking the logarithm of Eq.~\eqref{eq:Xratio_V0} yields the potential-height scaling law
\begin{equation}
\ln\,\bigl(-\ln T_{S}(V_0)\bigr)
\approx \ln\,\bigl(-\ln T_{S}(V_0')\bigr)
- 2\ln\,\left(\frac{V_0'}{V_0}\right).
\label{eq:Vscaling}
\end{equation}
This equation shows that changing the potential height shifts the double-logarithmic transmission curve vertically by $-2\ln(V_0'/V_0)$, while the resonance structure encoded in the product of CPs remains unchanged.\\
\indent
Figure~\ref{fig:Vscaling} confirms the analytical prediction presented above. Plot~\ref{fig:Vscaling}(a) shows $\ln(-\ln T_{S}(V_0))$ versus the Fermi energy $E$ for two potential heights, $V_0 = 60$~meV and $V_0 = 200$~meV, in unscaled form; here the curves share the same resonance structure but are vertically offset. Plot~\ref{fig:Vscaling}(b) shows the same data after rescaling according to Eq.~\eqref{eq:Vscaling}, where the curves collapse onto a common envelope over a wide energy range, confirming the predicted quadratic scaling with the potential height.\\
\paragraph{Scaling with initiator lengths.}
\begin{figure}[t]
\centering
\begin{overpic}[width=0.495\linewidth]{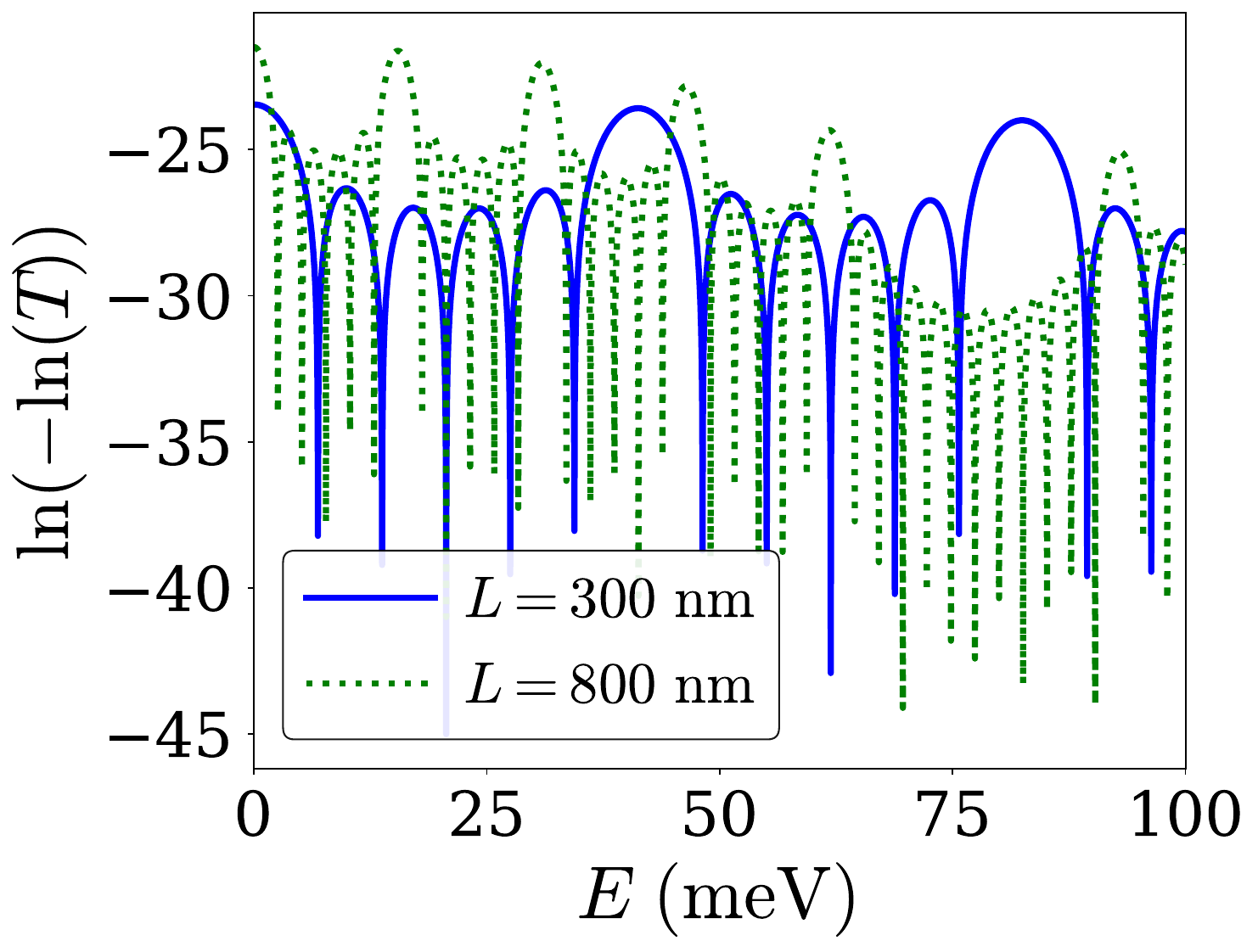}
\put(90,1.5){\bfseries (a)}
\end{overpic}%
\begin{overpic}[width=0.495\linewidth]{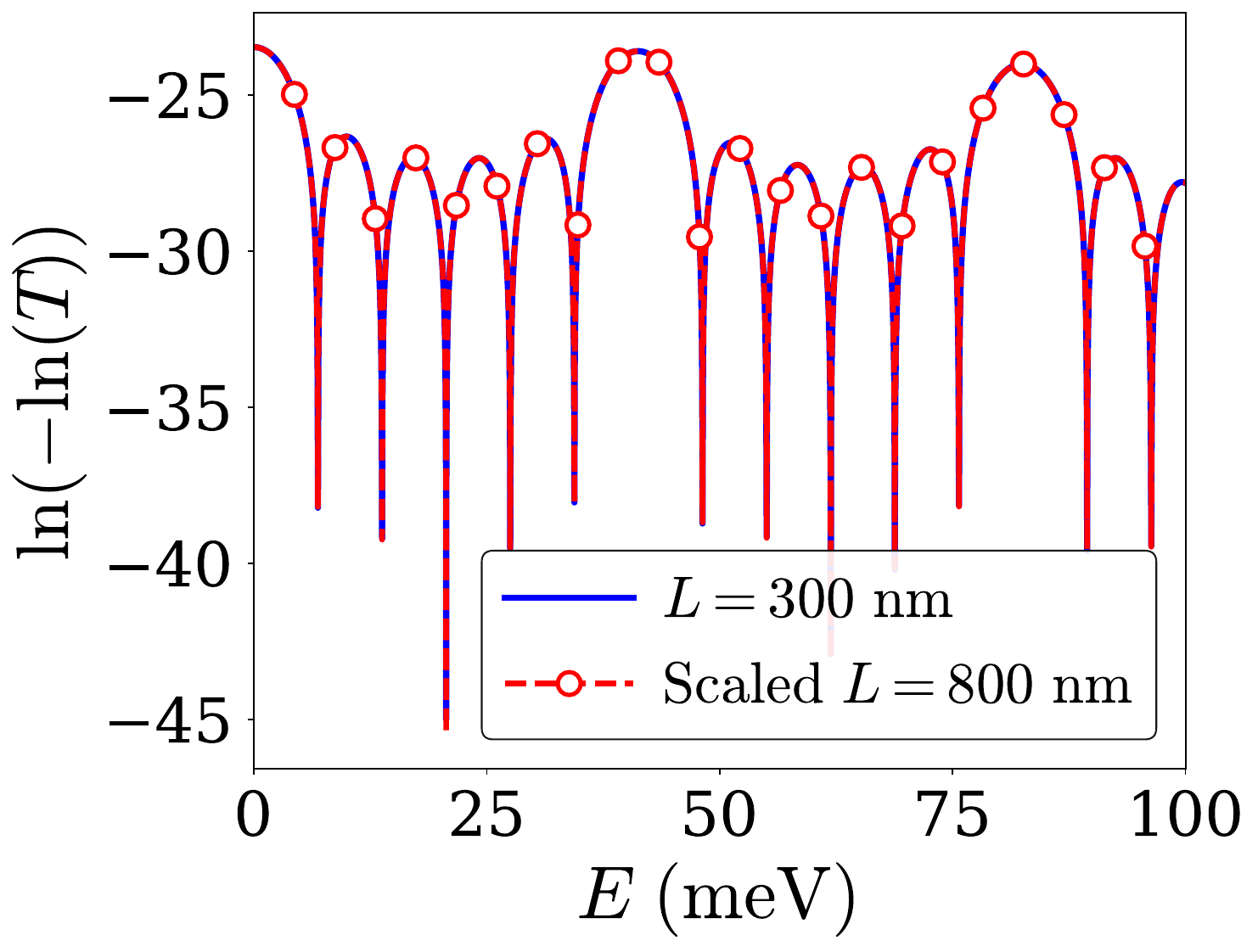}
\put(90,1.5){\bfseries (b)}
\end{overpic}
\caption{Double-logarithmic scaling of the transmission with the initiator length $L$. Plots (a) and (b) correspond to $\phi=30^\circ$, $N=6$, $V_0=60$~meV, $S=9$, and $\rho=6.25$. Plot (a) shows the unscaled quantity $\ln(-\ln T_L)$ as a function of the Fermi energy $E$ for two initiator lengths, $L=300$~nm and $L=800$~nm, for which the curves do not coincide. Plot (b) shows the same data after rescaling according to Eq.~\eqref{l_scaled}, which produces a collapse of the curves onto a common envelope.}
\label{fig:Lscaling}
\end{figure}
We now investigate the scaling behavior of the transmission probability with respect to the initiator length $L$. In this case, the potential height $V_0$ and the angle of incidence $\phi$ are held fixed, while the initiator length $L$ is varied.
From Eq.~\eqref{eq:XSdef}, it is evident that the deviation from perfect transmission $X_S$ depends on the initiator length through the single-barrier element $|M_{12}^{(S)}|^2$ and the Chebyshev arguments $\Gamma_q^{(S)}$. In contrast to the potential height, however, the initiator length enters both factors. The single-barrier (unit cell) element follows directly from Eq.~\eqref{eq:M12_V0sq}, in which $|M_{12}^{(S)}|^2$ is proportional to the square of the barrier width $b_S$. Since $b_S \propto L$, the single-barrier element scales quadratically with the initiator length,
\begin{equation}
|M_{12}^{(S)}|^2 \propto b_S^2 \propto L^2.
\label{eq:M12_Lsq}
\end{equation}
The Chebyshev arguments depend on the initiator length through the accumulated phase $q_x r_q^{(S)}$ and the superperiodic distance $r_q^{(S)}$. Since $q_x \propto E$ and $r_q^{(S)} \propto L$, this phase depends on $E$ and $L$ only through the product $E L$. A change of the initiator length $L \to L'$ therefore alters the Chebyshev arguments, unless it is accompanied by a rescaling of the Fermi energy $E \to E' = E\,(L/L')$ that holds the accumulated phase $q_x r_q^{(S)}$ fixed. Under this combined change, the product of CPs is unchanged,
\begin{equation}
\prod_{q=1}^{S} U_{N-1}^2(\Gamma_q^{(S)})\Big|_{L',E'}
= \prod_{q=1}^{S} U_{N-1}^2(\Gamma_q^{(S)})\Big|_{L,E}.
\label{eq:CS_Linv}
\end{equation}
Combining Eq.~\eqref{eq:M12_Lsq} with Eq.~\eqref{eq:CS_Linv}, the deviation from perfect transmission satisfies
\begin{equation}
\frac{X_S(L,E)}{X_S\bigl(L',E'\bigr)} = \frac{L^2}{L'^{\,2}},
\qquad E' = E\,\frac{L}{L'}.
\label{eq:Xratio_L}
\end{equation}
Since $-\ln T_S \approx X_S$ in the near-transparent regime, taking the logarithm of Eq.~\eqref{eq:Xratio_L} yields the parent-length scaling law
\begin{align}
\ln\bigl(-\ln T_S(L,E)\bigr)
&\approx \ln\left[-\ln T_S\left(L',E\,\frac{L}{L'}\right)\right]
\nonumber\\
&\quad - 2\ln\left(\frac{L'}{L}\right).
\label{l_scaled}
\end{align}
This equation shows that changing the initiator length shifts the double-logarithmic transmission curve vertically by $-2\ln(L'/L)$, provided the Fermi energy is rescaled by the inverse factor $L/L'$. The rescaling of the energy keeps the accumulated phase $q_x r_q^{(S)}$ fixed (since $q_x r_q^{(S)} \propto E L$), so that the resonance structure encoded in the product of CPs is preserved.\\
\indent
Figure~\ref{fig:Lscaling} confirms this analytical prediction. Plot~\ref{fig:Lscaling}(a) shows $\ln(-\ln T_L)$ as a function of the Fermi energy for two initiator lengths, $L=300$~nm and $L=800$~nm, before rescaling, where the corresponding curves do not coincide. Plot~\ref{fig:Lscaling}(b) displays the same data after rescaling according to Eq.~\eqref{l_scaled}, for which the curves collapse onto a common envelope over a wide energy range, confirming the predicted quadratic scaling with the initiator length.\\
%
%
\paragraph{Scaling with the angle of incidence.}
\begin{figure}[t]
\centering
\begin{overpic}[width=0.495\linewidth]{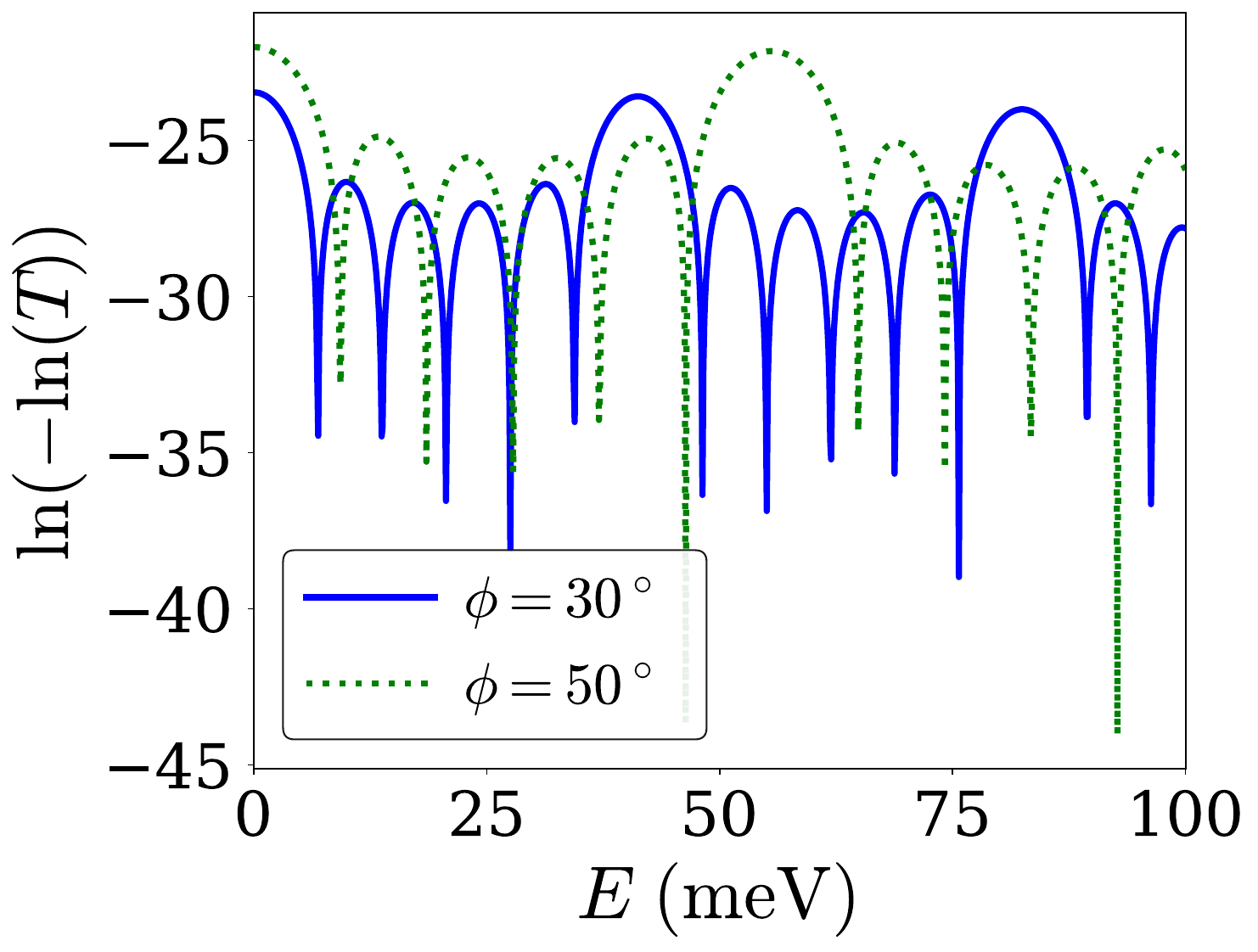}
\put(90,1.5){\bfseries (a)}
\end{overpic}%
\begin{overpic}[width=0.495\linewidth]{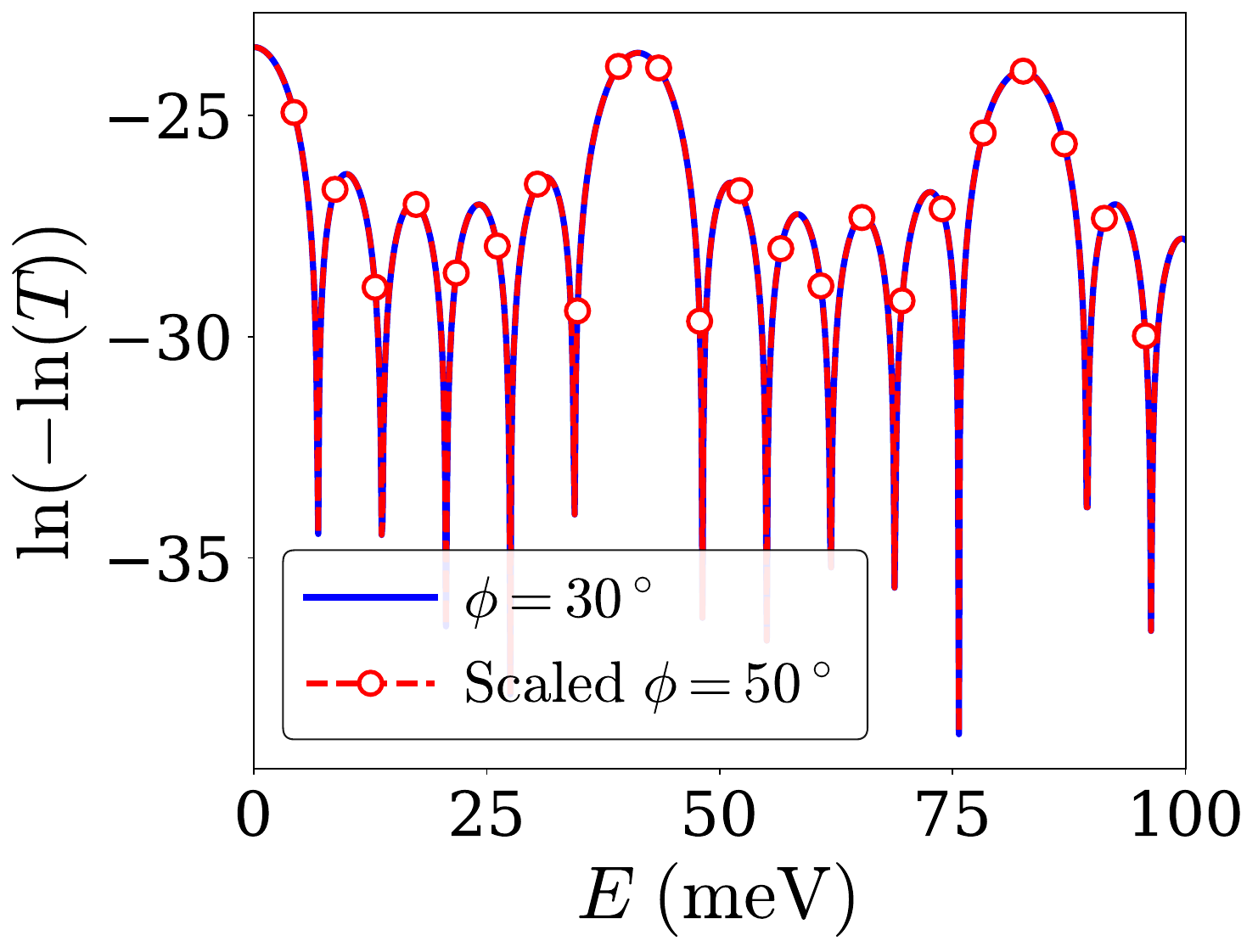}
\put(90,1.5){\bfseries (b)}
\end{overpic}
\caption{Double-logarithmic scaling of the transmission with the angle of incidence $\phi$. Plots (a) and (b) correspond to $L=300$~nm, $N=6$, $V_0=60$~meV, $S=9$, and $\rho=6.25$. Plot (a) shows the unscaled quantity $\ln(-\ln T_\phi)$ as a function of the Fermi energy $E$ for two angles of incidence, $\phi=30^\circ$ and $\phi=50^\circ$, for which the curves do not coincide. Plot (b) shows the same data after rescaling according to Eq.~\eqref{eq:phi_scaling}, which produces a collapse of the curves onto a common envelope.}
\label{fig:phiscaling}
\end{figure}
We finally investigate the scaling behavior with respect to the angle of incidence $\phi$, holding the potential height $V_0$ and the initiator length $L$ fixed. The analysis parallels that of the initiator length: the angle of incidence enters both the single-barrier element $|M_{12}^{(S)}|^2$ and the Chebyshev arguments $\Gamma_q^{(S)}$, and a compensating rescaling of the Fermi energy is again required. From Eq.~\eqref{eq:M12_V0sq}, it follows that the single-barrier element exhibits the scaling behavior $|M_{12}^{(S)}|^2 \propto \tan^2\phi$ at fixed Fermi energy. The Chebyshev arguments depend on $\phi$ through the accumulated phase $q_x r_q^{(S)}$, with $q_x = q_f\cos\phi$ and $q_f = E/\hbar v_f$; since $r_q^{(S)} \propto L$, this phase depends on $E$, $L$, and $\phi$ only through the combination $E L\cos\phi$. A change $\phi \to \phi'$ therefore leaves the product of CPs unchanged provided the Fermi energy is rescaled as
\begin{equation}
E' = E\,\frac{\cos\phi}{\cos\phi'},
\label{eq:Ephi_scaling}
\end{equation}
which keeps $E L\cos\phi$ fixed. Under this combined change, the variation of $X_S$ comes entirely from the single-barrier element, and the deviation from perfect transmission satisfies
\begin{equation}
\frac{X_S(\phi,E)}{X_S(\phi',E')} 
= \frac{\tan^2\phi}{\tan^2\phi'},
\qquad E' = E\,\frac{\cos\phi}{\cos\phi'}.
\label{eq:Xratio_phi}
\end{equation}
By taking the logarithm of Eq.~\eqref{eq:Xratio_phi}, one obtains the incident-angle scaling law as
\begin{align}
\ln\bigl(-\ln T_S(\phi,E)\bigr)
&\approx \ln\left[-\ln T_{S}\!\left(\phi',
E\,\frac{\cos\phi}{\cos\phi'}\right)\right] \notag\\
&\quad - 2\ln\left(\frac{\tan\phi'}{\tan\phi}\right).
\label{eq:phi_scaling}
\end{align}
Changing the angle of incidence thus shifts the double-logarithmic transmission curve vertically by $-2\ln(\tan\phi'/\tan\phi)$, while the energy rescaling preserves the resonance structure. As $\phi \to 0$ the factor $\tan\phi$ vanishes and the scaling law breaks down, consistent with the perfect transmission of normally incident Dirac electrons. Figure~\ref{fig:phiscaling} validates this analytical observation. The unscaled curves for $\phi = 30^\circ$ and $\phi = 50^\circ$ do not coincide as shown in Fig.~\ref{fig:phiscaling}(a), whereas after rescaling according to Eq.~\eqref{eq:phi_scaling} they collapse onto a common envelope as illustrated in Fig.~\ref{fig:phiscaling}(b).\\
\begin{figure}[htb]
\centering
\begin{overpic}[width=0.495\linewidth]{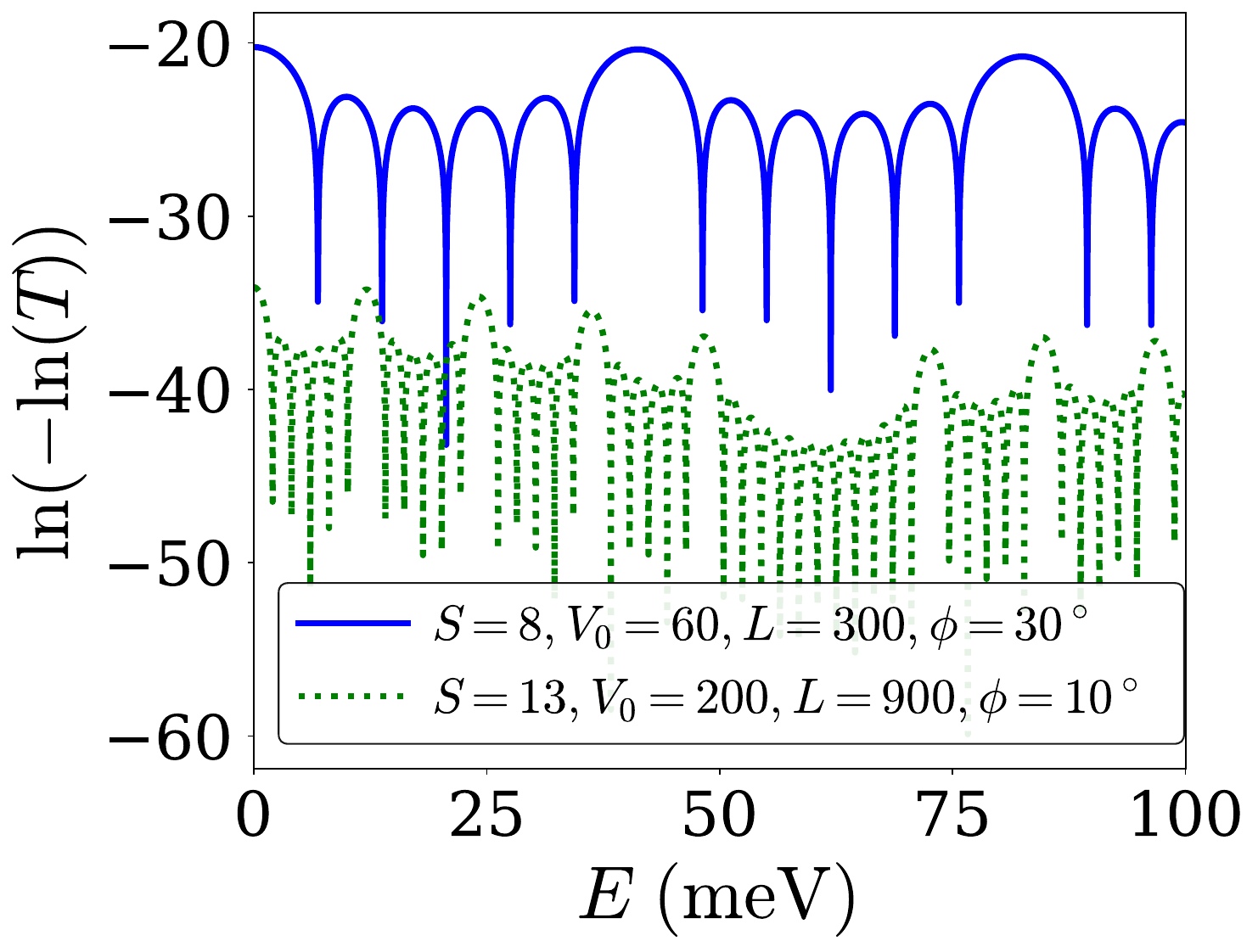}
\put(90,1.5){\bfseries (a)}
\end{overpic}%
\begin{overpic}[width=0.495\linewidth]{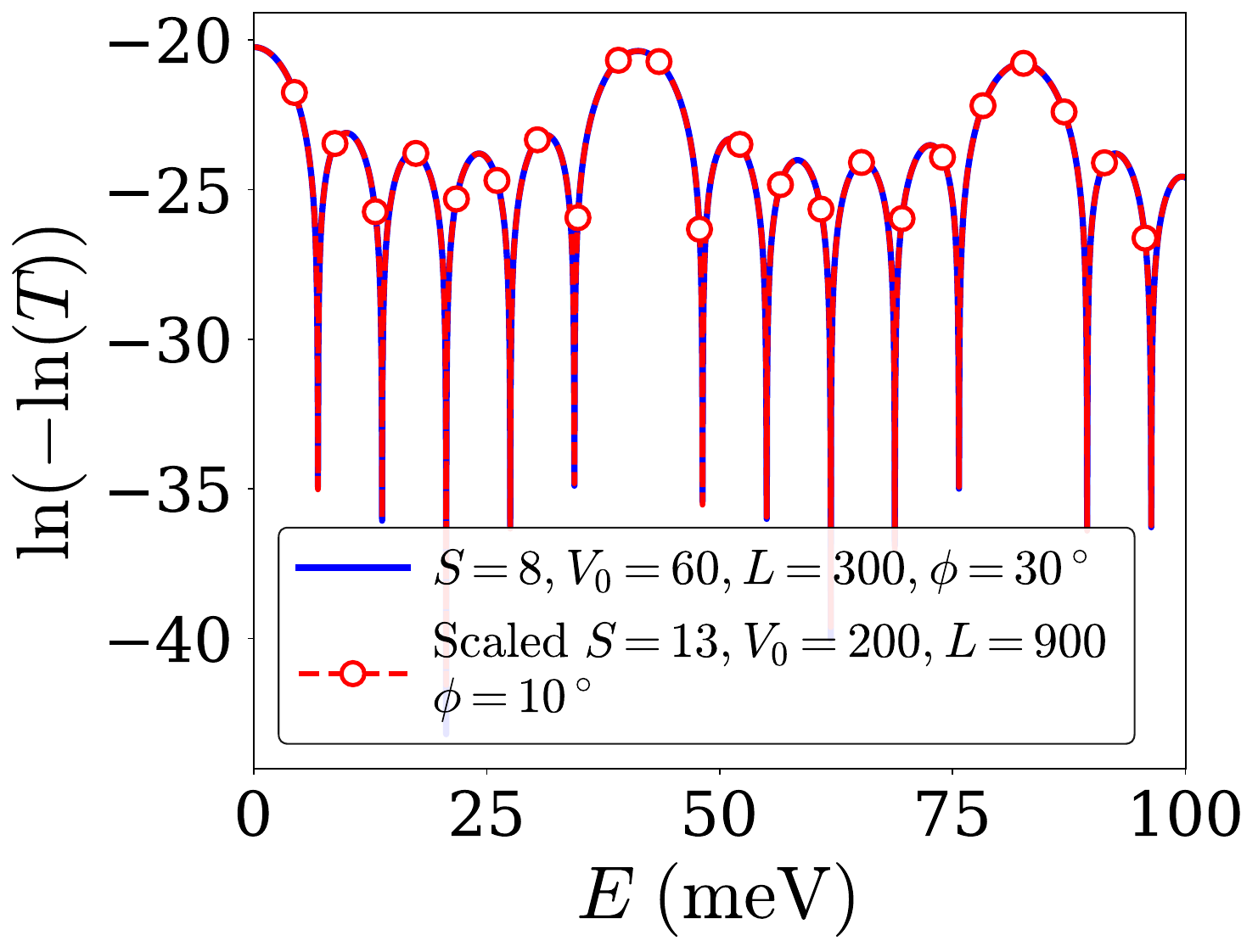}
\put(90,1.5){\bfseries (b)}
\end{overpic}
\caption{%
Unified double-logarithmic scaling of the transmission. Plots (a) and (b) correspond to $N=6$ and $\rho=6.25$. Plot (a) shows the unscaled quantity $\ln(-\ln T)$ as a function of the Fermi energy $E$ for two parameter sets differing in the Cantor stage, the potential height, the initiator length, and the angle of incidence, for which the curves do not coincide. Plot (b) shows the same data after rescaling according to Eq.~\eqref{eq:unified_scaling}, which produces a collapse of the curves onto a common envelope.
}
\label{fig:unified}
\end{figure}
\paragraph{Unified multi-parameter scaling law.}
We now combine the four scaling laws derived above into a single multi-parameter scaling law. We consider two systems that differ simultaneously in the Cantor stage, the potential height, the initiator length, and the angle of incidence, and relate their transmission in the near-transparent regime. As evident from Eq.~\eqref{eq:XSdef}, the deviation from perfect transmission $X_S$ is governed by the unit-cell transfer matrix element $|M_{12}^{(S)}|^2$ and the product of CPs. The transfer-matrix element collects the power-law dependencies established above. From Eqs.~\eqref{eq:M12scaling},~\eqref{eq:M12_V0sq}, and~\eqref{eq:M12_Lsq}, it scales as
\begin{equation}
|M_{12}^{(S)}|^2 \propto 
\frac{V_0^{2}\,L^{2}\tan^{2}\phi}{(N\lambda)^{2S}}.
\label{eq:M12_unified}
\end{equation}
The dependence of the product of CPs on $L$, $E$, and $\phi$ enters only through the accumulated phase $q_x r_q^{(S)} \propto EL\cos\phi$ and can therefore be removed by rescaling the Fermi energy so that this phase is held fixed,
\begin{equation}
E L\cos\phi = E' L'\cos\phi'.
\label{eq:phase_inv}
\end{equation}
The dependence on the Cantor stage, however, cannot be removed by rescaling. As established in Eq.~\eqref{eq:prodratio}, each additional stage contributes a factor $N^2$ to the product of CPs, yielding an overall factor $N^{2(S_j-S_i)}$ between the two systems. Combining this with Eq.~\eqref{eq:M12_unified}, the deviation from perfect transmission satisfies
\begin{equation}
\frac{X_{S_i}(V_0,L,\phi;E)}{X_{S_j}(V_0',L',\phi';E')}
= \frac{(N\lambda)^{2(S_j-S_i)}}{N^{2(S_j-S_i)}}\,
\frac{V_0^{2}}{V_0'^{\,2}}\,
\frac{L^{2}}{L'^{\,2}}\,
\frac{\tan^{2}\phi}{\tan^{2}\phi'},
\label{eq:Xratio_unified}
\end{equation}
with $E'$ given by Eq.~\eqref{eq:phase_inv}. 
By taking the logarithm of the above equation, one obtains the unified multi-parameter scaling law as
\begin{widetext}
\begin{equation}
\begin{aligned}
\ln\bigl(-\ln T_{S_i}(V_0,L,\phi;E)\bigr) \approx\;&
\ln\bigl(-\ln T_{S_j}(V_0',L',\phi';E')\bigr) \\
&+ \beta\,(S_j-S_i)
- 2\ln\!\left(\frac{V_0'}{V_0}\right)
- 2\ln\!\left(\frac{L'}{L}\right)
- 2\ln\!\left(\frac{\tan\phi'}{\tan\phi}\right),
\end{aligned}
\label{eq:unified_scaling}
\end{equation}
\end{widetext}
where $\beta$ is given by Eq.~\eqref{eq:beta} and the energy is rescaled according to Eq.~\eqref{eq:phase_inv}. Each term in Eq.~\eqref{eq:unified_scaling} reproduces one of the single-parameter scaling laws derived above, namely the scaling with the Cantor stage (see Eq.~\eqref{eq:scaling_law}), potential height (see Eq.~\eqref{eq:Vscaling}), initiator length (see Eq.~\eqref{l_scaled}), and angle of incidence (see Eq.~\eqref{eq:phi_scaling}). The stage term carries the opposite sign to the other three, since increasing the Cantor stage drives the system toward perfect transmission, whereas increasing the potential height, the initiator length, or the angle of incidence increases the deviation from it.\\
\indent
Figure~\ref{fig:unified} confirms the multi-parameter scaling law. Figure~\ref{fig:unified}(a) shows $\ln(-\ln T)$ for two parameter sets differing in all four control parameters, in unscaled form, where the curves do not coincide. Figure~\ref{fig:unified}(b) shows the same data after rescaling according to Eq.~\eqref{eq:unified_scaling}; the curves collapse onto a common envelope over a wide energy range, confirming that the four scaling laws combine into a single multi-parameter law. Quantum transport through graphene subjected to PCP is therefore governed by a unified self-similar mechanism: the recursive Cantor geometry determines the resonance structure through the accumulated phase $q_x r_q^{(S)}$, while the Cantor stage, the potential height, the initiator length, and the angle of incidence control the scaling of the transmission spectrum.
%
%
\section{Summary and Conclusion}
\label{section_04}
In this work, we have systematically investigated the transmission of Dirac electrons in monolayer graphene subjected to a polyadic Cantor-structured electrostatic barrier, addressing both the transmission characteristics and the scaling behavior that emerges from the hierarchical geometry of the potential. Using the SPP formalism~\cite{hasan2018super}, we obtained a closed-form expression for the transmission probability. The transmission spectra in the $(E,\rho)$ and $(\phi,\rho)$ planes show how the recursive Cantor construction shapes quantum transport. With increasing Cantor stage, the $(E,\rho)$ maps evolve from a sparse superlattice resonance structure at $S=1$, with a broad suppressed window near $E=V_0$, to a near-transparent background punctuated by narrow interference dips at higher stages (Figs.~\ref{fig:01} and~\ref{fig:01n_5}). The angular response (Fig.~\ref{fig:02}) exhibits three distinct regimes at $S=1$: transmission over a comparatively wide angular window in the $n$--$p$--$n$ configuration, a much narrower window in the $n$--$n$--$n$ case, and, at the Dirac point, transmission confined to a narrow angular window by evanescent-mode tunneling. These distinct angular signatures progressively disappear as the Cantor stage increases and the recursive geometry comes to dominate the transport. Throughout, the polyadic order $N$ controls the rate at which the near-transparent regime is reached, without altering the overall character of the spectra.\\
\indent
A central result of this work is the identification of robust double-logarithmic scaling laws for the transmission. We showed that the quantity $\ln(-\ln T)$ obeys a simple scaling relation with respect to each of four independent control parameters: the Cantor stage $S$, the potential height $V_0$, the initiator length $L$, and the angle of incidence $\phi$ (Figs.~\ref{fig:scalingcollapse},~\ref{fig:Vscaling},~\ref{fig:Lscaling}, and~\ref{fig:phiscaling}). In each case the deviation from perfect transmission factorizes into a single-barrier transfer matrix element and a product of CPs, and the parameter enters as a multiplicative rescaling of the transmission envelope while the resonance structure remains unchanged. For the initiator length and the angle of incidence, this rescaling is accompanied by a compensating rescaling of the Fermi energy that holds the accumulated phase $q_x r_q^{(S)} \propto E L \cos\phi$ fixed.\\
\indent
Combining the individual scaling results, we derived a unified multi-parameter scaling law [Eq.~\eqref{eq:unified_scaling}] that relates two systems differing simultaneously in all four parameters (Fig.~\ref{fig:unified}). The transmission is controlled by the combination $EL\cos\phi$, which fixes the resonance structure, such that the four control parameters enter only through multiplicative rescaling factors. The unified law reduces to each single-parameter scaling law when only one parameter is varied while the other three are held fixed, and provides a compact and predictive description of transport through graphene with polyadic Cantor systems.\\
\indent
Our findings show that the hierarchical self-similarity of the potential plays a central role in shaping quantum transport through fractal-inspired electrostatic barriers. The scaling laws are governed by the polyadic order $N$ and the scaling parameter $\rho$ through the geometric factor $\lambda = \rho/(\rho-(N-1))$, which directly connects the fractal geometry to the transport behavior. These results may also be relevant for graphene-based nanoelectronic devices~\cite{berger2004ultrathin}, where polyadic Cantor structures could be exploited for tunable energy filtering, angular selectivity, and scalable control of transmission.\\
\indent
The present framework may be extended to include the effects of disorder, magnetic fields~\cite{sun2010transport, rodriguez2017transmission}, strain~\cite{wang2021global, shekhar2025klein}, or interactions, as well as to other Dirac and Weyl materials. The unified scaling approach developed here provides a natural starting point for such generalizations and for the design of graphene-based devices employing hierarchical and fractal-inspired potentials.
%
\begin{acknowledgments}
S.S. acknowledges NBCFDC, New Delhi, India, for the JRF fellowship. MU sincerely acknowledges Prof. P. Senthilkumaran, Optics and Photonics Centre (OPC), Indian Institute of Technology Delhi (IIT Delhi), New Delhi, for his constant encouragement, valuable guidance, and support in fostering research activities. BPM acknowledges the PDF research grant for faculty member under the IoE Scheme for the year 2025--26 of Banaras Hindu University. 
\end{acknowledgments}
\bibliography{References}
\end{document}